%% file: newEDE_PRD.tex
\documentclass[aps,showpacs,floatfix,preprintnumbers,nofootinbib,11pt,prd]{revtex4-2}

\usepackage{graphicx}
\usepackage{color}
\usepackage{natbib}
\usepackage{multirow}
\usepackage{mathtools}
\RequirePackage{hyperref}
\usepackage{epsfig}
\usepackage{graphicx}
\usepackage[justification=centerlast, singlelinecheck=false, caption=false]{subfig}
\usepackage{amsfonts}
\usepackage{amssymb}
\usepackage{latexsym}
\usepackage{amsmath}

\def\ee{\end{equation}}
\def\ba{\begin{eqnarray}}
\def\ea{\end{eqnarray}}

\def\bq{\begin{quote}}
\def\eq{\end{quote}}

 at 10truept

\newcommand{\kmsMpc}{\, \text{km}\,\text{s}^{-1}\, \text{Mpc}^{-1}}

\newcommand{\beq}{\begin{equation}}
\newcommand{\eeq}{\end{equation}}
\newcommand{\beqa}{\begin{eqnarray}}
\newcommand{\eeqa}{\end{eqnarray}}
\newcommand{\bea}{\begin{eqnarray}}
\newcommand{\eea}{\end{eqnarray}}

 \newcommand{\be}{\beta}

\def\lesssim{~\mbox{\raisebox{-.6ex}{$\stackrel{<}{\sim}$}}~}

\def\ltap{\ \raise.3ex\hbox{$<$\kern-.75em\lower1ex\hbox{$\sim$}}\ }
\def\gtap{\ \raise.3ex\hbox{$>$\kern-.75em\lower1ex\hbox{$\sim$}}\ }
\def\gl{\ \raise.5ex\hbox{$>$}\kern-.8em\lower.5ex\hbox{$<$}\ }
\def\roughly#1{\raise.3ex\hbox{$#1$\kern-.75em\lower1ex\hbox{$\sim$}}}

\newcommand\CLASS{{\tt CLASS}}

\begin{document}

\title{Resolving the Hubble Tension with New Early Dark Energy} 
\author{Florian Niedermann}
\email{niedermann@cp3.sdu.dk}
\author{Martin S. Sloth} 
\email{sloth@cp3.sdu.dk}
\affiliation{CP$^3$-Origins, Center for Cosmology and Particle Physics Phenomenology \\ University of Southern Denmark, Campusvej 55, 5230 Odense M, Denmark}

\preprint{\tt CP3-Origins-2020-09}

\begin{abstract}

New Early Dark Energy (NEDE) is a component of vacuum energy at the electron volt scale, which decays in a first-order phase transition shortly before recombination~\cite{Niedermann:2019olb}. The NEDE component has the potential to resolve the tension between recent local measurements of the expansion rate of the Universe using supernovae (SN) data and the expansion rate inferred from the early Universe through measurements of the cosmic microwave background (CMB) when assuming $\Lambda$CDM. We discuss in depth the two-scalar field model of the NEDE phase transition including the process of bubble percolation, collision, and coalescence. We also estimate the gravitational wave signal produced during the collision phase and argue that it can be searched for using pulsar timing arrays. In a second step, we construct an effective cosmological model, which describes the phase transition as an instantaneous process, and derive the covariant equations that match perturbations across the transition surface. Fitting the cosmological model to CMB, baryonic acoustic oscillations and SN data, we report $H_0 = 69.6^{+1.0}_{-1.3} \, \kmsMpc$ $(68 \%$  C.L.) without the local measurement of the Hubble parameter, bringing the tension down to $2.5\, \sigma$. Including  the local input, we find $H_0 = 71.4 \pm 1.0 \, \kmsMpc$ $(68 \%$  C.L.) and strong evidence for a non-vanishing NEDE component with {a~$\simeq 4\, \sigma$ significance}.

 \end{abstract}

\maketitle

\tableofcontents

\section{Introduction}
\label{sec:intro}
Over the past decades our cosmological standard model has been put to the test with ever-increasing precision. Thanks to measurements of the cosmic microwave background (CMB) its parameters are now constrained at the percent level and provide us with a consistent description of our Universe starting shortly after the Big Bang and ending with a phase of accelerated expansion where length scales are stretched at an exponential rate. It is a remarkably successful model because it resolves physics between extreme high and low curvature regimes with only a small handful of parameters. One of them is the Hubble parameter, $H_0$, denoting the present expansion rate of space. Recently, its measurement has given rise to a profound observational and theoretical crisis that is threatening the success of the $\Lambda$CDM model. In this paper (and a recent companion letter~\cite{Niedermann:2019olb}), we propose a  novel and simple resolution to the Hubble crisis in terms of a first-order phase transition that takes place in a dark sector before recombination. 

\subsection{The Hubble Tension}

The problem arises due to two competing sets of measurements that infer different values of $H_0$. First, there are \textit{early-time} measurements. The most precise value comes from the European Space Agency satellite, {\sl Planck}, that measures the temperature fluctuations, polarization and lensing in the CMB radiation, with the newest data release finding~\cite{Aghanim:2018eyx} $H_0 = 67.36 \pm 0.54 \,\kmsMpc$.\footnote{For this value the {\sl Planck} TT, TE, EE and lensing dataset have been combined. {Here and henceforth, if not stated otherwise, the uncertainties are at $68 \%$  C.L.}}  This observation depends on the expansion history before \textit{and} after recombination and, thus, is highly sensitive to our assumed cosmological model. For example, increasing $H_0$ while keeping all other $\Lambda$CDM parameters fixed will shift the whole expansion history to higher values of {$H(z)$} and, therefore, affects not only the photon distribution in the primordial plasma but also the free-streaming of photons towards us after they have decoupled from the plasma. Another example of an early-time measurement is provided by observations of baryonic acoustic oscillations (BAO), which are sensitive to physics \textit{before} recombination (as well as $H_0$ directly) as they {depend on} the sound horizon that has been imprinted in the visible matter distribution briefly before photons decouple. 

On the other hand, there is a qualitatively different set of measurements that is operative at \textit{late times} in a low-redshift patch around our position in space. These \textit{local} observations are almost entirely insensitive to changes in the expansion history and, hence, our cosmological model~\cite{Dhawan:2020xmp}. They use methods like the cosmic distance ladder to reach out to redshifts of about $z \simeq 0.4$. For example,  the SH$_0$ES Collaboration~\cite{Riess:2019cxk} measured  $H_0 = 74.03 \pm 1.42 \, \kmsMpc$ by using Type Ia supernovae (SN) calibrated through Cepheids as standard candles. Comparing both measurements yields a difference which is $4.4$ $\sigma$ in significance. The H$0$LiCoW experiment, which uses strong-lensing time delays, finds ~\cite{Wong:2019kwg}  $H_0 = 73.3^{+1.7}_{-1.8} \, \kmsMpc$, corresponding to a similar discrepancy. When combined with SH$0$ES it {raises} the tension with the early-time measurement above the $5$ $\sigma$ level~\cite{Verde:2019ivm}. A special role is played by the analysis in~\cite{Freedman:2019jwv,Freedman:2020dne}. Unlike SH$_0$ES, it uses the tip of the red giant branch method (rather than Cepheids) to calibrate the distance ladder, and reports a value which is still compatible with both early- and late-time probes, $H_0 = 69.6 \pm 1.9 \,\kmsMpc$ (see, however,~\cite{Yuan:2019npk}). There are many other measurements which up to this point have larger uncertainties, though. For example, interferometric and spectroscopic observations of water megamasers in accretion disks of active galactic nuclei yield~\cite{Pesce:2020xfe} $H_0 = 73.9 \pm 3 \,\kmsMpc$, and gravitational wave measurements give rise to~\cite{Abbott:2019yzh} $H_0 = 68^{+14}_{-7} \,\kmsMpc$. The fact, however, remains, there is a statistically significant tension between direct late-time and $\Lambda$CDM-inferred early-time measurements of $H_0$, which {is also} known as the Hubble tension (for reviews see \cite{Knox:2019rjx,Bernal:2016gxb}). 

Of course, a frequent concern is that the local probes have systematic errors that are hard to control as they rely on {notoriously difficult} distance measurements {in our small-$z$ neighborhood}. Many potential sources for errors have been studied in the past. For example, in \cite{Davis:2019wet}, it was found that redshift uncertainties cannot account for the discrepancy, and, in \cite{Wojtak:2013gda, Odderskov:2017ivg,Wu:2017fpr}, a similar conclusion was drawn with regard to the local Hubble flow. In \cite{Rigault:2018ffm,Rigault:2014kaa} the Cepheid calibration of the SH$0$ES Collaboration was criticized; however, these corrections have been argued  not to be sufficient to resolve the tension in later work \cite{Rose:2019ncv,Jones:2018vbn}. In the meanwhile, the Cepheid discussion in the observational community continues with vigor~\cite{Shanks:2018rka,Riess:2018kzi,Breuval:2019anj}. {The hope, of course, is that this controversy will be resolved once other local measurements, like the ones based on gravitational waves, will have reduced their uncertainties.}

The upshot we can take home from this debate is that systematic errors might still have a say in the matter; however, as of yet there is no clear indication that they could resolve the tension. We believe this makes it imperative from a theory standpoint to also think about alternative ways of addressing the problem. The crucial observation is that the CMB and BAO measurement have a strong dependence on our cosmological model {through their dependence on acoustic oscillations in the photon-baryon plasma}. This means that by changing the expansion history we can hope to be able to {raise the BAO and CMB inferred} value up to the locally measured one, which itself remains unaffected by the change (see, for example, \cite{Bernal:2016gxb,Mortsell:2018mfj}). 

This reasoning has already set in motion a lot of activity in the theory community, and different proposals can be divided in two classes. First there are attempts to modify the background evolution at a late stage after recombination. This is usually done by modifying the expansion history at low redshift by introducing an \textit{ad hoc} modification of $H(z)$, for example parametrized in terms of a spline model~\cite{Bernal:2016gxb,Keeley:2019esp,Raveri:2019mxg}. However, so far all of these attempts have been plagued by creating a tension with BAO measurements (or by introducing a lot of free parameters)~\cite{Bernal:2016gxb,Aylor:2018drw,Verde:2016ccp}. The sound horizon $r_s$, which characterizes oscillations in the primordial plasma before recombination, is key to understanding this failure~\cite{Knox:2019rjx,Arendse:2019itb}: BAO measurements are anchored at early times through their dependence on~$r_s$. Used in combination with the {\sl Pantheon} dataset they constrain the combination $r_s H_0$ and, therefore, require us to lower $r_s$ if we want to increase $H_0$ in order to have agreement with the local measurement. However, if we use $\Lambda$CDM to describe early-time physics, lowering the sound horizon $r_s$ has been found to be incompatible with raising $H_0$, as this would lead to {an unacceptable deformation of the} CMB power spectrum~\cite{Knox:2019rjx}.
In fact, the authors in~\cite{Aylor:2018drw} argued that any solution to the Hubble tension has most likely to be operative \textit{before} recombination to avoid this BAO trap. This, of course, cannot exclude the possibility that late-time modifications play a partial role in resolving the tension. Also, as pointed out in \cite{Krishnan:2020obg}, new low-redshift data might shed further light on the relevance of late-time approaches.

This brings us to the second direction which attempts to modify  $\Lambda$CDM at early times. At first sight, this seems to be the least favorable option, as the CMB is a very clean probe. After all, changing the dynamics in the primordial plasma is expected to directly influence the CMB power spectrum, which is highly constrained by {\sl Planck} data. So it is quite intriguing that {a} promising resolution of the Hubble tension  modifies the $\Lambda$CDM model exactly where we would have thought it is constrained the most: closely to matter-radiation equality where the composition of the energy budget affects the relative and absolute heights of the first peaks in the power spectrum. This is done by introducing an \textit{early dark energy} (EDE) component~\cite{Poulin:2018dzj,Poulin:2018cxd,Smith:2019ihp,Lin:2019qug,Kaloper:2019lpl,Alexander:2019rsc,Hardy:2019apu,Knox:2019rjx,Sakstein:2019fmf,1798362} which contributes a sizable fraction to the energy budget around matter-radiation equality, just before it starts to decay subsequently in order not to overclose the Universe.\footnote{To be more precise, here we refer to specific {axion setups that have been} devised to address the Hubble tension.  The general idea of having an early dark energy component is much older though and was, for example, studied in \cite{Wetterich:2004pv,Doran:2006kp,Pettorino:2013ia,Calabrese:2011hg} and more recently in \cite{Chamings:2019boh}.} So far this has been realized in terms of a single scalar field that transitions from a slow-roll to an oscillating {(or fast-roll)} phase via a \textit{second-order phase transition}. During the slow-roll the field energy (playing the role of EDE) mimics a cosmological constant, and, during the oscillating phase, it decays as a power law which is controlled by the shape of the field's \textit{anharmonic} potential. This mechanism, therefore, employs the single-field approach to {slow-roll} inflation at much lower energies set by the eV rather than a high-energy scale close to GUT physics.  It is safe to say that until now EDE is {one of the more promising ways} to address the Hubble tension, as other early-time proposals have suffered from introducing new tensions not present in the $\Lambda$CDM model. One such alternative example is \textit{dark radiation} (DR), which relies on introducing additional relativistic species -- often modeled as sterile neutrinos -- to lift $H_0$ to higher values and has long been regarded a consistent resolution \cite{Wyman:2013lza,Dvorkin:2014lea,Leistedt:2014sia,Ade:2015rim}. However, increasingly precise CMB measurements have recently disfavored this possibility \cite{Raveri:2017jto,Aghanim:2018eyx} (not completely ruled out, though). 

The previous implementation of EDE is not free of problems either, as it requires shallow \textit{anharmonic} potentials to be efficient in addressing the Hubble tension. While monomial potentials, $\phi^{2n}$ with $n \geq 2$, were claimed not to be sufficient, mainly due to {a worsening of the fit to high-$\ell$ polarization data}~\cite{Agrawal:2019lmo}, specific terms of the non-perturbative axion potential, $\propto \left[ 1 - \cos\left( \phi / f\right) \right]^n$ with $n = 3$, yield a significantly better fit to data~\cite{Karwal:2016vyq,Poulin:2018dzj,Poulin:2018cxd,Smith:2019ihp}. This, however, constitutes a {tuning} where leading terms in an instanton approximation have been {set} to zero~\cite{Kaloper:2019lpl}. It is unclear whether these potentials can be obtained from a UV theory, and, in particular, from the `string-axiverse' as suggested by the authors, in a technically natural way. After all, the generic expectation in axion dark energy physics is to obtain a shallow \textit{harmonic} potential~\cite{Hill:1988vm,Frieman:1995pm,Kim:2002tq}.

\subsection{New Early Dark Energy}

In this work, we discuss a novel solution to the Hubble tension, called \textit{New Early Dark Energy} (NEDE),  which is based on a \textit{first-order phase transition} that {occurs} shortly before recombination in a dark sector at zero temperature~\cite{Niedermann:2019olb}. Effectively, it describes a jump in the vacuum energy from the $\text{eV}$ down to the $\text{meV}$ scale, in accordance with the spacetime curvature observed today. It builds up on the phenomenological success of the EDE proposal but uses different physics to do so. In fact, we think NEDE provides a more natural framework to resolve the Hubble tension while also providing new unique experimental signatures. 

Both single-field EDE and NEDE  share two defining properties, which are crucial for their phenomenological success.  First, there is an additional energy component, not present in $\Lambda$CDM, which comes to contribute an $\sim 10 \%$ fraction to the energy budget at some point $t_*$ close to matter-radiation equality. Second, that component starts to decay at least as fast as radiation after $t_*$. {How can this be} realized in a first-order phase transition {of NEDE} at zero temperature? The main challenge is to prevent the phase transition from happening too early, in which case the sound horizon and, hence, also $H_0$ would be largely unaffected. This, however, is not enough. We also need the phase transition to occur on a timescale $1/\bar{\beta}$ that is short compared to the Hubble expansion, i.e., $H/\bar{\beta} \ll 1$. This avoids the {premature} nucleation of bubbles of true vacuum that would grow {too large} before they collide with their smaller cousins. This would lead to large-scale anisotropies which would have imprinted themselves in the CMB. {These challenges are overcome by NEDE within a} \textit{two-field} scalar model in a dark sector that features a built-in trigger mechanism.  This is an adaption of an old idea that has been studied in the context of hybrid inflation at much higher energies~\cite{Linde:1990gz,Adams:1990ds,Copeland:1994vg,Cortes:2009ej}.\footnote{In a non-zero temperature scenario the phase transition could also be triggered due to the cooling of the Universe. For a corresponding late Universe scenario see \cite{Dutta:2009ix}.}
Here, the field starts its evolution  deep  inside a  potential valley with an almost flat bottom, corresponding to an ultra-light mass $m \sim 10^{-27} \text{eV}$, and very steep walls of mass $M \sim \text{eV}$. Initially, the field is stuck at its initial sub-Planckian field value due to the large Hubble friction, and its false vacuum energy provides the NEDE component. It is, however, protected from tunneling towards the true minimum of the potential {by the high walls of the false vacuum valley}. When $H \lesssim m$, the Hubble drag is being released and the field starts rolling down the valley. Now things move fast. Within less than a Hubble time, the valley opens up, effectively shrinking the barrier between the field and its true minimum, which then naturally triggers the phase transition {(a sketch is provided in Fig.~\ref{fig:plot3d})}. This leads to the sudden nucleation of a large number of bubbles of true vacuum that start to expand and quickly approach the speed of light. As this process fills space very rapidly, bubbles start to collide before they reach cosmological scales.  This condensate of colliding bubble walls is {complicated to describe,} however, at large observable scales, we expect it to be dominated by kinetic energy that redshifts away even quicker than radiation. This resonates with the observation that anisotropic stress, which comprises a sizable fraction of the colliding bubble condensate on short scales $\lesssim \bar{\beta}^{-1} $, is known to redshift like $1/a^6$ in the simpler case of a homogeneous system. Another part of the condensate it radiated away through gravitational waves and other model-dependent decay channels. In other words, the decay of NEDE at large scales is driven by the short-scale dynamics of colliding bubble walls and their microscopic decay. The corresponding model details, including a derivation of the percolation timescale $\bar{\beta}$ in terms of fundamental parameters as well as a first discussion of gravitational wave signatures, is provided in Sec.~\ref{sec:NEDE}

Before we can test our model against data, we have to find an effective description valid on cosmological scales. This is done in Sec.~\ref{sec:cosmo_model} in terms of an ideal fluid with a discontinuous equation of state parameter. It starts out in the very early Universe as $-1$, describing the vacuum energy of our scalar field, and then at redshift $z_* \, (\sim 5000)$ it jumps upwards to $w^*_{NEDE} > 1/3$, in accordance with a quickly decaying fluid. Again, this cosmological model is based on the assumption that the phase transition occurs on a short timescale $1/\bar{\beta} \ll 1/H$ and, therefore, can be described as an instantaneous process when solving cosmological equations. The trigger mechanism is implemented in terms of a single ultra-light scalar field $\phi(t)$ with mass $m$. In terms of our microscopic two-field model it parametrizes the almost flat direction along the bottom of the potential valley and promotes the tunneling probability to a function $\Gamma(\phi)$. The transition is then triggered when the ratio $H/m$, and, hence, $\phi / \phi_\text{ini}$, drops below a certain threshold value, which for a generic choice of fundamental model parameters falls in the range $0.18 < H_*/m \lesssim 0.21$, where the lower bound corresponds to the point of maximal  tunneling probability and the upper bound makes sure that that oscillations of the trigger field around the true vacuum are suppressed. In other words, for a given mass $m$, the effective parameter $H_*/m$ (implicitly) determines the {decay time $t_*$ (or redshift $z_*$, equivalently)}. Things become more interesting at the perturbation level. Adiabatic perturbations $\delta \phi$ of the trigger field lead to spatial variations of the decay time. These, in turn, provide the initial conditions for  the fluctuations of the decaying fluid after the transition. We use Israel's junction conditions~\cite{Israel:1966rt} to covariantly match perturbations across the transition surface. If the phase transition were to occur much earlier, we would not need to worry about this subtlety because all observable modes would safely reside outside the horizon uniquely determined as adiabatic modes.  However, in our case, modes that are observable in the CMB on short scales have already entered the particle horizon, and, as a result, we find that they are sensitive to the details of the phase transition, in particular, the ratio $H_*/m$.  

In Sec.~\ref{sec:param_extraction}, we will finally perform a full cosmological parameter extraction. Our base model introduces two additional parameters (beyond the six $\Lambda$CDM parameters), the mass of the trigger field, $\log_{10}(m/m_0)$, and the fraction of NEDE at  decay time,  $f_\text{NEDE}$. The other parameters such as the ``trigger parameter'', $H_*/m$, the equation of state parameter of the NEDE fluid, $w_\text{NEDE}$, or its rest-frame sound speed $c_s^2$, are fixed based on our theory expectations, but allowed to vary when we investigate extensions of our base model subsequently. 
Our joint analysis includes CMB data from {\sl Planck} 2018 (including temperature, polarization and lensing data), {BAO data}, constraints on the growth of large-scale structure, the {\sl Pantheon} dataset and the locally determined value of $H_0$. The latter is excluded from a subset of our runs in order to infer the evidence for NEDE without using `the statistical pull' of $H_0$. Special care is taken to avoid sampling volume effects that are known to lead to unphysical artifacts within EDE-type models when not properly accounted for~\cite{Smith:2019ihp,Lin:2019qug}. We also provide a detailed comparison of NEDE with its early-time competitors dark radiation (DR), acoustic dark energy (ADE)~\cite{Lin:2019qug} and single-field early dark energy (EDE)~\cite{Poulin:2018cxd}. On a technical level, we implement each model in a Boltzmann code and use a \textit{Monte Carlo Markov chain} (MCMC) algorithm to perform the parameter extraction.

\subsection{Summary of Results}

Here we list our most important results.\footnote{{We work units in which $\hbar=c=1$. For our cosmological perturbation theory we widely adopt the notation and conventions of~\cite{Ma:1995ey}.}}

\begin{itemize}
\item There is a consistent and radiatively stable choice of microscopic parameters that leads to a very quick first-order phase transition with percolation timescale $\bar{\beta}^{-1} \ll  H^{-1}$ around $z = 5000$. For a radiatively stable choice of coupling $\tilde \lambda$ between the two scalar fields, the two-field tunneling problem can be effectively reduced to a one-field tunneling problem, making our scenario computationally accessible.
\item Taking the limit $\bar{\beta}^{-1}   \to H^{-1}$ renders the expected gravitational wave signal produced by the phase transition marginally compatible with the claimed peak sensitivity of the \textit{Square Kilometre Array}~\cite{Carilli:2004nx,Janssen:2014dka,Bull:2018lat} for frequencies at around $f = 10^{-9}\, \text{Hz}$. At the same time, this limit leads to the prediction of distinct anisotropic structures in the CMB arising from the colliding bubble wall condensate.
\item We worked out the full covariant set of cosmological matching equations across a generic space-like hypersurface valid for sub- and super-horizon modes. We then used it to study both curvature and isocurvature modes within NEDE {and} identified the region in parameter space where the latter are sufficiently suppressed to be neglected.

\item The base model MCMC analysis without the local measurement of $H_0$ results in $f_\text{NEDE} = 7.7^{+3.8}_{-4.0} \, \%$ corresponding to a $1.9\, \sigma$ evidence for NEDE provided we account for volume sampling effects in the limit $f_\text{NEDE} \to 0$. Moreover, the model introduces an approximate degeneracy in the $r_s$--$H_0$ plane  satisfying $H_0 r_s \simeq const$ as required by BAO measurements. Correspondingly, the uncertainties in the determination of $H_0 = 69.6^{+1.0}_{-1.3}\, \kmsMpc $ become relatively large bringing the tension down to $2.5\, \sigma$ (from $4.3 \, \sigma$ within $\Lambda$CDM when the same dataset combination is chosen).

\item When the local measurement is included, we report for our base model an increased Hubble parameter of $H_0=71.4 \pm 1.0 \kmsMpc$ and an evidence for NEDE at  {$ 4.3\, \sigma$} $[f_\text{NEDE} = 12.6^{+3.2}_{-2.9} \, \%]$. The overall fit improves by $\Delta \chi^2(\text{total}) = -15.6$.  At the same time, the so-called $S_8$ tension {increases slightly to} the $2.7 \, \sigma$ level, but not by much compared to $\Lambda$CDM for which we find $1.9 \, \sigma$ (w/ SH$_0$ES) and $2.3 \, \sigma$ (w/o SH$_0$ES).

\item { Different extensions of our base model confirm the expectations arising from our microscopic description.  In particular, we obtain $H_* / m = 0.206^{+0.013}_{-0.022}$ in good agreement with the theoretically viable range, as well as $w_{NEDE}^* = 0.70^{+0.12}_{-0.16}$, consistent with a fluid dominated by kinetic energy and small-scale anisotropic stress. On the other hand, we find no evidence for large-scale anisotropies.}

\item We compare NEDE with single-field EDE. While we find that both models share their phenomenological success, they use different physics to do so. Most notably, NEDE relies on a $k$ independent rest-frame sound speed and is compatible with a wider range of  initial field values $ \phi_\text{ini} / M_{pl}  > 10^{-4}$.  Moreover, its potential adheres to the usual low-energy rules and does not require a fine-tuning. 
\end{itemize}

\section{New Early Dark Energy} \label{sec:NEDE}

In the following we will outline a field theoretical model of NEDE. {To that end}, we will briefly review how a first-order phase transition depends on the shape of a generic quartic potential. In a second step, we consider a generalized two-field model {with} a clock field that triggers the transition at a given time $t_*$ in the early expansion history somewhat close to recombination.

\subsection{One-Field Vacuum Decay}

Here we briefly review the vacuum decay of a single scalar field at zero temperature. We are particularly interested in  how the decay rate depends on the shape of the potential, which will become crucial in devising a dynamical two-field model in the next section. We will broadly follow the discussion in \cite{Adams:1993zs}.

We consider a scalar field $\psi$ with general renormalizable potential $V(\psi)$,
\begin{align}\label{eq:pot1D}
V(\psi) = \frac{\lambda}{4} \, \psi^4 - \frac{1}{3}\alpha M\, \psi^3 + \frac{1}{2}\beta M^2 \, \psi^2 + \gamma M^3 \, \psi + d  \,,
\end{align}
where $M$ is a generic mass scale and $\alpha$, $\beta$ and $\gamma$ are dimensionless constants. We assume the potential to be bounded from below, implying $\lambda>0$.  The constant $d$ corresponds to an irrelevant shift that can be absorbed in a cosmological constant term. Furthermore, we can perform a field translation to set $\gamma=0$ and switch the sign of $\psi$ if $\alpha<0$. We can therefore make the choice $d=\gamma=0$ and $\alpha>0$ without any loss of generality. After rewriting the potential in terms of the dimensionless field variable $\bar{\psi} = 3\, \lambda \, \psi / (\alpha \, M) $, we have 

\begin{align} \label{eq:action1}
\bar{V}(\bar{\psi}) =  \frac{1}{4} \bar{\psi}^4 - \bar{\psi}^3  + \frac{\delta}{2} \, \bar{\psi}^2 \,,
\end{align}
where we introduced the dimensionless potential $\bar{V} = 81 \lambda^3 V / (\alpha^4 \, M^4) $ as well as 
\begin{align}\label{eq:delta}
\delta  = 9 \, \frac{\lambda \beta }{\alpha^2}\,.
\end{align}
\begin{figure*}[t]
	\subfloat[Both vacua: $\bar{\psi}_1$ (left) and $\bar{\psi}_2$ (right).\label{fig:pot1}]{
		{\includegraphics[width=7.8cm]{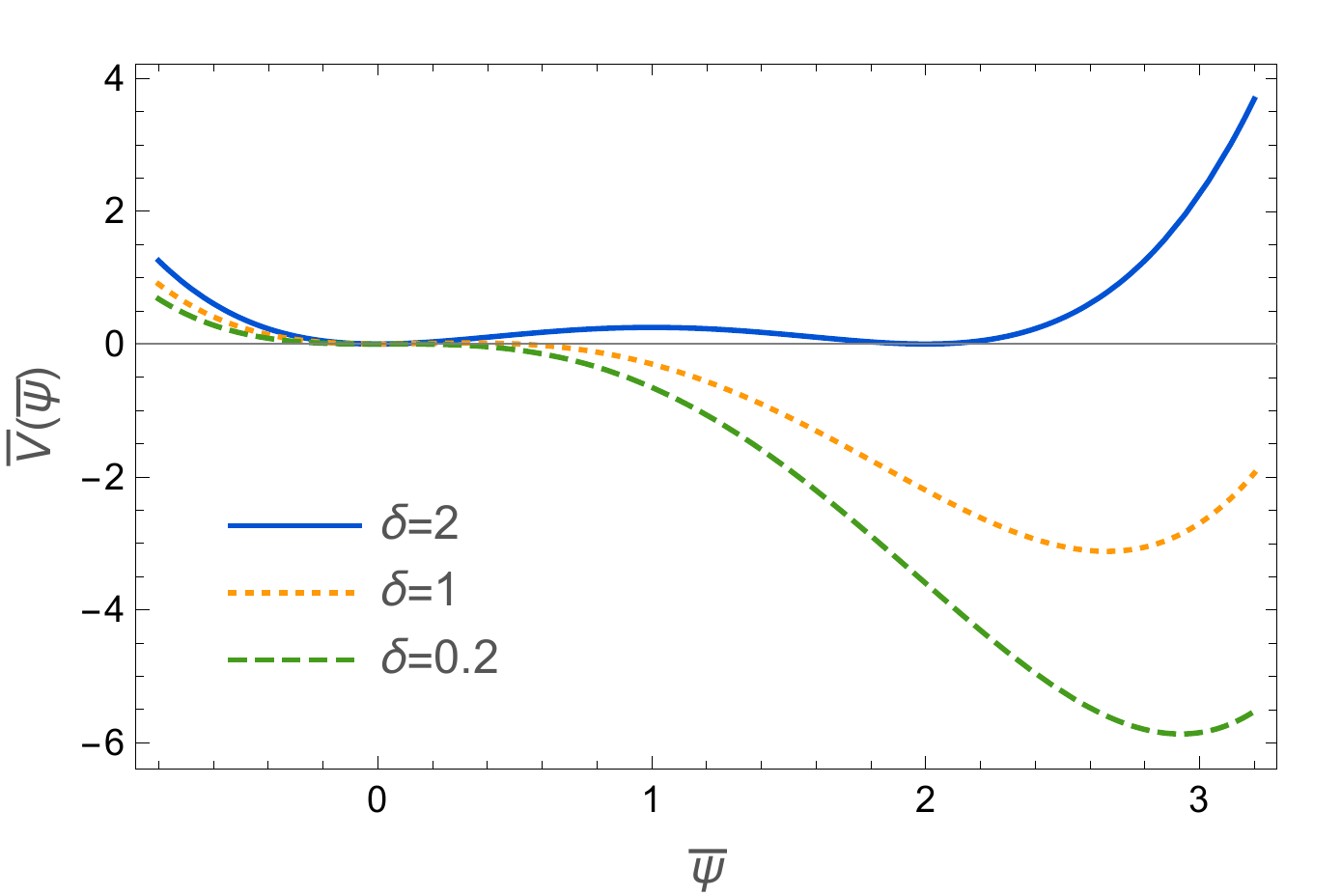}}\qquad
	}%
	\subfloat[Zoomed in on vacuum $\bar{\psi}_1 = 0$.\label{fig:pot2}]{
		{\includegraphics[width=8 cm]{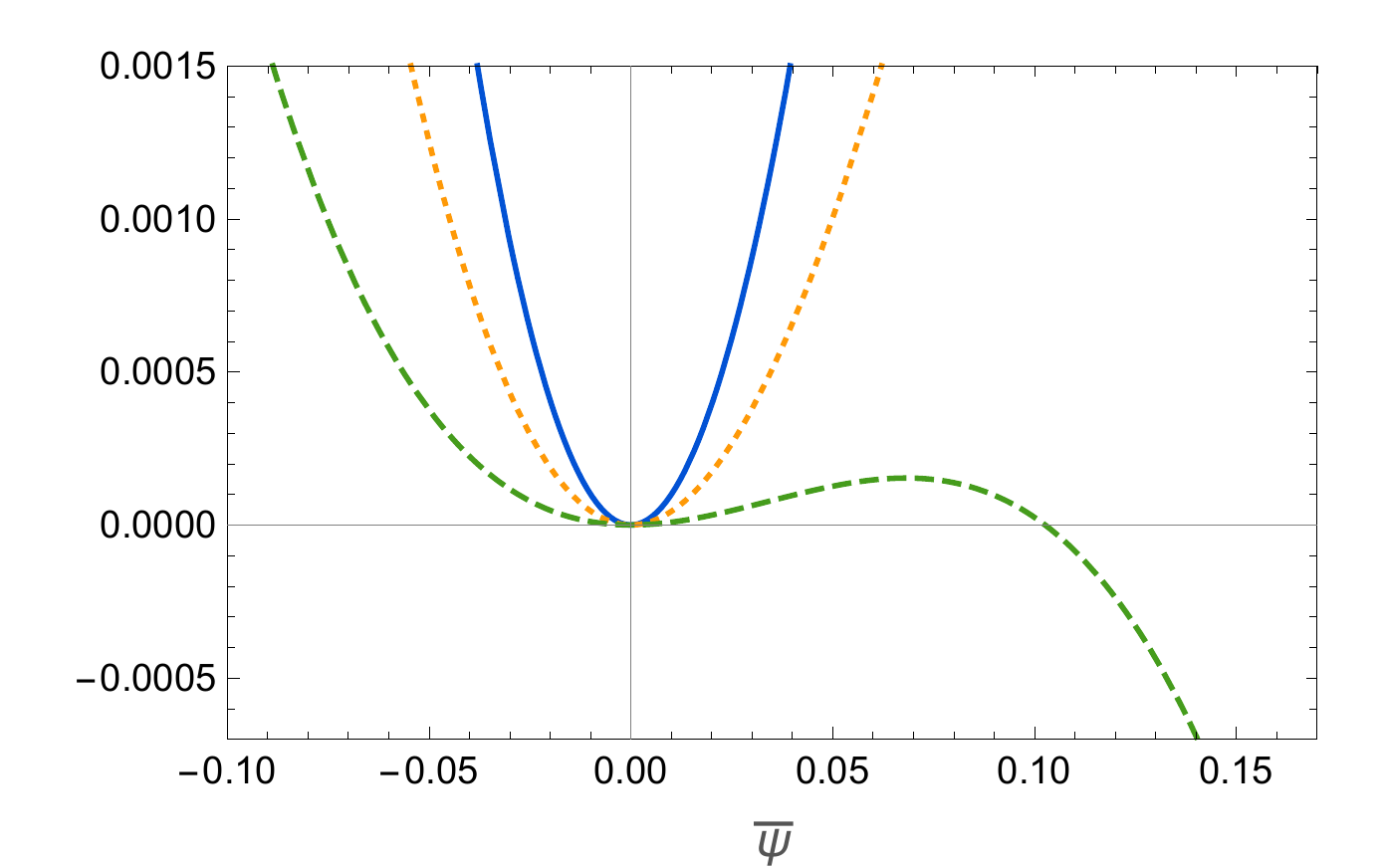}}%
	}%
	\caption{Dimensionless potential $\bar{V}$ for different choices of $\delta$. The vacua become degenerate as $\delta \to 2$. For $\delta \to 0$, the barrier height vanishes and the true vacuum is deepest.} \label{fig:pot}
\end{figure*}

So we see that the properties of the potential are conveniently captured by the dimensionless parameter~$\delta$. For $0 < \delta < 9 / 4$, the potential exhibits two minima at 
\begin{align}\label{eq:psi_min}
\bar{\psi}_{1} = 0 \quad \text{and} \quad \bar{\psi}_{2}= \frac{1}{2}\left( 3 + \sqrt{9 - 4 \delta }\right) \,,
\end{align}
as well as a maximum at\footnote{The case $\delta \leq 0$ (or $\beta \leq 0$ equivalently), for which the maximum is at $\bar{\psi}=0$, is more akin to scenarios of hybrid inflation where the field undergoes a second-order phase transition. We provide a preliminary discussion of such a scenario in Sec.~\ref{sec:HEDE}.}  $\bar{\psi}_{max} = \left( 3 - \sqrt{9 - 4 \delta }\right)/2$. 
We now assume $\bar{\psi}=0$ initially, which for  $0 < \delta < 9 / 4$ corresponds to the false vacuum.
There are two interesting limiting cases. 
\newline
\paragraph*{\textbf{Thin-wall limit:}} For $\delta = 2 - \epsilon$ and $\epsilon \ll 1$ both minima are degenerate corresponding to the thin-wall limit (see solid, blue line in Fig.~\ref{fig:pot}). Specifically, the true vacuum sits at
\begin{subequations}
\begin{align}
\bar{\psi}_2 \simeq 2 \,,
\end{align}
and the difference in energies is 
\begin{align}
\bar V(\bar \psi_1)- \bar V(\bar \psi_2) \simeq 2 \epsilon\,.
\end{align}
\end{subequations}
The barrier height becomes maximal with $\bar V(\bar{\psi}_{max})-\bar V(\bar{\psi}_1) \simeq 1/4 $. 
\newline
\paragraph*{\textbf{Deep-well limit:}} For $\delta = \epsilon \ll 1$, on the other hand, the height of the barrier becomes tiny, $\bar V(\bar{\psi}_{max})-\bar V(\bar{\psi}_1) \simeq \epsilon^3 / 54 $, and the true vacuum at
\begin{subequations}
\begin{align}\label{eq:psi2}
\bar{\psi}_2 \simeq 3 
\end{align}
 is deepest (see dashed, green line in Fig.~\ref{fig:pot}),
\begin{align}\label{eq:V_diff}
\bar V(\bar{\psi}_1)-\bar V(\bar{\psi}_2) \simeq \frac{27}{4} \, .
\end{align}
\end{subequations}
We will see that the case most interesting for our application lies between these extremes with $0<\delta \lesssim 1.5$.

In order to infer the corresponding tunneling probabilities, we need the Euclidian action evaluated at the $O(4)$ symmetric ``bounce solutions'' $\bar{\psi}_\delta(r)$, where $r$ is a four-dimensional radial coordinate in Euclidian space. These configurations are defined as stationary points of the Euclidian action
\begin{align}
S_{E} &= 2 \pi^2 \int_0^{\infty} d r \, r^3 \left[ \frac{1}{2} \left( \frac{d \psi}{d r} \right)^2 + V(\psi) \right] \nonumber\\
&=  \frac{2 \pi^2}{\lambda} \int_0^{\infty} d \bar{r} \, \bar{r}^3 \left[ \frac{1}{2} \left( \frac{d \bar{\psi}}{d \bar{r}} \right)^2 + \bar{V}(\bar{\psi}) \right] 
\,,
\end{align}
where $\bar{r} = \alpha M r / (3 \sqrt{\lambda})$. Correspondingly, they fulfill the Euclidian equations of motion,
\begin{align}\label{eq:eom_psi1}
\bar{\psi}''(r) +  \frac{3}{r}\, \bar{\psi}'(r) = \bar{\psi}^3 - 3 \, \bar{\psi}^2 + \delta \, \bar{\psi } \,,
\end{align}
and are subject to the boundary conditions $d \bar{\psi}_\delta / d \bar{r} = 0$ at $\bar{r} =0 $ (to ensure regularity at the origin) and $\bar{\psi}_\delta = 0$ as $\bar{r} \to \infty$ (for the false vacuum to lie outside the bubble). {A fully analytic solution exists only in the thin-wall limit~\cite{Coleman:1977py}. In general, a numerical shooting method needs to be applied. For the model in \eqref{eq:action1} a sufficiently accurate approximation for $S_E$ was derived in \cite{Adams:1993zs}, yielding}\footnote{We checked the result with our own implementation of the shooting method. Moreover, in the limit $\delta \ll 1$ we derive $ \lambda \, S_{E} \simeq  91/4 \, \delta$, which agrees with the result in \cite{Linde:1981zj} when we identify $\delta_\text{there}^2 = 9 \lambda M^2 / \delta $. And in the opposite limit, it agrees with the analytical result obtained from the thin-wall approximation.} 
\begin{align}\label{eq:SE}
S_{E}\Big|_{\bar{\psi}=\bar{\psi}_\delta} \simeq \frac{4 \, \pi^2}{3 \lambda} \left( 2 - \delta\right)^{-3} \left(\alpha_1 \delta + \alpha_2 \delta^2 + \alpha_3 \delta ^3  \right)
\end{align}  
where
\begin{align}
\alpha_1 = 13.832\,, && \alpha_2 = -10.819\,, && \alpha_3 = 2.0765 \,.
\end{align}
{As a result, }$S_{E}$ diverges for $\delta \to 2$ (thin-wall limit) and vanishes linearly {as} $ \delta \to 0$ (deep-well limit). For moderate values in the range $\delta \in [0.01,0.5]$, we find $\lambda \, S_E \in [0.2, 17.4 ]$. The {tunneling} rate per unit volume {is then defined as} 
\begin{align}\label{eq:def_Gamma}
\Gamma = K \exp{\left( -S_E \right)}\,.
\end{align}
{In~\cite{Callan:1977pt,Linde:1981zj}, it was argued that the determinant factor $K \sim M^4$, provided the dimensionless parameters in \eqref{eq:pot1D} do not introduce large hierarchies. }This formula is applicable as long as $S_E / \hbar \gg 1$ for the underlying semi-classical calculation to be valid. This can be satisfied for a sufficiently weak coupling, at least $\lambda < 0.1$, provided $ \delta$ is not tuned too small. {As a numerical example, for $\lambda = 0.01$ and $\delta = 0.1$, we obtain $S_E \simeq 245 \gg 1$.} {In short}, the tunneling rate vanishes in the thin-wall limit and monotonically increases for {decreasing} $\delta$. 

\subsection{Triggered Vacuum Decay}\label{sec:2-field}

If the nucleation rate is constant, we will encounter {an analogy of the} ``big bubble'' problem, which was discussed already in the context of old inflation~\cite{Copeland:1994vg}. 
{Even if a constant nucleation rate is initially small compared to the growing Hubble rate, there will be a small but non-vanishing probability to form bubbles early in the expansion history of the Universe, and, since these bubbles will expand with the speed of light and rapidly become very big, the true vacuum volume will be dominated by a small number of very large bubbles \cite{Copeland:1994vg}. {This is problematic for two reasons: First, }it will lead to unacceptably large inhomogeneities in the Universe. {Second, the transition would happen too slow to have the desired impact on the sound horizon.} To solve this problem, one needs to make sure that the bubble nucleation is turned on quickly. }
This can be achieved by introducing an additional field, which acts as a trigger for the bubble nucleation.

The aim of this section is, therefore, to find a field theoretical model that promotes $\delta$ to a function of another field $\phi$, i.e., $\delta \to \delta_\text{eff}(\phi)$. By controlling the slow-roll dynamics of $\phi$, we can scan over different values of $\delta_\text{eff}$ and switch on the phase transition in $\psi$ (and $\phi$) at any given time.  The difference between the false and the true vacuum energy stored in the $\psi$ field then provides most of the NEDE component (with only a sub-dominant contribution arising from $\phi$). As NEDE is stable at early times, we must construct a model that initially guarantees $\delta_\text{eff} \gg 2$ in order to sufficiently suppress bubble nucleation. At a later stage, however, we want NEDE to start its decay. This means that we look for a dynamical mechanism that sends  $\delta_\text{eff} \to \delta_\text{eff}^* < 2$, and, in turn, switches on the nucleation of bubbles of true vacuum. {This is then followed by a coalescence phase where the colliding bubble wall condensate redshifts away while also decaying to radiation.} This idea has been studied in the context of hybrid inflation in \cite{Linde:1990gz,Adams:1990ds,Copeland:1994vg,Cortes:2009ej} where the authors employed a field theoretical trigger mechanisms to end inflation by tunneling (and overcome the ``graceful exit'' problem in earlier first-order single-field constructions). We will apply a similar setup as a simple way of realizing our NEDE scenario. To that end, we generalize the single field model to 
\begin{align}\label{eq:action2}
V(\psi, \phi) = \frac{\lambda}{4} \, \psi^4  + \frac{1}{2}\beta M^2 \psi^2 - \frac{1}{3} \alpha M \psi^3 
+ \frac{1}{2} m^2 \phi^2 +\frac{1}{2} \tilde{\lambda} \, \phi^2 \psi^2 \,,
\end{align}
where again everything is expressed in terms of a single mass scale $M$ and positive, dimensionless parameters $\lambda$, $\tilde{\lambda}$, $\alpha$ and $\beta$. The kinetic terms for $\psi$ and $\phi$ are canonically normalized.
This potential can be recovered from the previous one in \eqref{eq:pot1D} if we set $\gamma = 0 = d$, replace $\beta$ by a function of $\phi$,
\begin{align}
\beta \to \beta + \tilde{\lambda} \frac{\phi^2}{M^2}\, ,  
\end{align}
and add a mass term for $\phi$.

\begin{figure}
 \centering 
{\includegraphics[width=8.cm]{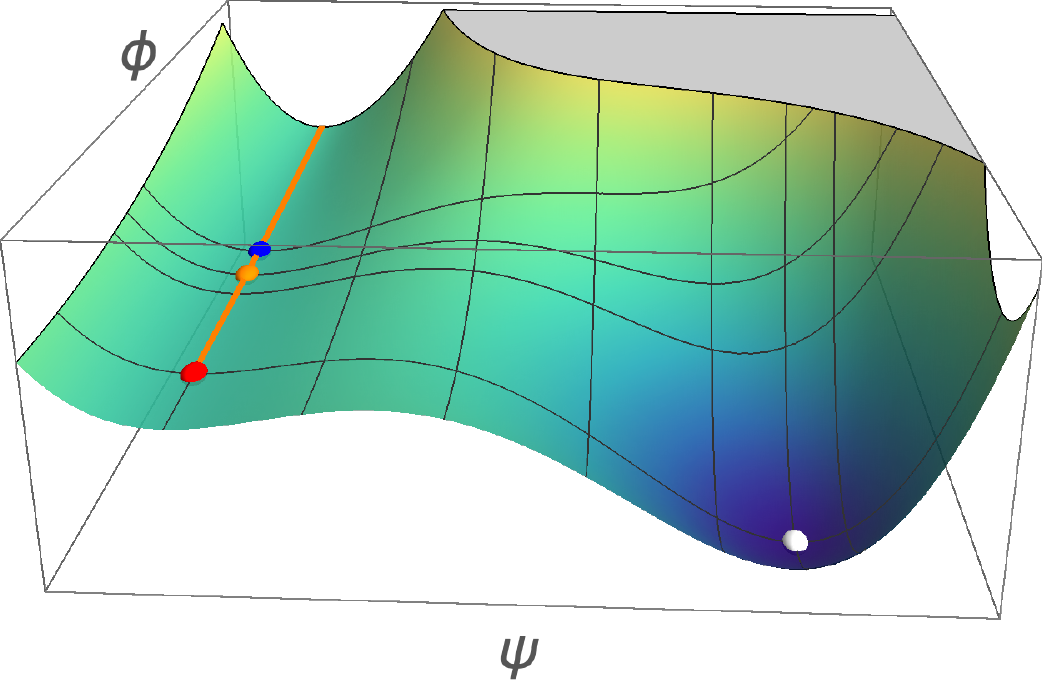}}
 \caption{{Illustration of the} potential in \eqref{eq:action2}. {At early times, the field is frozen high up the false vacuum valley (above the blue dot) and protected against tunneling to the true vacuum (white dot) by a high potential barrier. Tunneling is turned on when the field rolls past the orange dot. Shortly after that, bubble percolation overcomes the expansion of space and reaches its maximal efficiency at the red dot.}}
\label{fig:plot3d}
\end{figure}

The general mechanism is simple: Initially, $\psi = 0$ and $\phi$ is frozen at $\phi \simeq \phi_\text{ini}$. {Once the Hubble drag is released, i.e., when $H \lesssim m$, $\phi$ rolls down the potential}.  {Now, if we require} 
\begin{align}\label{eq:cond1}
\alpha^2 > 4 \, \beta \lambda \,,
\end{align}
{or $\delta < 9/4$ equivalently, a second (global) minimum exists.} {As a result, when} $\phi$ drops below a certain threshold $\phi_*$ on its way to {$\phi=0$}, the field configurations with $\psi=0$ (representing a valley in the two-dimensional potential) become unstable against quantum tunneling and NEDE starts to decay. For a schematic plot of the potential see Fig.~\ref{fig:plot3d}.

We are interested in the tunneling rate as a function of the rolling field $\phi(t)$ and, hence, cosmological time $t$. To that end we derive a semi-analytic expression for the Euclidian action in the limit where the evolution of $\phi$ is sufficiently slow and $\tilde \lambda / \lambda \ll 1$. As before, it is useful to introduce a dimensionless potential 
\begin{subequations}
\begin{align}\label{eq:pot2d}
\bar{V}(\bar{\psi},\bar{\phi}) = \frac{1}{4} \bar{\psi}^4 - \bar{\psi}^3  + \frac{\delta_\text{eff}(\bar{\phi})}{2} \, \bar{\psi}^2 + \frac{1}{2} \kappa^2 \bar{\phi}^2\,,
\end{align}
where we used the dimensionless variables
\begin{align}
\bar{V} = \frac{81 \lambda^3}{\alpha^4 \, M^4} \, V \,, && \bar{\psi} = \frac{3 \lambda}{\alpha M} \psi \,, && \bar{\phi} = \frac{3 \sqrt{\lambda \tilde{\lambda}}}{\alpha M} \phi 
\end{align}
and introduced
\begin{align}\label{eq:kappa}
\kappa = \frac{3 \lambda}{\alpha\, \sqrt{ \tilde \lambda}} \frac{m}{M} \;.
\end{align}
\end{subequations}
As advertised, the parameter $\delta$ in \eqref{eq:delta} is promoted to a function of $\bar{\phi}$,\footnote{We assume as before $\beta > 0$ which ensures the positivity of $\delta$ and hence $\delta_\text{eff}$. It would also be interesting to study the case $\beta < 0$, which for a suitable choice of parameters gives rise to a second-order phase transition like in hybrid inflation (see for example \cite{Copeland:1994vg}). As far as we know this has not yet been studied in the context of EDE proposals.}
\begin{align}\label{eq:delta_phi}
\delta_\text{eff}(\bar{\phi}) = \delta + \bar{\phi}^2 \,,
\end{align}
controlling the shape of the potential for $\bar{\psi}$. In particular, it shows that for $\delta < 0.2$ we cycle through the different configurations depicted in Fig.~\ref{fig:pot} when $\bar{\phi} \to 0$.

The problem of calculating the tunneling probability is rather complicated in the two-field case. First, finding the bounce solution requires us to solve the coupled system of Euclidian equations of motion for $\phi$ and $\psi$. Second, due to the rolling of $\phi$, the boundary condition in Euclidian time is not $O(4)$ symmetric (or equivalently, the initial configuration before the bubble nucleation is not Lorentz invariant in real time). This means that the bounce solutions, in general, depend on $t_E$ and the three-dimensional radial coordinate $\mathbf{x}^2 =\delta_{ij} x^i x^j$ explicitly [rather than the $O(4)$ symmetric combination $r= \sqrt{t_{E}^2 + \mathbf{x}^2}$]. We would therefore need to solve a coupled system of partial (rather than ordinary)  differential equations to infer the evolution of the bubbles after their nucleation. This difficulty can, however, be avoided if we assume $\kappa \ll 1$ (we will check the validity of this assumption \textit{a posteriori}). In that case, we can neglect the last term in \eqref{eq:pot2d} and the corresponding Lorentz-invariant vacuum solution becomes $\bar{\psi} = 0$ and $\bar{\phi} \simeq const$. This limit corresponds to the case where the field $\phi$ evolves very slowly on scales $1/ M $ (relevant for the bounce solution).

The Euclidian action for $O(4)$-symmetric configurations then reads
\begin{align}
S_E = \frac{2 \pi^2}{\lambda} \int_0^{\infty} d \bar{r} \, \bar{r}^3 \left[ \frac{1}{2} \left( \frac{d \bar{\psi}}{d \bar{r}} \right)^2 + \frac{1}{2}\frac{\lambda}{\tilde \lambda} \left(\frac{d \bar{\phi}}{d \bar{r}} \right)^2 + \bar{V}(\bar{\psi}) \right] \,,
\end{align}
leading to the equations of motion
\begin{subequations}
\begin{align}
\bar{\psi}'' + \frac{3}{\bar{r}} \bar{\psi}'  &= \bar{\psi}^3 - 3 \bar{\psi}^2 + \delta_\text{eff}(\bar{\phi})\bar{\psi } \,, \label{eq:eom_psi2} \\
\bar{\phi}'' + \frac{3}{\bar{r}} \bar{\phi}'  &= \frac{\tilde \lambda}{\lambda} \left[\bar{\psi}^2  \bar{\phi} +\mathcal{O}\left(\kappa^2\right) \right] \,. \label{eq:eom_phi}
\end{align}
\end{subequations}
They are subject to the boundary conditions
\begin{subequations}
\begin{align}
\lim_{\bar{r} \to \infty} \bar{\psi}&=0\,, & \lim_{\bar{r} \to \infty}\bar{\phi} &= \bar{\phi}(t_*)  \,,\\
\lim_{\bar{r} \to 0}\bar{\psi}'&=0\,, &  \lim_{\bar{r} \to 0}\bar{\phi}'&=0   \,, \label{eq:bdry_phi}
\end{align}
\end{subequations}
where $\bar{\phi}(t_*)$ is the value of $\bar{\phi}$ at the time of tunneling. The important observation is that in the limit $\tilde \lambda / \lambda \ll 1$ both equations decouple and \eqref{eq:eom_phi} is trivially solved by $\bar{\phi}(\bar{r}) = \bar{\phi}(t_*) =const$. In that case, \eqref{eq:eom_psi2} becomes formally equivalent to \eqref{eq:eom_psi1}. As a result, the Euclidian action is simply given by its one-field expression \eqref{eq:SE} subject to the replacement $\delta \to \delta_\text{eff}(\bar{\phi})$,
\begin{align}\label{eq:SEeff}
S_{E} \simeq \frac{4 \, \pi^2}{3 \lambda} \left( 2 - \delta_\text{eff} \right)^{-3} \left(\alpha_1\, \delta_\text{eff} + \alpha_2 \,\delta_\text{eff}^2 + \alpha_3\, \delta_\text{eff}^3  \right) \,.
\end{align} 
The time dependence of $\bar{\phi}(t)$ (relevant on cosmological timescales) then determines the time dependence of the tunneling rate, $\Gamma(t)$, given by~\eqref{eq:def_Gamma}. 

\begin{figure*}[t]
 	\subfloat[Radial profiles for $\bar{\psi}$ (solid lines) and $\bar{\phi}$ (dotted lines) for $\delta_\text{eff} = 1$. As  $\tilde \lambda / \lambda \to 0 $ the profile of $\bar \phi$ approaches a constant.]  
	{\includegraphics[width=7.8 cm]{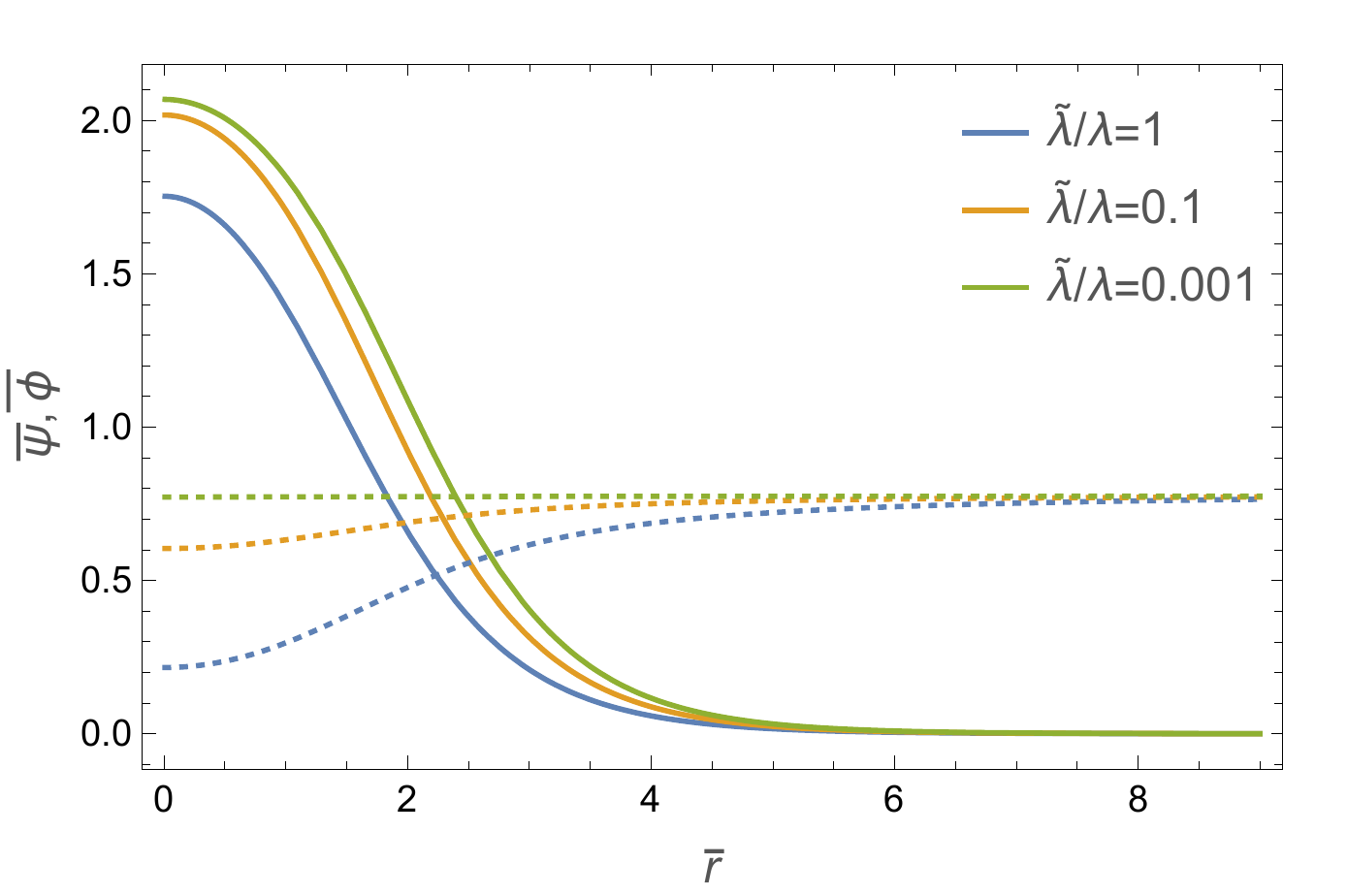}
	\label{fig:profile}
	}\quad
	\subfloat[Euclidian action as a function of $\delta_\text{eff}$ for $\delta = 0.4$. The dashed line plots the semi-analytic result \eqref{eq:SE} derived in the one-field case. We find that it is approached as $\tilde \lambda / \lambda \to 0 $. ]   
	{\includegraphics[width=7.8 cm]{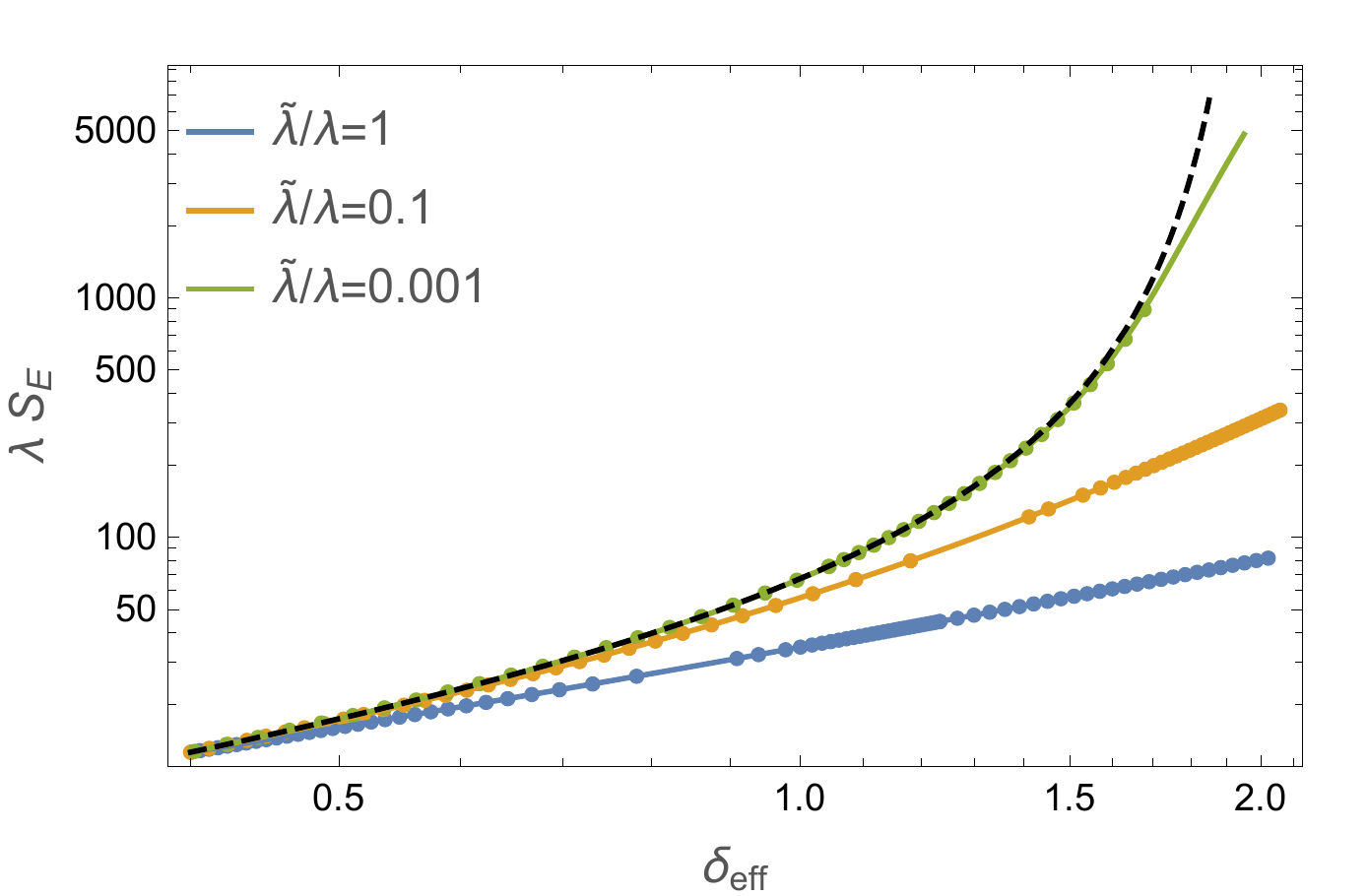}
	\label{fig:SE}}
 	\caption{Result of the two-field shooting method. Each dot corresponds to one numerical integration. Different colors correspond to different 	choices of $\tilde \lambda / \lambda$.  \label{fig:Bounce}}
	
\end{figure*}

We also studied the system away from the limit $\tilde \lambda / \lambda \ll 1$ by solving the coupled system of equations numerically, yet still assuming $\kappa \ll 1$. The result is depicted in Fig.~\ref{fig:Bounce} for different values of $\tilde{\lambda} / \lambda$ and $\delta = 0.4$. We see in Fig.~\ref{fig:profile} that $\bar{\phi}$ indeed approaches a constant profile as $\tilde \lambda / \lambda \to 0$. {Moreover, Fig.~\ref{fig:SE} shows} that the result in \eqref{eq:SE} is applicable at least up to values $\lambda S_E \simeq 1000 $ provided $\tilde{\lambda} / \lambda < 0.001$. This is enough to study our tunneling mechanism, which, {as we will see,} is triggered only when $S_E$ drops below $\sim 250$ and bubble percolation becomes more efficient than the expansion of space. {For larger values, on the other hand, it is extremely unlikely that a bubble is either nucleated within our Hubble volume or enters from outside the horizon in case it has been nucleated in a different patch before.}  {In addition, }we also find that our trigger mechanism would still work if $\tilde \lambda \sim  \lambda  $.  In particular, percolation would cease for sufficiently large values of $\delta_\text{eff}$. Unfortunately, we could not find an analytic expression for $S_E$ in this more general case, but later we will anyway find that it is disfavored by naturalness considerations. 

Finally, for $\delta_\text{eff} > 2$, all three curves in Fig.~\ref{fig:SE} asymptote to a power law. This is different from the one-field case (corresponding to the dashed line), where $S_E$ diverges for $\delta \to 2$. The reason is that there is always a non-vanishing probability that the field tunnels along a diagonal direction in the $\phi-\psi$ plane, albeit highly suppressed.  

\subsection{Bubble Percolation}\label{sec:percolation}
We can now use the expression in \eqref{eq:SEeff} to study the percolation phase within our model. In particular, we will derive the range of model parameters for which percolation happens quickly on cosmological timescales to avoid phenomenological problems with bubbles growing to cosmological scales. 

Before we do so, we derive estimates for the two mass scales $M$ and $m$ controlling the $\psi$ and $\phi$ sector, respectively. The scale $M$ is fixed by the requirement that NEDE gives an $\sim 10\, \%$ contribution around the time of the phase transition. Otherwise, its effect on the expansion history would be negligible. To that end, we introduce the NEDE fraction
\begin{align}\label{eq:def_frac}
f_\text{NEDE} 	&\equiv \frac{\bar{\rho}_\text{NEDE}^*}{\bar{\rho}_*} \nonumber\\
				&=\frac{V(\psi_*,\phi_*) - V(\psi_2,0)  }{\bar{\rho}_*} \,,
\end{align}
where $\bar{\rho}$ is the total background energy density and we normalized $\bar{\rho}^*_\text{NEDE}=V(\psi_*,\phi_*) - V(\psi_2,0)$ with respect to  the true vacuum at $\psi=\psi_2$ and $\phi = 0$. Here and henceforth, we use an asterisk as a shorthand for evaluation at percolation time $t_*$, i.e., $A_* = A^* \equiv A(t_*)$. We can use the results of the last section to further evaluate \eqref{eq:def_frac}, 
\begin{align}\label{eq:def_EDE}
\bar{\rho}_\text{NEDE}^* = \frac{c_\delta}{12}  \frac{\alpha^4 \, M^4}{\lambda^3}  + \frac{1}{2} m^2 \phi_*^2\,.
\end{align}
where 
\begin{align}
{c_\delta = \frac{1}{216} \left( 3 + \sqrt{9 -4 \delta}\right)^2 \left( 3 - 2 \delta + \sqrt{9 -4 \delta}\right) \, ,}
\end{align}
evaluating to $c_\delta \in [1,0.41]$ for $\delta_\text{eff} \in [0,1]$.
In addition, we require the trigger field $\phi$ to be sub-dominant before the phase transition. A sufficient condition reads
\begin{align}\label{eq:cond_phi}
m^2 \phi_*^2 < m^2 \phi_\text{ini}^2 \ll \frac{c_\delta}{6} \frac{\alpha^4 M^4}{\lambda^3} \;.
\end{align}
We then obtain from \eqref{eq:def_EDE}
\begin{align}\label{eq:cond3}
M^4 \simeq \frac{12}{c_\delta} \, \frac{\lambda^3 \, f_\text{NEDE}}{\alpha^4} \, \bar{\rho}_*\,. 
\end{align}
Provided $\alpha$ is not hierarchically small, we see that the mass scale $ M$ is bounded by the energy scale of the cosmic fluid at decay time $t_*$. Assuming that we are still in the radiation-dominated epoch when the transition takes place, we have $\bar{\rho}_* \simeq \bar{\rho}^*_{rad}/(1-f_\text{NEDE}) $, {where the radiation density is } 
\begin{align}
\rho_{rad}^* = \frac{\pi^2}{30} g_*(T) \left(1+z_*\right)^4 \left(\frac{g_s(T_0)}{g_s(T_*)} \right)^{4/3} T_0^4\,.
\end{align}
{Here, $g_*(T)$ and $g_s(T)$ are the total number of relativistic and the effective number of entropy carrying degrees of freedom at temperature $T$, respectively. As our transition occurs rather late in the early Universe, we have $g_s(T_*) = g_s(T_0) $ , where $T_0$ is the CMB temperature today. Plugging in numerical values, we find}
\begin{align}\label{eq:M}
M^4 \simeq (0.4 \, \text{eV})^4 \, \frac{1}{c_\delta}\, \left(\frac{\lambda^3\alpha^{-4}}{0.01} \right) \left( \frac{f_\text{NEDE} / (1-f_\text{NEDE})}{0.1} \right) \left( \frac{g_*(T_*)}{3.9} \right) \left( \frac{z_*}{5000} \right)^4  \,.
\end{align}
We will use the estimate $M \sim \text{eV} $ throughout the remainder of this work.

At late times, after bubble percolation has been completed, we have to make sure that the remaining constant energy density stored in the $\psi$ vacuum is not over-closing the Universe, explicitly
\begin{align}
\frac{c_\delta}{12} \frac{\alpha^4}{\lambda^3} M^4+ \rho_{\lambda} \lesssim \left( \text{meV} \right)^4 \,.
\end{align}
Here, we introduced the cosmological constant contribution to the energy density, $\rho_{\lambda} = M_{pl}^2 \Lambda$. This relation can always be satisfied by tuning the value of $\Lambda$, which is nothing else but the usual cosmological constant problem (see, for example, \cite{Weinberg:1988cp}). 

Next, we turn to the mass of the trigger field $\phi$, which is frozen at early times at $\phi \simeq \phi_\text{ini}$. From \eqref{eq:cond_phi} and \eqref{eq:cond3}, we derive the upper bound
\begin{align}\label{eq:cond_4}
\frac{\phi_\text{ini}}{M_{pl}} \ll   \sqrt{6\, f_\text{NEDE}}\, \frac{H_*}{m} \lesssim 1\;,
\end{align}
which is sufficient to ensure that the contribution of $\phi$ to the total energy density is sub-dominant and smaller than the one arising from $\psi$ (at least during its slow-roll regime before the phase transition). It also shows that  $\phi$ is always sub-Planckian. 
Once $m \sim H_*$, slow-roll ends and $\phi \to 0$ for the first time, which, in turn, triggers the phase transition as $\delta_\text{eff} \to \delta^*_\text{eff}$. Using $\rho_* = 3 M_{pl}^2 H_*^2$ together with \eqref{eq:cond3} and \eqref{eq:M}, this yields an expression for the mass, 
\begin{align}\label{eq:m}
m = 1.8 \times 10^{-27} \, \text{eV} \, (1-f_\text{NEDE})^{-1/2}\left( \frac{g_*(T_*)}{3.9} \right)^{1/2} \left( \frac{z_*}{5000} \right)^2 \, \left(\frac{0.2}{H_*/m}\right) \,.
\end{align}
where we introduced the ``trigger parameter'' $H_*/m$. Later we will show that for a typical parameter choice the decay is triggered at $H_*/m \simeq 0.2$, close to the zero crossing of $\phi$.
As an example, for the transition to take place at redshift $z \sim 5000$, we need an extremely tiny scale of the order of $m \sim 10^{-27} \text{eV} \ll M \sim \text{eV}$. We will see that this enormous hierarchy between $m$ and $M$ is key in making the bubble percolation efficient (and avoiding potential problems with large inhomogeneities arising from colliding bubble wall dynamics), and in Sec.~\ref{radstab} we demonstrate that the hierarchy is technically natural.

We quantify the efficiency of the bubble nucleation in terms of the percolation parameter $p = \Gamma / H^4$. Provided $p  > 1$, at least one bubble can be expected to be nucleated within one Hubble patch and Hubble time. To make the phase transition an instantaneous event on cosmological timescales and avoid phenomenological problems with large bubbles, we impose the stronger condition $p \gg 1$ during bubble percolation. On the other hand, if $p \ll 1$, the percolation cannot keep up with the expansion of space and a typical Hubble patch does not contain any bubble. This is the condition we want to realize before the transition.

In order to express $p$ in terms of our model parameters, we need an estimate for the determinant factor $K$ introduced in \eqref{eq:def_Gamma}, which is notoriously difficult to calculate. However, as we are interested only in the magnitude of $p$ and our $\psi$ potential is controlled by a single scale $M$ -- recall that the dimensionless parameters $\alpha$, $\beta$ and $\lambda$ can be taken to be $\mathcal{O}(0.1 - 1)$ and the $\psi$-$\phi$ interaction term is also of the order of $\sim M$ when nucleation becomes possible due to the vev of $\phi$  -- we can use dimensional analysis to identify $K \sim M^4 (\gg H_*^4)$. In other words, bubble nucleation in this regime is a one-scale problem, a corresponding argument is, for example, provided in \cite{Linde:1981zj,Callan:1977pt}. In any case, we have
\begin{align}\label{eq:percolation_param}
p 	\sim \frac{M^4}{H_*^4} \exp{\left(-S_E \right)} \sim \exp{\left(250-S_E \right)} \,,
\end{align}
where we assumed $m \sim H_* \sim 10^{-27} \text{eV}$ (corresponding to a phase transition around $z \sim 10^4$)  and $M \sim \text{eV}$. We distinguish two phases:

\paragraph*{\textbf{Quasi-stable phase ($ p \ll1$):}} Bubble nucleation is irrelevant. This is fulfilled throughout the slow-roll phase when $H \gg m$ and $\phi$ is frozen close to its initial value $\phi_\text{ini}$; it still holds when $m \sim H $ and $\phi$ starts to roll  towards zero, provided $S_E > S_E^* \simeq 250 $ [corresponding to an exponential suppression of the right side of \eqref{eq:percolation_param}]. The discussion of the last section shows that this is guaranteed if $\delta_\text{eff} \geq 2 $ (irrespective of the value of $\tilde \lambda$). Because of \eqref{eq:delta_phi}, a sufficient condition for the stability of the false vacuum at early times then reads 
\begin{align}\label{eq:cond2}
\frac{\lambda\, \tilde{\lambda}}{\alpha^2}  \frac{\phi^2_\text{ini}}{M^2} > \frac{2}{9} \,.
\end{align}
{Note that this condition is also sufficiently suppressing bubble nucleation at the earliest possible times when $H \sim M$. \textit{A priori}, these  very old bubbles are problematic because they would have had a lot of time to grow and could have entered our horizon before the time $t_*$, defying the whole idea of NEDE and leading to large-scale anisotropies. To avoid this problem, we simply demand that there is a sufficiently suppressed probability that a single bubble would have been nucleated into \textit{any} of the Hubble patches that will have entered our particle horizon by the time $t_*$. This requires that the percolation parameter at that early time is $p|_{H \sim M} < 10^{-56}$ which translates to  $S_E \gtrsim 130$ which is indeed  a weaker condition than the one derived before. }

\paragraph*{\textbf{Percolation phase ($ p \geq 1 $)}:} Bubbles of true vacuum are being nucleated. This happens for our numerical example when the condition $S_E < 250$ is met. {The corresponding percolation time $t_*$ is then implicitly determined through $p(t_*) \simeq 1$,  which also defines the critical value $\delta_\text{eff}^* = \delta_\text{eff}(t_*)(< 1.5)$.} This phase should be short on a scale $1/H_*$ to {prevent} bubbles {from growing} to cosmological size. In other words, $p$ quickly increases to {values $ p  \gg p(t_*) \simeq 1$, describing} a strong burst of nucleation events. For example, for $S_E \to 10$ as $\phi \to 0$, the percolation rate {could} grow as large as $p \sim 10^{104}$, {which is} an extremely high rate.\footnote{This huge percolation parameter sets our proposal apart from similar scenarios in an inflationary context where the scales $M$ and $H_* \simeq m$ are less clearly separated \cite{Copeland:1994vg,Cortes:2009ej}.} However, we will see that in our case the percolation is completed before that maximal value is reached.

In order to obtain a more quantitative picture, we first derive the time evolution of $\phi$. Since its dynamical equation,
\begin{align}\label{eq:eom_clock}
\ddot \phi + 3 H \dot \phi + m^2 \phi =0\,,
\end{align}
is controlled by a single timescale, $ 1/H \sim 1/m$, its first zero crossing takes place roughly within one Hubble time after it has dropped out of slow-roll. In fact, \eqref{eq:eom_clock} can be solved analytically for a radiation-dominated Universe in terms of Bessel functions,
\begin{align}\label{eq:phi_attr_parabolic}
\phi(x) = \sqrt{2} \, \Gamma(5/4) \, \phi_\text{ini} \, x^{-1/4} \,  J_{1/4}\left(x / 2\right) \,,
\end{align}
where we introduced $x=m / H$ and $\Gamma$ is the Gamma function. Here, we have already chosen the attractor branch which we identified by expanding $\phi(x)$ in the limit $x \ll 1$. The normalization is such that $\phi \to \phi_\text{ini}$ as $x \to 0$.
A quick numerical evaluation of \eqref{eq:phi_attr_parabolic} yields the zero of $\phi$ at
\begin{align}\label{eq:cond_percolation}
H_c / m \simeq 0.18 \,,
\end{align}
where $H_c \lesssim H_*$ is the value of $H$ at zero crossing time $t_c$.  There is a critical value $\delta_\text{eff}^*\, (< 1.5)$ for which the phase transition becomes efficient, implicitly defined by $p|_{\delta^*_\text{eff}} \simeq 1 $ (corresponding to $S_E(\delta_\text{eff}^*)\simeq 250$ in our numerical example). We can use \eqref{eq:delta_phi} to infer the corresponding value of $\phi$,
\begin{align}\label{eq:phi_star}
\phi^2_* = M^2 \frac{\alpha^2}{9 \lambda \tilde \lambda} \, \left( \delta_\text{eff}^*  - \delta\right) \;.
\end{align}
Using \eqref{eq:cond2}, we derive an upper bound on $ \phi_* / \phi_\text{ini}$,
\begin{align}\label{eq:bound_phi_star}
\frac{ \phi_*^2}{\phi^2_\text{ini}} < \frac{1}{2} \left( \delta_\text{eff}^* - \delta\right) < 1 \,.
\end{align}
The corresponding value of $H_*$ can then be inferred numerically from \eqref{eq:phi_attr_parabolic}. In particular,  for $ 0 < ( \delta_\text{eff}^* - \delta) \lesssim 1.5$,  we find  $0.18 < H_* / m \lesssim 0.6$. {Later we will further tighten the upper bound to ensure a short percolation phase.} The slope of $\phi$ {at $t_*$} varies between  $-2.2 \, \phi_\text{ini} <\dot \phi_* / H_* <-0.5 \, \phi_\text{ini}$. {In this work, we are mostly interested in the regime where $\phi_* \simeq 0$. This then translates to $\dot \phi_* / H_* \simeq - 2.2 \, \phi_\text{ini}$, which will be used for our later estimates. It would be straightforward to discuss the more general case, albeit being of minor significance for our order one estimates.}

Now we have gathered all ingredients to calculate the duration of the percolation phase.  Approximating $\Gamma(t)$ as a linear exponential, $\Gamma(t) \propto \exp(\bar\be t)$, around $t=t_*$, we obtain an estimate for the inverse duration,
\begin{align}
\bar{\beta} \equiv - \frac{d S_E}{dt} \simeq \frac{\dot \Gamma}{\Gamma}  \,,
\end{align}
where we assumed that we can neglect the time dependence of $K$. We further use that $S_E$ is only a function of $\delta_\text{eff}$ for $\tilde \lambda / \lambda \ll 1$, which allows us to further evaluate $\bar{\beta}$,

\begin{align}\label{eq:beta_bar}
H_* \bar{\beta}^{-1} & \simeq - H_* \left( \frac{d S_E}{d \delta_\text{eff}} \frac{d \delta_\text{eff}}{dt} \right)^{-1} \nonumber \\
					& \simeq \left[ \frac{d (\lambda S_E)}{d \delta_\text{eff}}\right]^{-1} \frac{\alpha^2}{36 \tilde \lambda} \left(\frac{M}{\phi_\text{ini}}\right)^2 \, \frac{\phi_\text{ini}}{\phi_{*}} \,.
\end{align}
This expression can be simplified by replacing $\phi_*$ defined in \eqref{eq:phi_star} and using the inequality~\eqref{eq:cond2},
 \begin{align}\label{eq:beta_bar2}
H_* \bar{\beta}^{-1} < \left[ \frac{d (\lambda S_E)}{d \delta_\text{eff}}\right]^{-1} \frac{1}{4 \sqrt{2}} \left(\delta^*_\text{eff}-\delta \right)^{-1/2} \lambda \,.
\end{align}
Physically, this constitutes a bound on the maximal time bubbles have to grow before they start colliding. As we want the phase transition to be completed at least within one Hubble time, we demand $H_* \bar{\beta}^{-1} < 1$. A sufficient condition for this to be true is then
\begin{align}\label{eq:cond_lambda00}
\lambda < (4 \sqrt{2})  \sqrt{\delta^*_\text{eff}-\delta} \left[ \frac{d (\lambda S_E)}{d \delta_\text{eff}}\right]\,.
\end{align}
This constitutes only a very mild restriction on the allowed values of $\lambda$. In fact, a numerical evaluation shows that $d (\lambda S_E) / d \delta_\text{eff} \in [10, 10^{3}]$ for $\delta^*_\text{eff} \in [0, 1.5]$. This implies that unless $\sqrt{\delta^*_\text{eff}-\delta}$ is not hierarchically suppressed, the bound \eqref{eq:cond_lambda00} can be fulfilled for any weakly coupled theory with $\lambda < 1$.

As our phase transition occurs in a dark sector at zero temperature there is no plasma which would slow down the propagation of the bubble walls. We can therefore assume that they expand at the speed of light. As a result, $\bar{\beta}^{-1}$ is the maximal size a bubble can reach. When this maximal bubble collides and dissipates it will lead to perturbations within the dark sector of the same size. These structures can leave an imprint on the photon fluid {through their gravitational interactions} and that way -- at least in principle -- lead to directly observable structures in the CMB. These structures subtend an angle 
\begin{align}
\theta_\text{NEDE} = \frac{1}{\bar{\beta} \, a_* D_\text{rec}}\,,
\end{align}
where we introduced the comoving angular diameter distance at the time of recombination,
\begin{align}\label{eq:D_rec}
D_\text{rec} = \int_{0}^{z_\text{rec}}\!\! dz \, \frac{1}{H(z)}\,.
\end{align}
We can approximate it based on a fiducial cosmology (more precision is not needed at this stage),
\begin{align}
D_\text{rec} \simeq \frac{1}{H_0} \int_0^{1100} dz \,\left[ 0.3 \,(1+z)^3 + 0.7\right]^{-1/2} \simeq 3.2/H_0 \,,
\end{align}
which then yields
\begin{align}
\theta_\text{NEDE} \simeq 0.4^\circ  \times H_* \bar{\beta}^{-1} \left( \frac{g_*(T_*)}{3.9} \right)^{-1/2} \, \left(\frac{5000}{z_*} \right) \, \left(\frac{h}{0.7}\right) \sqrt{1- f_\text{NEDE}} \,,
\end{align}
where $H_0 \equiv\, 100 h \, \kmsMpc$.
In particular, for $ H_* \bar{\beta}^{-1} \simeq 1$ and $z_* \simeq 5000$ the bubbles give rise to structures\footnote{The CMB cold spot is about 5 degrees on the sky for comparison. It is an intriguing possibility that such a feature in the CMB could be related to NEDE~\cite{Vielva:2003et,Cruz:2004ce}} that correspond to an angle of half a degree, which is just larger than the maximal angular resolution of the CMB observation of\footnote{This value is compatible with~\cite{Liddle:1991tr} where it is argued that the CMB is sensitive only to structures with co-moving size greater than about $10 h^{-1}\, \text{Mpc} $.} $\sim 2 \pi / 2500 \simeq 0.14^\circ $. We leave the question as to whether the amplitude of the corresponding structures would be large enough to leave an observable imprint for future work. After all, we can impose the bound 
\begin{align}\label{eq:constraint_beta}
H_* \bar{\beta}^{-1} < 0.4 \times \left( \frac{z_*}{5000} \right)\,  \left( \frac{g_*(T_*)}{3.9} \right)^{1/2} \, \left(\frac{0.7}{h}\right) \,  \left(1- f_\text{NEDE}\right)^{-1/2} \, ,
\end{align}
which then prevents bubbles from growing to observable size.\footnote{Note that NEDE still influences the CMB \textit{indirectly}, because it changes the background evolution and supports adiabatic perturbations that source the gravitational potential also ``felt'' by the photons.} 
This also tightens the upper bound on $\lambda$ in  \eqref{eq:cond_lambda00},
\begin{align}\label{eq:cond_lambda0}
\lambda < 2.3 \times \left( \frac{z_*}{5000} \right)\,  \left( \frac{g_*(T_*)}{3.9} \right)^{1/2} \, \left(\frac{0.7}{h}\right) \sqrt{\frac{\delta^*_\text{eff}-\delta}{1-f_\text{NEDE}}} \left[ \frac{d (\lambda S_E)}{d \delta_\text{eff}}\right]\,.
\end{align}
which evidently is still compatible with any value $\lambda < 1$ and, hence, does not constitute a meaningful parameter restriction. 

In fact, we will mostly impose the even stronger bound  $H_* \bar{\beta}^{-1} < 10^{-2}$, which can be easily achieved if $\lambda < 0.1$.  {We do this for different reasons.} First, a percolation phase that is significantly shorter than $1/H_*$ justifies the use of an effective description where the phase transition happens instantaneously on cosmological timescales (we will discuss such an effective model in Sec.~\ref{sec:cosmo_model}). {Second}, it justifies the approximation of $\Gamma(t)$ as a linear exponential, used to derive $\bar{\beta}$ in \eqref{eq:beta_bar}.  Third, as argued after \eqref{eq:def_Gamma}, it also ensures the applicability of the semi-classical approximation used to calculate the tunneling rate $\Gamma$.  In the following, we will therefore assume that $\lambda < 0.1 $. However, we also have to make sure that $\lambda$ is not hierarchically small to still be compatible with both our estimate of $K$ used to derive \eqref{eq:percolation_param} and the condition $\tilde{\lambda} \ll \lambda$. 

Finally, having a small value of $\lambda$ also simplifies the expression for $S_E$. To be precise, provided $\lambda < 0.02$, we have $S_E \lesssim 250$ (as needed for an efficient percolation) for values of $\delta_\text{eff}$ below $ \simeq  0.2$, which then justifies using a linear approximation for \eqref{eq:SEeff},
\begin{align}\label{eq:SE_linear}
\lambda \, S_E \simeq \frac{\pi^2}{6} \alpha_1 \delta_\text{eff}\,.
\end{align}
This, in turn, allows us to approximate $\delta^*_\text{eff}$ as
\begin{align}\label{eq:delta_eff_star}
\delta^*_\text{eff} &\simeq  \frac{6}{\alpha_1 \, \pi^2} \lambda \, S_E^* \nonumber \\
				& \simeq 	0.11 \times  \left( \frac{\lambda}{0.01}\right) \left(\frac{S_E^*}{250} \right) \,.
\end{align}

In short, percolation becomes very efficient for $\lambda < 0.1 $ lasting only a fraction of a Hubble time before it has covered the entire space with bubbles of true vacuum. We can therefore {describe} the phase transition {on cosmological scales} as an instantaneous process at time $t_*$ implicitly determined by $p(t_*) \simeq 1$. 
During this short burst the percolation parameter $p$ increases exponentially until bubble percolation is completed after the time $\bar{\beta}^{-1} $ (before $p$ would reach its peak value at $\phi = 0$).  

\subsection{Bubble Coalescence and Decay}\label{sec:coalesence}
Before we turn to the bubble dynamics, there is an additional decay channel for the energy stored in the (false) vacuum of $\psi$ besides the bubble wall condensate. Part of it goes into oscillations of $\phi$ around the true vacuum. The reason is that $\phi$ after the phase transition does, in general,  not have a vanishing vev, which becomes apparent from Fig.~\ref{fig:profile} in the limit $r \to 0$. If we want to calculate this part, we have to take into account that the mass of $\phi$ increases due to the non-vanishing vev of $\psi$. In other words, oscillations in the $\phi$ direction around the true vacuum experience a steeper potential than the ones around the false vacuum. From \eqref{eq:action2},
\begin{align}
m^2 &\to m^2 + \tilde{\lambda} \, \psi_2^2 \nonumber \\
		& =m^2 + \mathcal{O}(1)\times \frac{\tilde{\lambda} \, \alpha^2}{\lambda^2} M^2
\end{align}
where we used  $\psi_2 = \mathcal{O}(1) \alpha M /\lambda$, which follows from \eqref{eq:psi_min}, to derive the second line. We further assume that the vev of $\phi$ remains constant during the phase transition,\ $\phi(r) \simeq \phi_*$. As we have seen, this is a good approximation for $\tilde \lambda / \lambda \ll 1$.  After the transition, i.e., for $t_* + \bar{\beta}$, we then get 
\begin{align}\label{eq:rho_phi}
\bar{\rho}_{\phi}|_{t_{*} + \bar{\beta}^{-1}} 	&= \frac{1}{2}\left(m^2 + \mathcal{O}(1)\frac{\tilde{\lambda} \, \alpha^2}{\lambda^2} M^2 \right) \phi_*^2 \nonumber \\
									&= \mathcal{O}(1) \times \left( \delta^*_\text{eff} - \delta \right)f_\text{NEDE} \, \bar{\rho}_* \,,
\end{align}
where we neglected the term proportional to $m^2$ and used \eqref{eq:phi_star} and \eqref{eq:cond3}. This shows that for a generic choice of model parameters the potential energy in $\phi$ contributes a sizeable fraction to the energy budget even though it was negligible before the transition. 
In this work, however, we focus on the special case where most of NEDE is converted into kinetic {and gradient} energy of the bubble walls. 
Explicitly, if we require $\bar{\rho}_\phi / \bar{\rho}_* \lesssim 10^{-2} $, which is enough suppression to ignore oscillations in the $\phi$ sector; this translates to a rather mild parameter tuning through \eqref{eq:rho_phi},\footnote{We intend to relax this condition in future work. This opens the way towards providing a fraction of dark matter (DM) in terms of decaying NEDE.}
\begin{align}\label{eq:cond8}
 \delta^*_\text{eff} - \delta \lesssim \frac{10^{-2}}{ f_\text{NEDE}} \sim 10^{-1}\,,
\end{align}
where $\delta_\text{eff} = \mathcal{O}(0.1)$ as in \eqref{eq:delta_eff_star} and $\delta < \delta_\text{eff}$.
In particular, this is only weakly affecting the bound on $\lambda$ in \eqref{eq:cond_lambda0}. 

While the space is being filled with {true vacuum} bubbles {their walls quickly accelerate to the speed of light} and start to collide. {As bubble} percolation becomes efficient very quickly, i.e., {it raises to values $p \gg p(t_*) \simeq 1$ in a small fraction of a Hubble time}, {we can describe this as an instantaneous process}. {The subsequent collision phase is complicated to describe except for the idealized case of two equally sized bubbles~\cite{Hawking:1982ga}.}  {From a phenomenological perspective, this coalescence phase risks polluting the CMB with non-scale-invariant density perturbations arising from the colliding wall structures.} For example, exactly these perturbations strongly constrain models that use a first-order phase transition to end inflation, giving rise to {an analogy of} the ``Big Bubble constraint'' (see, for example,~\cite{Copeland:1994vg}). In our case, however, this {would be happening} in a dark sector and it is not clear whether the imprint in the CMB would be strong enough. In any case, {in the two-field model,} as we have argued before, this problem (or feature) can be avoided by imposing the constraint \eqref{eq:constraint_beta}, preventing bubbles from growing to observable sizes. 

In the next step, the condensate of colliding bubble walls decays. In fact, there are several channels for dissipating {its} energy. If the phase transition is effectively happening in vacuum in the dark sector decoupled from the visible sector, the energy released by the bubbles is converted into kinetic and gradient energy of the bubble walls. To keep the discussion as general as possible, we will consider three generic ways how the gradient energy in the bubble walls can dissipate. The  first two are model independent, and the last {one} is model-dependent. 

\paragraph{\textbf{Gravitational Waves:}} Some of the gradient energy will dissipate in gravitational radiation. To estimate the fraction of the gradient energy that dissipates into gravitational radiation, we will assume that the transverse and traceless part of the energy momentum tensor, which sources gravitational waves, is of the same order as the released vacuum energy $\Pi_\psi \sim \rho_\text{NEDE}$ \cite{Binetruy:2012ze}, suppressing the tensor structure for simplicity. A typical dimensional estimate of the amount of gravitational waves produced is then obtained by observing that, from the linearized Einstein equation, we have for the gravitational wave amplitude $\ddot h \sim 16 \pi G \Pi_\psi$, and, assuming the duration of the phase transition to be parametrized by $\bar\be$ as in the previous subsection, we find $\ddot h \sim \bar\be \dot h\sim \bar\be^2 h$. The energy density radiated away in gravitational waves is then $\rho_{GW} \sim \dot h^2/(32 \pi G)\sim 8\pi G\Pi_\psi^2/\bar\be^2$, or in other words, only a fraction of $\rho_\text{NEDE}$ is typically radiated away in gravitational waves
\beq
\frac{\rho_{GW}}{\rho_\text{NEDE}} \sim 3~(H_*\bar\be^{-1})^2 ~f_\text{NEDE} ~\ll 1~.
\eeq
\paragraph{\textbf{Cosmic fluid:}}  The remaining and dominant part of the gradient energy will redshift away. The details of this process are complicated to describe because the corresponding state of colliding bubbles is highly inhomogeneous and anisotropic, characterized by structures whose density contrast  $ \delta_{\psi}(t, \mathbf{x})  =  \rho_{\psi}(t, \mathbf{x}) /  \bar{\rho}_\psi (t) - 1$ exceeds unity.   The crucial observation is that the scale of these non-linearities,  $\bar{\beta}^{-1}  \ll 1/H$, stays below cosmological length scales [this is a consequence of demanding that the bubble condensate does not leave any direct imprint in the observed CMB power spectrum; cf.\ Eq.~\eqref{eq:constraint_beta}]. This raises the question how these structures, which provide a sizable fraction of the energy budget, affect the expansion history.  Asked differently, is there an effective fluid description of a colliding bubble condensate valid on cosmological scales $\gg \bar{\beta}^{-1}$? To our knowledge this question has not been answered before, and  a complete discussion would require a fully \textit{relativistic} coarse-graining procedure, which goes beyond the scope of our work.\footnote{A quantitative statement can be made before bubbles start to collide. Taking the thin-wall limit, it was shown that the effective equation of state parameter approaches $1/3$ due to the relativistic movement of the bubble walls~\cite{Watkins:1991zt}. However, within our model percolation is very efficient and bubbles start to collide well within a Hubble time, making this result inapplicable. Besides, the thin-wall approximation is not valid for the values of $\delta^*_\text{eff}\, (< 1.5)$ we are considering.}  However, we can sharpen our intuition by looking at the \textit{non-relativistic} case. There,  small-scale non-linearities are known to lead to a \textit{positive} excess pressure on large scales, corresponding to an ideal fluid with a non-vanishing equation of state parameter~\cite{Peebles:2009hw,Baumann:2010tm},
\begin{align}\label{eq:w_eff}
w_\text{eff} \simeq \frac{1}{3} \langle v_\psi^2 \left( 1 + \delta_{\psi} \right)\rangle \,,
\end{align}
where $v_\psi (t, \mathbf{x})$ is the velocity field associated with $\psi$ and $\langle \circ \rangle$ denotes the smoothing procedure over small-scale non-linearities as introduced in~\cite{Baumann:2010tm}.  We note that \eqref{eq:w_eff} assumes a negligible contribution from the gravitational potential, which requires a highly non-virialized system (for virialized systems  $w_\text{eff} = 0$). The important point is that $w_\text{eff}$ is not zero as one would naively expect for non-relativistic matter; it rather receives a positive contribution that scales with $\delta_{\psi}$. Moreover, any anisotropic stress, $\Pi_\psi$, averages to zero, which resonates with the observation that there is no preferred direction on sufficiently large scales. Of course, these results hold only for $v^2_\psi \ll 1 $, whereas for a realistic bubble wall condensate $v^2_\psi \to 1$ as the walls quickly accelerate to relativistic speeds. Nevertheless, based on the non-relativistic result, we also expect in the relativistic case the emergence of an excess pressure on large scales, whose value will depend on the precise tunneling potential (which controls the size of non-linearities within the bubble wall domains). To be precise, rather than an equation of state parameter of $1/3$, valid for a radiation fluid in linear perturbation theory, we expect its value to move towards stiffness, i.e., $1/3 < w_\text{eff} < 1 $. 

We can also provide a heuristic argument, which relies on the observation that on small scales the bubble wall condensate gives rise to a sizable amount of anisotropic stress, $\Pi_\psi$. In homogeneous setups, this component  is known to dilute as $1/a^6$~\cite{Barrow:1981pa,Turner:1986tc}, mimicking a stiff fluid with $w=1$. The tendency of the large-scale fluid towards stiffness, $w_\text{eff} > 1/3$, can then be understood as a consequence of the anisotropic contributions on small scales. 

\paragraph{\textbf{ Additional channels:}} There could be additional model-dependent decay channels for the $\psi$ fluid, so that the remaining fluid, which has not yet been converted into gravitational waves, could decay into radiation. In principle, one could have an almost instant decay of the $\psi$ fluid into some dark or visible radiation component immediately after the phase transition. {In addition, }one might consider the simple possibility that the $\psi$ condensate decays into $\phi$ particle excitations. However, this is suppressed by a very stringent bound on the coupling scale $\tilde \lambda$ following from naturalness considerations, which we will derive in the next section. 

To keep the discussion as general as possible, we will in our phenomenological treatment of NEDE consider all three possible decay channels and model NEDE after the phase transition as an effective fluid with an equation of state parameter $1/3 < w^*_\text{NEDE} <1$. Of course, $w^*_\text{NEDE}$ is  generically time-dependent; however, as the impact of NEDE on cosmological observables is sharply localized around its decay time, we can approximate it as a constant. For example, in our baseline model we will fix it to $w^*_\text{NEDE} = 2/3$, which we believe to be a natural choice, but we will also allow it to vary within its physical range when we discuss {different model} extensions of our baseline model.

\subsection{Radiative Stability}\label{radstab}
So far, the only restrictions on our model parameters {arose} from \eqref{eq:cond1}, \eqref{eq:cond_lambda0} and the tuning in \eqref{eq:cond8} (if we want to study a regime for which $\phi$ oscillations are suppressed after the decay). In particular, $\alpha$ and $\beta$ can still be $\mathcal{O}(1)$. This means that our model is controlled only by the coupling constants, $\tilde \lambda$ and $\lambda$, satisfying $\tilde \lambda \ll \lambda < 0.1$, as well as the two mass scales, $m$ and $M$, satisfying $m \ll M \ll M_{pl}$. The hierarchy of the latter is fixed by \eqref{eq:cond3}, 
\begin{align}\label{eq:hierarchy}
\frac{M^2}{m^2} = \mathcal{O}(1)  \times \frac{\sqrt{\lambda^3 \,f_\text{NEDE}}}{ \alpha^2} \frac{M_{pl} }{ H_*} \gg 1 \,,
\end{align}
which evaluates to $\sim 10^{53}$ if we plug in phenomenologically plausible numbers. This raises the question of how we can stabilize it against quantum corrections. After all, the ultralight mass $m$ will receive quantum corrections from its coupling to the much heavier $\psi$ field. Working perturbatively around the $\psi = 0$ background, the dominant contribution comes from the diagram

\begin{align}
\delta m^2 &=
\begin{gathered}
{\includegraphics[width=2.5cm]{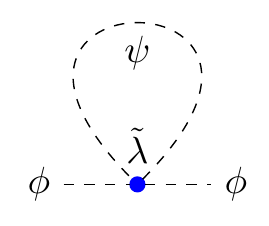}}
\end{gathered}
\nonumber \\
&= \frac{\tilde \lambda}{2} \frac{1}{(4 \pi)^{d/2}} \, \Gamma ( 1-d/2 ) \left( \beta M^2 \right)^{d/2-1} \,,
\end{align}
where we used dimensional regularization. After properly renormalizing this expression, we find a correction (suppressing logs but keeping factors of $\pi$) of the order of $\tilde{\lambda} \, \beta \, M^2 / (32 \pi^2) $. If we allow for a mild tuning of $m^2$ of one order of magnitude, we obtain an upper bound on $\tilde \lambda$:
\begin{align}\label{eq:cond5}
\tilde \lambda \leq \mathcal{O}(1) \times 10^3 \frac{m^2}{\beta M^2} \ll 1 \;.
\end{align}
This is an extremely small coupling bounded from above by $\sim 10^{-50}$ {when} using the estimates suggested in \eqref{eq:M} and \eqref{eq:m}. While this ensures that our assumption $\tilde \lambda \ll \lambda$ can be easily justified it also raises concerns about the effectiveness of our trigger mechanism. In fact, if the coupling scale is too small, it can lead to a violation of condition \eqref{eq:cond2}, which would mean that bubble nucleation is not shutting off in the very early Universe. We therefore assume that $\tilde \lambda$ saturates its naturalness bound, which then via \eqref{eq:cond3} {and \eqref{eq:cond2} yields} a lower bound on $\phi_\text{ini}$,
\begin{align}\label{eq:phi_bound}
\mathcal{O}(1) \times 10^{-9} \, \frac{\delta}{0.1} \,\left(\frac{\lambda}{0.01}\right) \, \left(\frac{H_*/m}{0.2}\right)^2\, \left(\frac{f_\text{NEDE}}{0.1}\right)  < \frac{\phi_\text{ini}^2}{M_{pl}^2} \ll  1 \,,
\end{align} 
where the upper limit followed from { condition \eqref{eq:cond_4}, ensuring that $\phi$ is } sub-dominant before the decay. We also find that we can easily have   $\phi^2_\text{ini} = 10^{-8} M^2_{pl}$ without introducing any additional parameter tunings, making sure that we safely stay away from any quantum gravity regime. Having a small coupling also raises the question as to whether our assumption $\kappa \ll 1$ used to derive the Euclidian action was justified as $\kappa \propto 1/ \sqrt{\tilde \lambda}$ (see the discussion in Sec.~\ref{sec:2-field}). Again, using the upper bound on the coupling, we derive from \eqref{eq:kappa}
\begin{align}
\kappa^2 = \mathcal{O}(1) \times 10^{-6}  \,  \left(\frac{\lambda}{0.01} \right) \left(\frac{\delta}{0.1} \right)\,,
\end{align}
which indeed is sufficiently small. We also note that diagrams with more vortices are suppressed by additional powers of $\tilde{\lambda}$ and, hence, always sub-dominant even for huge (but sub-Planckian) values of the external momenta.

Moreover, having an explicit expression for $\tilde \lambda$ allows us to further simplify the inverse percolation time $\bar{\beta}$ in \eqref{eq:beta_bar}; using \eqref{eq:phi_star}, it evaluates to
\begin{align}\label{eq:percol_efficiency}
H_* \bar{\beta}^{-1} & \simeq \left[ \frac{d (\lambda S_E)}{d \delta_\text{eff}}\right]^{-1} \frac{\alpha}{12 } \frac{\sqrt{\lambda}}{\sqrt{\tilde \lambda}}\, \frac{M}{\phi_\text{ini}} \left( \delta_\text{eff}^* - \delta \right)^{-1/2}.
\end{align}
We again substitute the upper bound on $\tilde \lambda$, approximate $S_{E}(\delta_\text{eff})$ as in \eqref{eq:SE_linear} and use \eqref{eq:cond3}, which then yields
\begin{align}\label{eq:beta_final}
H_* \bar{\beta}^{-1} & = \mathcal{O}(1) \times \, 10^{-3}\, \left(\frac{\delta}{0.1}\right)^{1/2}\,\left(\frac{f_\text{NEDE}}{0.1}\right)^{1/2}\, \left(\frac{{0.1}}{\delta^*_\text{eff}-\delta}\right)^{1/2} \left(\frac{M_{pl}/\phi_\text{ini}}{10^4} \right)\, \left(\frac{H_*/m}{0.2} \right)\, \left( \frac{\lambda}{0.01}\right)^{3/2}\,,
\end{align}
where we assumed parameter values in accordance with our previous discussion, and the $\mathcal{O}(1)$ factor accounts for the uncertainty in the upper bound on $\tilde \lambda$. The lesson from this is that we can use $\lambda$ as a dial to realize percolation phases that can be extremely short, lasting $\ll 1/H_*$, {when} $\lambda \ll 1$ but also rather long, up to $\sim 1/H_*$, when $\lambda \sim  1$. Note, however, that the last limit has to be treated with care, as different approximations we used rely on $\lambda$ being somewhat small. Alternatively, we can achieve a longer duration by demanding $\delta_\text{eff}^*-\delta < {0.1}$, which, however, constitutes a parameter tuning. As we will see in the next section, the value of $H_* \bar{\beta}^{-1}$ controls the strength of the gravitational wave signal, which becomes stronger the longer the phase transition lasts.

We are now able to provide for illustration an explicit example for a consistent set of model parameters corresponding to a phase transition at $z_* = 5000$ with $f_\text{NEDE} = 10 \%$ and $H_*/m =0.2$. When we confront our model with data in Sec.~\ref{sec:param_extraction}, scanning over all consistent choices of parameters, we will see that this choice falls within the $95 \%$ C.L.\  of the model's mean cosmological parameters. We can further take $\lambda = 0.01$, $\beta \simeq \alpha \simeq 1$, which {in turn} determines $\delta = 9 \lambda \beta / \alpha \sim 0.1$ as well as the mass scales $M=0.04 \, \text{eV} $ and $m \sim 2.0 \times 10^{-27} \, \text{eV}$ from \eqref{eq:M} and \eqref{eq:m}, respectively. From \eqref{eq:percolation_param} we then obtain the percolation condition as $S_E^* \simeq 230$, which via \eqref{eq:delta_eff_star} fixes $\delta_\text{eff}^* \simeq 0.1$. {By imposing a very mild one-digit tuning  of $\beta$ (or $\alpha$), we can then ensure}  $\delta - \delta_\text{eff}^* \simeq {0.1}$ in accordance with \eqref{eq:cond8}. In order to keep loop corrections under control, we choose $\tilde \lambda \simeq 10^{3} m^2 / M^2 \simeq 2.3 \times 10^{-48}$ saturating the upper bound in \eqref{eq:cond5}. With these choices the microscopic system is fully determined. It amounts due to \eqref{eq:beta_final} to a percolation time $H_* \bar{\beta}^{-1} \sim 10^{-3}$, corresponding to a quick transition on cosmological timescales.

\subsection{Gravitational Waves}

\begin{figure}
 \centering 
{\includegraphics[width=11cm]{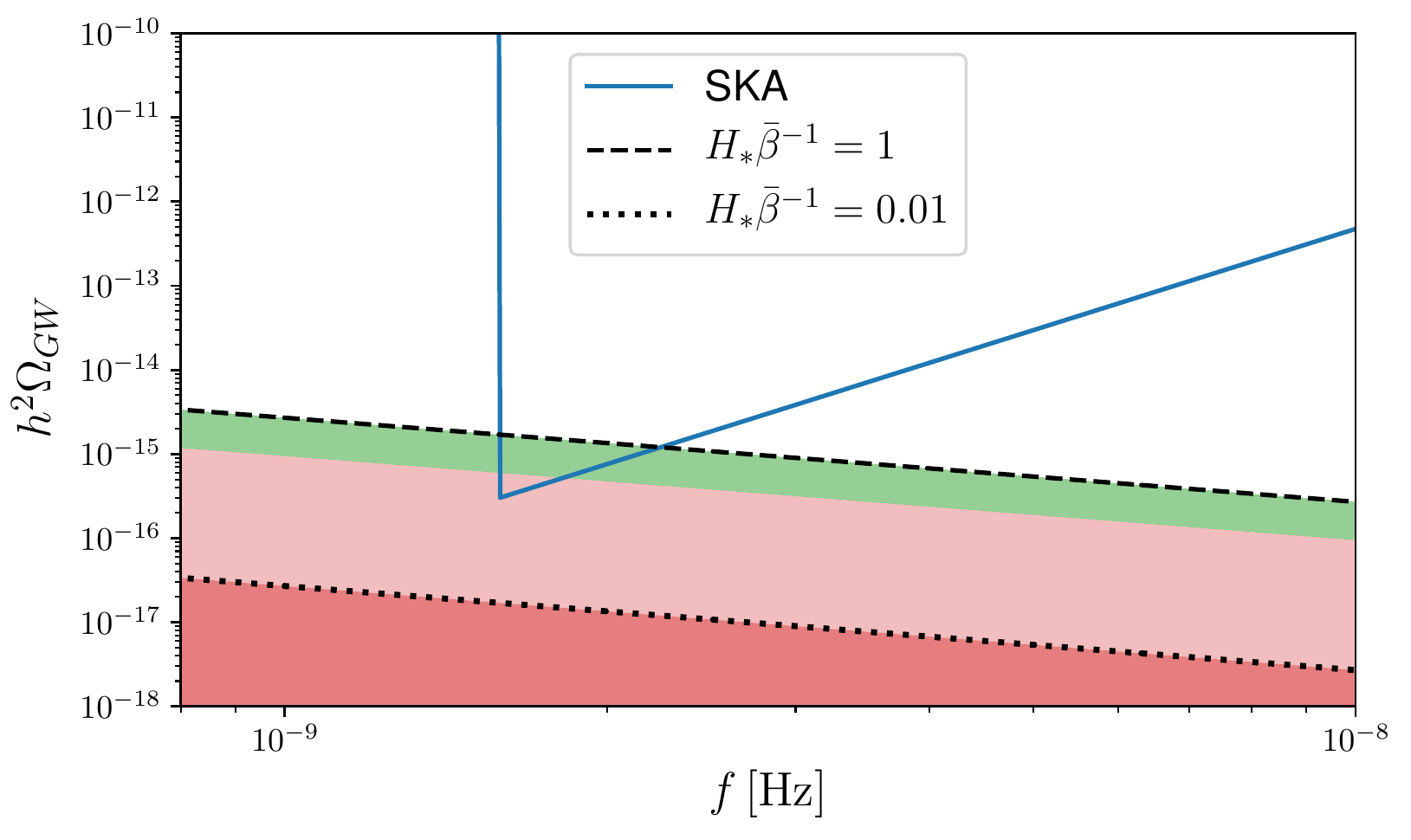}}
 \caption{Sensitivity curve (blue solid) of the \textit{Square Kilometre Array} (SKA) in terms of the dimensionless energy density of gravitational waves. Below the black dotted line (dark red) we can describe the phase transition as an instantaneous process on cosmological scales, making it accessible to our effective description in Sec.~\ref{sec:cosmo_model}. On the other hand, pushing $\bar{\beta}$ towards cosmological scales, i.e., $\bar{\beta}^{-1} \to H^{-1}$, might lead to an observable gravitational wave signal. In this limit, we also expect the colliding bubble walls to leave a direct imprint in the CMB (green shaded region). We used $f_\text{NEDE} = 0.14$ and $z_* = 5300$ as of Tab.~\ref{tab:means_NEDE_LCDM}.}
\label{fig:GW}
\end{figure}

A unique feature of our model, setting it apart from the second-order scenario, is the production of a stochastic background of gravitational waves. Here, we {estimate} the corresponding spectrum (for a review see e.g.\ \cite{Caprini:2018mtu}). This task is greatly simplified by the fact that the transition occurs in a dark sector at zero temperature. So there is no plasma slowing down the expansion of the  bubble walls, and gravitational waves are  produced only during the collision of bubbles of characteristic size $1/\bar{\beta}$. The reason is that colliding bubbles, as opposed to a single expanding bubble, produce a non-vanishing anisotropic stress which, in turn, sources gravitational waves (for seminal papers see \cite{Kosowsky:1991ua,Kosowsky:1992rz}). The spectrum in terms of the dimensionless energy density can be calculated by using the ``envelope approximation'' first introduced in \cite{Kosowsky:1992vn}, resulting in\footnote{We cite from~\cite{Caprini:2018mtu}. With regard to their notation, we set $\kappa_\phi = 1$, $\kappa_v = 0$ and $v_w =1$ (as there is no coupling to any plasma) and identified $\alpha=f_\text{NEDE}/(1-f_\text{NEDE})$ (assuming that the transition {still} takes place during the radiation-dominated epoch) as well as $\beta = \bar{\beta}$.}
 \begin{align}\label{eq:spectrum_gw}
 h^2 \Omega_{GW} = 3.8 \times 10^{-8} \left(H_* \bar{\beta}^{-1} \right)^2 \left( \frac{f_\text{NEDE}}{0.1} \right)^2 \left( \frac{3.9}{g_*(T_*)}  \right)^{1/3} S_{GW}(f)\,,
 \end{align}
with spectral shape 
\begin{align}
 S_{GW}(f) \simeq \frac{3.8\left( f/f_* \right)^{2.8}}{1 + 2.8 \left( f/f_*\right)^{3.8}}\,.
\end{align}
The peak frequency as measured today is extremely small,
\begin{align}\label{eq:peak_f}
f_* \simeq 2.6 \times 10^{-17} \,  \text{Hz} \, \frac{1}{\left(H_* \bar{\beta}^{-1}\right)} \left(\frac{1+z_*}{5000} \right) \left( \frac{g_*(T_*)}{3.9}  \right)^{1/6}\,.
\end{align}
The spectrum grows as $f^3$ for $f \ll f_*$ and falls off as $f^{-1}$ for $f \gg f_*$. Only in the latter regime do we have the hope of detecting a signal once we reach the typical frequency of \textit{pulsar timing arrays} at $f \sim\, 10^{-9} \text{Hz}$~\cite{Moore:2014lga}. Substituting \eqref{eq:peak_f} into \eqref{eq:spectrum_gw}, we find
\begin{align}
 h^2 \Omega_{GW} \simeq   1.3 \times 10^{-15} \, \left(H_* \bar{\beta}^{-1} \right)\,\left(\frac{10^{-9}\, \text{Hz}}{f}\right)\,\left( \frac{f_\text{NEDE}}{0.1} \right)^2 \left( \frac{3.9}{g_*(T_*)}  \right)^{1/6}\,\left(\frac{1+z_*}{5000} \right)\,.
\end{align}
In Fig.~\ref{fig:GW}, we illustrate that for $\left(H_* \bar{\beta}^{-1} \right) \to 1$, which can be achieved for large enough values of $\lambda (< 1)$  or $\delta_\text{eff}^*-\delta < {0.1}$, we just reach the claimed peak sensitivity\footnote{{This has to be understood as a necessary condition for detection. A more complete analysis needs to consider the signal-to-noise ratio, too.}} of the \textit{Square Kilometre Array} at $h^2 \Omega_{GW} \sim 10^{-15} $~\cite{Carilli:2004nx,Janssen:2014dka,Bull:2018lat}. This limit also forces us into  the green shaded region where the CMB becomes potentially sensitive to structures produced by the bubble collisions (as discussed in Sec.~\ref{sec:percolation}). On the other hand, if we want the phase transition to be quick, as we do for the most part of this work, we are forced below the dotted line. In that case, the signal falls short of the experiment's claimed peak sensitivity.

The overall picture can change significantly if we look at different sophistications of our model. For example, by coupling the scalar field to the standard model sector, we expect to enhance the gravitational wave signal due to sound waves produced in the primordial fluid as well as turbulence effects. At the same time, we would be more sensitive to the perturbations produced during the collision phase, because they could be directly imprinted in the photon fluid. Also, the above derivation assumes a purely radiation-dominated Universe and does not take into account the admixture of the NEDE component in the cosmic fluid, which will chance the numerical value of the signal (not the order of magnitude, though). 

In summary, while we have provided a first discussion of the model's distinct signatures, both possibilities, gravitational waves and bubble wall imprints in the CMB, demand a more complete analysis.

\section{Cosmological Model} \label{sec:cosmo_model}

Simulating the percolation of bubbles with their subsequent collision and dissipation phase is a complicated task. We therefore build our analysis on different simplifying assumptions relating to the evolution of the background and the perturbations, separately.

\begin{enumerate}
\item Bubble nucleation happens instantaneously on cosmological timescales. In the previous section, we have argued that this is satisfied for our two-field model for a wide range of natural parameter choices, requiring only a mild suppression of $\lambda$. For example, for $\lambda \simeq 0.01$ we found in \eqref{eq:beta_final} that $H_* \bar{\beta}^{-1}\sim 10^{-3} $ [this choice, due to \eqref{eq:constraint_beta}, also ensures that the CMB observation is not resolving spatial structures created by the largest bubbles].

\item The condensate of colliding vacuum bubbles can be described as a fluid with effective equation of state parameter $ > 1/3$ on large scales $\gg \bar{\beta}^{-1}$ (see the discussion in Sec.~\ref{sec:coalesence}).  There is also a radiation component arising from its decay.  As the impact of both fluids on cosmological observables is localized around the decay time, we can approximate the two-component fluid as a single fluid with a constant equation of state parameter.  This assumption could be relaxed in future work by explicitly modeling a two-component fluid.

\item Motivated by our microscopic model \eqref{eq:action2}, we assume that the time $t_*$ at which the phase transition starts is determined by the mass $m$ of a sub-dominant trigger field $\phi$. It is initiated according to \eqref{eq:cond_percolation} around the time where $\phi$ enters its oscillatory stage and the percolation parameter $p>1$ for the first time.  

\item The fluid perturbations start with vanishing shear stress $\sigma_\text{NEDE}$ directly after the transition. This is motivated by the fact that perturbations created due to the collision dynamics have a wavelength too short to leave an imprint on the CMB power spectrum, provided condition \eqref{eq:constraint_beta} is fulfilled.  Otherwise, the perturbation dynamics is fully characterized by the background equation of state parameter $w_\text{NEDE}$, the rest-frame sound speed $c_s$, and the viscosity parameter $c_\text{vis}$.

\end{enumerate}

In accordance with these assumptions, we do not have to resolve the $\psi$ dynamics explicitly. We will rather introduce a single fluid that exhibits a \textit{sudden} change in its  equation of state, switching from $-1$ to $w^*_\text{NEDE}$ at time $t_*$. We note that we will refer to both phases of matter as the NEDE fluid. \textit{Before} the transition it models the (classical) vacuum energy of the field $\psi$ and \textit{after} the transition it describes the energy density stored in the bubble wall condensate and its decay products. In contrast, we will track the evolution of $\phi$ explicitly. As we will see, this is necessary to keep track of {spatial} fluctuations of the transition time. While these assumptions will allow us to relatively quickly arrive at a first phenomenological check of our model, we intend to relax and scrutinize them further in our subsequent work.

\subsection{Instantaneous Transition}
Our effective model parametrizes the NEDE equation of state parameter as

\begin{align}
w_\text{NEDE}(t) = 
\begin{cases}
-1   &\text{for} \quad t< t_* \,, \\ 
w_\text{NEDE}^*  &\text{for}\quad t \geq t_* \,.
\end{cases}
\end{align}
This describes a sudden change at time $t_*$ from a constant energy density to a decaying fluid with equation of state parameter $w_\text{NEDE}^*$. The energy density then scales as
\begin{align} \label{eq:rho_EDE_bg}
\bar{\rho}_\text{NEDE}(t) = \bar{\rho}^*_\text{NEDE} \left( \frac{a_*}{a(t)}\right)^{3[1+w_\text{NEDE}(t)]}\,,
\end{align}  
where $\bar{\rho}^*_\text{NEDE}$ has been introduced in \eqref{eq:def_frac} denoting the energy density of NEDE directly before the phase transition. It can be related to the potential energy of our two-field model via \eqref{eq:def_EDE} (strictly speaking this definition also contains a minuscule energy contribution arising from $\phi$). Also note that $\bar{\rho}_\text{NEDE}$ is continuous across the transition. {From a microscopic perspective} this is justified because the transition happens instantaneously and all of the field's potential energy is {ultimately converted to kinetic and gradient energy of the bubble walls}. Later, for our parameter extraction, we will use the fraction of NEDE before its decay, $f_\text{NEDE}$, as defined in \eqref{eq:def_frac}, rather than $\bar{\rho}_\text{NEDE}^*$ as a free parameter because it gives us a more intuitive idea about the amount of NEDE. 

{We denote the jump} of a function $f(t)$ across the phase transition at time $t_*$ {as}
\begin{align}\label{eq:disc_bracket}
\left[ f \right]_\pm = \lim_{\epsilon \to 0} \left[ f(t_*-\epsilon) - f(t_*+\epsilon) \right] \equiv f^{(-)} - f^{(+)}\;.
\end{align}
{The jump of the Friedmann equations then yields}
\begin{subequations}
\label{eq:matching_bg}
\begin{align}
\label{eq:JumpH}
\phantom{\left[ \dot H \right]_\pm} \left[ H \right]_\pm &= 0\;, \\
\label{eq:JumpHdot}
\left[ \dot H \right]_\pm &= 4 \pi G (1 + w^*_\text{NEDE}) \bar{\rho}_\text{NEDE}^* \, ,
\end{align}
\end{subequations}
where we used {the continuity of the background energy density}, 
\begin{align}\label{eq:Jump_rho}
[\bar{\rho}]_\pm=0 \,,
\end{align}
{and assumed the continuity of all other equation of state parameters (except for NEDE), i.e., $[w_i]_\pm = 0$. } In Appendix~\ref{sec:matching}, we present an alternative but equivalent derivation of \eqref{eq:JumpH} based on Israel's matching equations~\cite{Israel:1966rt}.

\begin{figure*}
 	\subfloat[Different fluid components. $\rho_\text{NEDE}$ decays quickly after the transition (dotted line), and $\rho_{\phi}$ is sub-dominant before and vanishing after the transition.]
	{\includegraphics[width=3in]{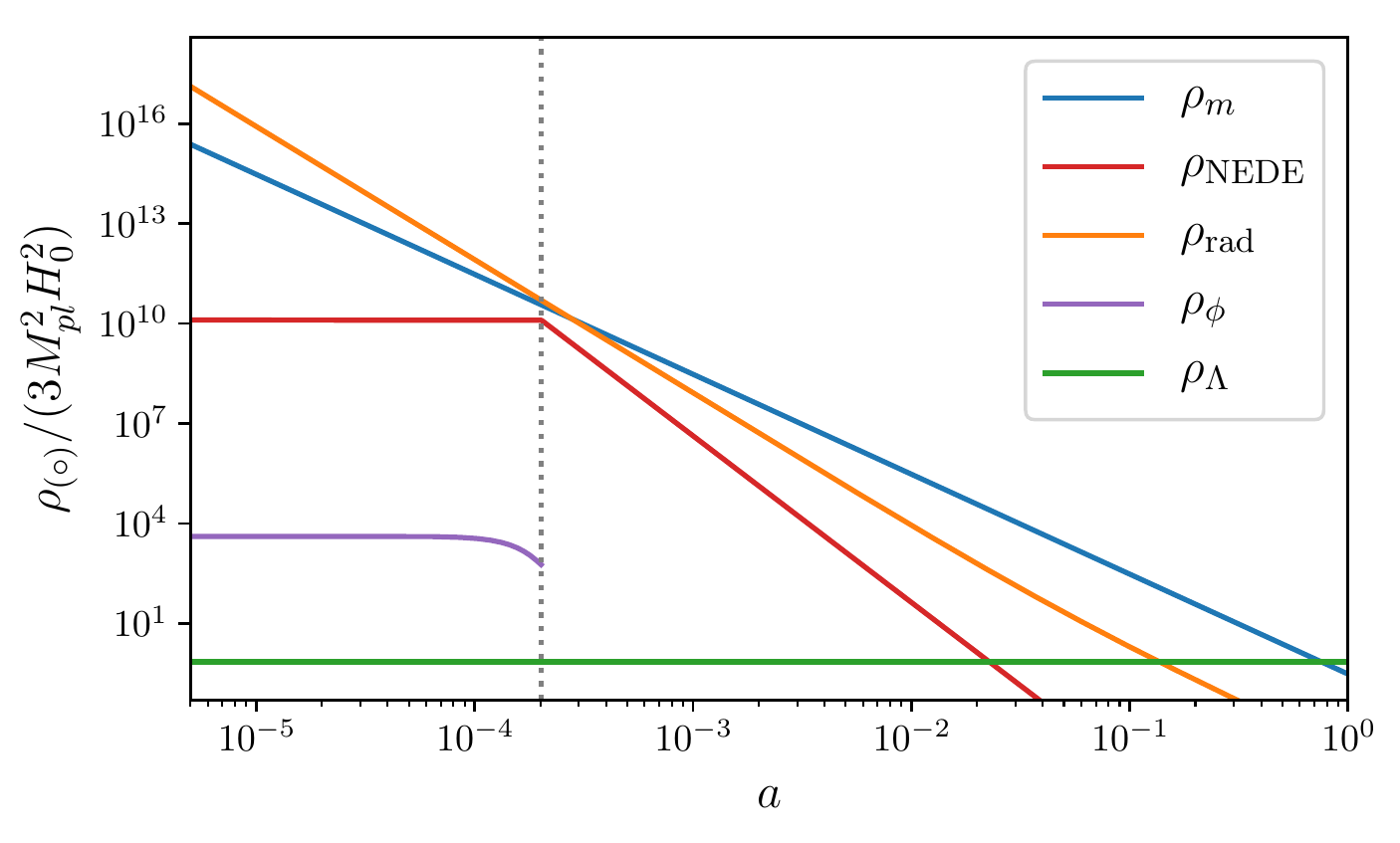}}
	 \label{fig:bg_fld} 	
	  	\quad
 	\subfloat[Effective equation of state parameter, $w_\text{eff}=p_\text{tot}/\rho_\text{tot}$. During radiation and matter domination the dashed and the solid line is approached, respectively. ]
	{\includegraphics[width=3in]{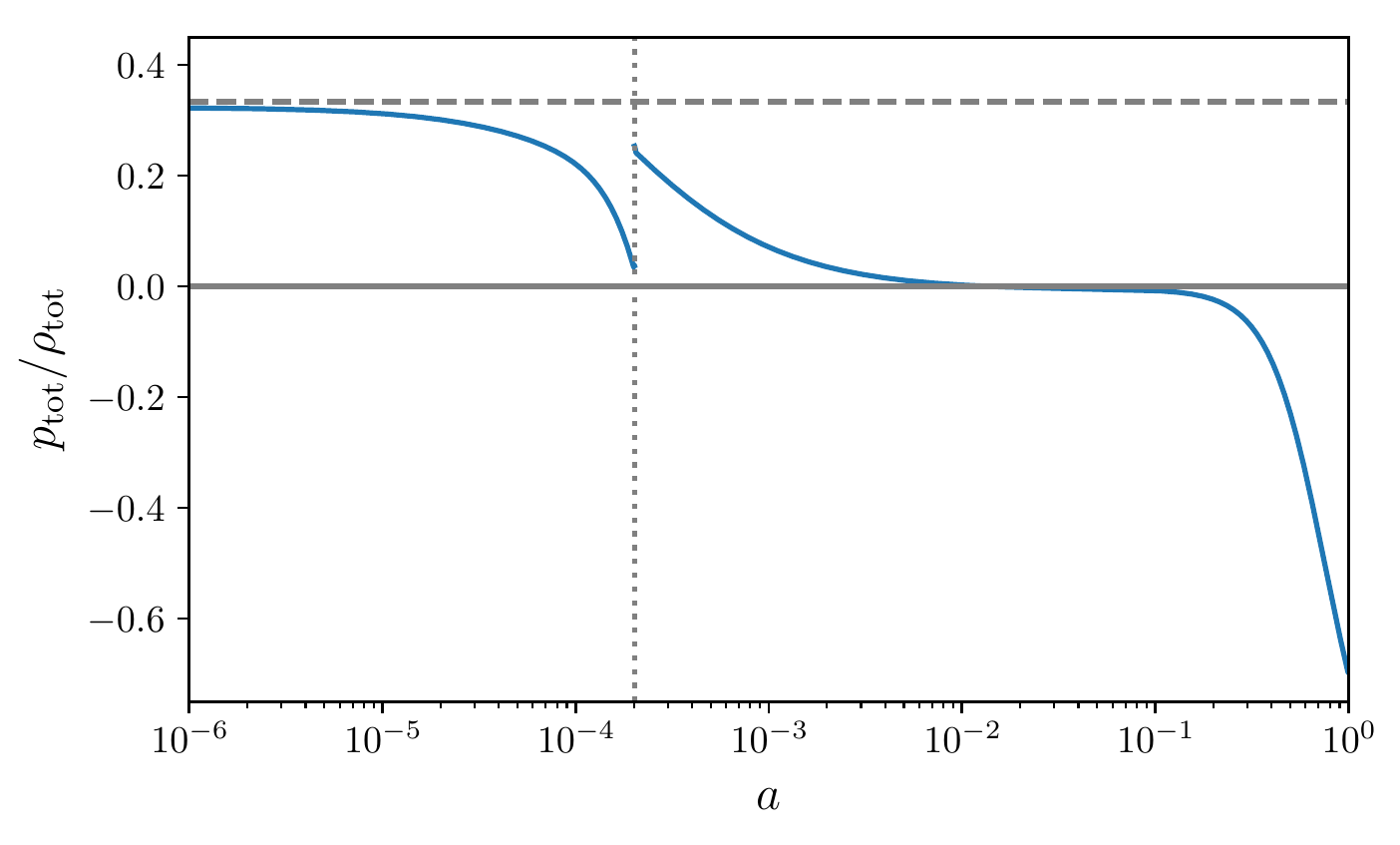}}%
	\label{fig:bg_fld_eos}
 \caption{Background evolution for the best fit model in Tab.~\ref{tab:means_NEDE_LCDM} (w/ SH$_0$ES) as a function of the scale factor $a$. The transition is characterized by a jump in the effective equation of state parameter (dotted line). $\rho_m = \rho_b + \rho_\text{cdm}$ and $\rho_\text{rad}$ are the matter and radiation density, respectively, and $\rho_\Lambda$ is the cosmological constant contribution. }
\label{fig:bg}
\end{figure*}

We also introduce the field  $\phi$  to {trigger} the phase transition. Its on-shell dynamics before the transition is controlled by \eqref{eq:eom_clock}. As it is a sub-dominant field component, its back-reaction on the geometry can be neglected, and it therefore quickly locks into the attractor solution valid for a radiation-dominated Universe [as derived later in \eqref{eq:phi_attr}]. On the background level its single purpose is to provide a time switch triggered by its first zero crossing at $\eqref{eq:cond_percolation}$. 
Since its vev does not change during the (almost instantaneous) phase transition, we have $\left[ \phi \right]_\pm =0$. Integrating its dynamical equation in \eqref{eq:eom_clock} over an infinitesimal time interval yields $ [\dot \phi ]_\pm=0$. As argued before, we are studying a regime where the scalar is also sub-dominant after the phase transition. In terms of our field theory model this corresponds to the limit \eqref{eq:cond8}. {A summary of the background evolution is provided in Fig.~\ref{fig:bg} (with values corresponding to our best-fit base model discussed in Sec.~\ref{sec:param_extraction}). The left panel shows the composition of the background energy density, $\bar{\rho} = \sum_i \bar{\rho}_i$, and the right panel depicts the discontinuity in the effective equation of state parameter.}

\subsection{Perturbation Matching}
{Before the phase transition, $\rho_\text{NEDE} = const$ and NEDE fluctuations can be consistently assumed to vanish. After the transition, on the other hand, they are necessarily produced due to the gravitational sourcing of their equations. We will also see that they are crucial for the phenomenological success of our first-order phase transition. However, their implementation involves a subtlety due to the fact that the transition occurs during the CMB epoch where observable modes have already entered the horizon. As a result, we cannot use the usual super-horizon approximation to initialize perturbations in the NEDE fluid after the phase transition. Instead, we have to derive the full set of matching equations describing how  cosmological perturbations evaluated directly before and after the transition at time $t_*$ can be related. As the central result of this section, we will express the initial NEDE density perturbation, $\delta \rho_\text{NEDE}^*(\mathbf{x}) \equiv \rho_\text{NEDE}(t_*, \mathbf{x}) - \bar{\rho}_\text{NEDE}(t_*)$, and the initial divergence of the fluid velocity, $\theta_\text{NEDE}^* (\mathbf{x})$, in terms of fluctuations of the space-like transition surface $\Sigma$ for both sub- and super-horizon modes. }

{On a technical level, }we introduce a function $q(t, \mathbf{x})$ to define $\Sigma$,
\begin{align}\label{eq:q}
q(t_*, \mathbf{x})|_\Sigma = const \;.
\end{align}
Later, we will identify $\phi(t, \mathbf{x}) = q(t, \mathbf{x})$, implying that the transition happens across a surface of constant trigger field. Fluctuations in $\phi$ will therefore lead to small spatial fluctuations of the decay time {around the background value $t_*$}, explicitly $\delta q^* (\mathbf{x}) \equiv q(t_*, \mathbf{x}) - \bar{q}(t_*) = \delta \phi(t_*,\mathbf{x})$, which, in turn, will give rise to adiabatic fluctuations in the EDE fluid {(provided we choose adiabatic initial conditions for $\phi$)}.

In the following, we first derive the general matching equations that are needed to consistently implement a generic transition in cosmological perturbation theory. After that, we apply the result to our particular two-field model. We work in the synchronous gauge, {for which the metric perturbations in momentum space take the form}
\begin{align}\label{eq:sync}
ds^2 =  - dt^2  +a(t)^2 \left(\delta_{ij} + h_{ij} \right)dx^i dx^j \,,
\end{align}
where 
\begin{align}
h_{ij} = \frac{k_i k_j}{k^2} h + \left(\frac{k_i k_j}{k^2} - \frac{1}{3} \delta_{ij}\right)6 \eta\,,
\end{align}
and $h=\delta^{ij} h_{ij}$. This choice leaves a residual gauge freedom, $x^0 \to x^0 + \bar{\xi\,}^0 $ and $x^i \to x^i + g^{ij}\partial_j \bar{\xi} $, with 
\begin{subequations} \label{eq:residual}
\begin{align} \label{eq:residual1}
\bar{\xi} &= - D_2 \, a^2 \int dt \frac{1}{a^2} + D_1 \, a^2 \, ,\\
\bar{\xi}_0 & = D_2\, ,
\end{align}
\end{subequations}
where we introduced $D_1=D_1(\mathbf{k})$ and $D_2 = D_2(\mathbf{k})$, which are time-independent functions of the spatial momenta. The  metric perturbations then transform as
\begin{subequations}
\label{eq:residual_metric}
\begin{align}
\Delta_{\bar{\xi}} h &= 2 \frac{k^2}{a^2} \bar{\xi} + 6 H \bar{\xi_0}, \\
\Delta_{\bar{\xi}} \eta &= - H \bar{\xi}_0 \,.
\end{align}
\end{subequations}
The initial perturbations in the NEDE fluid can be derived from the perturbed Einstein equations that are first-order in time derivatives~\cite{Ma:1995ey}:
\begin{subequations}
\label{eq:Einstein}
\begin{align}
 \frac{1}{2} H \dot h - \frac{k^2}{a^2} \eta &=  4 \pi G \delta \rho \,,  \\
 \frac{k^2}{a^2} \dot \eta&= 4 \pi G \left(\bar{\rho} + \bar{p} \right) \frac{\theta}{a}\, .
\end{align}
\end{subequations}
{We introduced the total divergence of the fluid velocity, $\left(\bar{\rho} + \bar{p} \right)\theta = \sum _i \left(\bar{\rho}_i + \bar{p}_i \right) \theta_i$, as well as the total energy density perturbation,  $\delta \rho = \sum_i \delta \rho_i$, and employed the same definitions as in \cite{Ma:1995ey}, in particular, $i k^i \delta T^0_{\,\,i} \equiv (\bar{\rho} + \bar{p}) a \, \theta  $ and $\delta \rho \equiv = - \delta T^0_{\,\,0} $}.\footnote{The factor $a$ arises from the conversion to cosmological time $t$.} Under the residual gauge freedom, they transform as
\begin{subequations}
\label{eq:residual_EMT} 
\begin{align}
\Delta_{\bar{\xi}} \delta \rho &= \dot{\bar{\rho}} \, \bar{\xi}_0 \;,\\
\Delta_{\bar{\xi}} \delta \theta &=  \frac{k^2}{a} \, \bar{\xi}_0 \;.
\end{align}
\end{subequations}
Using \eqref{eq:residual_metric} and \eqref{eq:residual_EMT}, it is then straightforward to show that the Einstein equations in \eqref{eq:Einstein} are indeed invariant. {In order to obtain a closed system, we need to impose} Israel's matching conditions~\cite{Israel:1966rt}, which we derive in full generality in Appendix~\ref{sec:matching}. They express the metric field after the transition, e.g., $\eta^{(+)}$, in terms of the fields right before it, e.g., $\eta^{(-)}$.  In the synchronous gauge, they read
\begin{subequations}
\label{eq:matching0}
\begin{align}
\left[ \eta \right]_\pm + H_* \left[ \frac{\delta q}{\dot{\bar{q}}} \right]_\pm &=0 \, , \label{eq:jump_eta}\\
\left[ \dot\eta \right]_\pm + \left[  \dot H \frac{\delta q}{\dot{\bar{q}}} \right]_\pm &=0 \, ,\\ 
\left[ h \right]_\pm + 6 [\eta]_\pm &=0 \,  ,\\  
\left[\dot h\right]_\pm + 6 [\dot \eta]_\pm  + 2 k^2 \left[ \frac{\delta q}{\dot{\bar{q}}} \right]_\pm &=0 \,   ,
\end{align}
\end{subequations}
where $\delta q(t,\mathbf{k})$ parametrizes fluctuations in the transition surface around some homogeneous background $\bar{q}(t)$. In the appendix, we also derive the matching for matter fluctuations of a given fluid component $i$ with a constant equation of state parameter during the transition (hence not applicable to NEDE):
\begin{subequations}
\begin{align}
\left[ \delta_i \right]_\pm + 3 H_* \left( 1 + w_i \right) \left[ \frac{\delta q}{\dot{\bar{q}}} \right]_\pm &= 0\,, \label{eq:jump_delta}\\
a_* \left[  \theta_i \right]_\pm  - k^2 \left[ \frac{\delta q}{\dot{\bar{q}}} \right]_\pm &= 0\,, 
\end{align}
\end{subequations}
where as usual $\delta_i = \delta \rho_i / \bar{\rho}_i$. 
Moreover, we show that higher momenta of the distribution functions are continuous across the surface; specifically, for the anisotropic stress we have \ $\left[\sigma_i\right]_\pm = 0$, and similarly for the higher moments in the Boltzmann hierarchy, $\left[F_{\nu l}\right]_\pm = \left[F_{\gamma l}\right]_\pm = 0$ for $l \geq 3$ {(relevant only for neutrinos $\nu$ and photons $g$)}. 

In general, our residual gauge freedom can be fixed independently before and after the transition, corresponding to four \textit{a priori} independent constants $D^{(\pm)}_1$ and $D^{(\pm)}_2$. However, the derivation of the above relations made the assumption that spatial diffeomorphisms $\partial_i \xi$ are continuous across the transition surface.\footnote{It is straightforward to relax this assumption, but it does not add any useful freedom to the calculation.} This requires $\left[ \xi \right]_\pm=0$, which due to \eqref{eq:residual1} {fixes} one of the four constants. Using the transformation of $\delta q$,
\begin{align}
\Delta_{\bar{\xi}} \delta q = \dot{\bar{q}}\, \bar{\xi}_0\, ,
\end{align}
together with the definition in \eqref{eq:residual}, it is then straightforward to show that the matching equations are invariant under the residual gauge freedom provided $\left[ \xi \right]_\pm=0$; in particular, we find
\begin{align}
\Delta_{\bar{\xi}}\left[ \frac{\delta q}{\dot{\bar{q}}} \right]_\pm  = \left[ D_2 \right]_\pm \equiv D_2^{(-)} - D_2^{(+)}.
\end{align}
We use this freedom to impose $\left[ \frac{\delta q}{\dot{\bar{q}}} \right]_\pm = 0$, which simplifies the matching equations significantly:
\begin{subequations}
\label{eq:matching}
\begin{align}
\left[ h \right]_\pm = \left[ \eta \right]_\pm =\left[ \delta_i \right]_\pm=\left[ \theta_i \right]_\pm  =  0 \,, \\
\left[ \dot h \right]_\pm = -6 \left[ \dot \eta \right]_\pm = 6 \left[\dot H \right]_\pm \frac{\delta q_*}{\dot{\bar{q}}_*}\,,
\end{align}
\end{subequations}
where $\left[\dot H \right]_\pm$ is specified in \eqref{eq:JumpHdot}.\footnote{The remaining two constants are then fixed as usual by requiring adiabatic super-horizon perturbations in the metric to scale as $h = C \, k^2 \tau^2$ (with $\tau$ the conformal time and $C$ constant) when fixing initial conditions during radiation domination (see for example \cite{Ma:1995ey}).} 

In summary, all perturbations (except for NEDE perturbations) are continuous, while their first derivatives, in general, are not. In particular, the discontinuity of $\dot \eta$ and $\dot h$ is {set by} the spatial variations of the transition surface $\delta q_* = \delta q(t_*, \mathbf{k})$. The choice of $\delta q_*$ is a physical one, and we will discuss different constructions later.  

The derivation of the junction conditions assumed that the equation of state of a particular matter component $i$ is not changing during the transition, so it {cannot be applied} to NEDE. {In fact, $\bar{\rho}_\text{NEDE} = const$} before the decay, which implies that, unlike with ordinary fluids, the gravitational source term is decoupled  from its perturbations [see the fluid perturbation equation in \eqref{eq:pert_dyn}], and it is consistent to assume
\begin{align}
\delta_\text{NEDE} = \theta_\text{NEDE} = \sigma_\text{NEDE} = F_{\text{NEDE}\, l} = 0 \quad \text{for} \quad t < t_*\;,
\end{align}
where $\sigma_\text{NEDE}$ denotes the anisotropic stress in the NEDE fluid and we allowed for the possibility of higher multipoles $F_{\text{NEDE}\, l}$ (as relevant when going beyond a fluid description). 
{This then allows us to further evaluate the discontinuity of the NEDE perturbations}
\begin{subequations}
\label{eq:matching_EDE}
\begin{align}
\left[ \delta_\text{NEDE} \right]_\pm &= -\delta_\text{NEDE}^{(+)} \equiv -\delta_\text{NEDE}^{*} \, ,\\
\left[ (\bar{\rho}_\text{NEDE} + \bar{p}_\text{NEDE} ) \theta_\text{NEDE} \right]_\pm &= 
 		- \bar{\rho}^*_\text{NEDE} (1 + w^*_\text{NEDE} ) \theta_\text{NEDE}^{(+)} 
 		\equiv -\bar{\rho}^*_\text{NEDE} (1 + w^*_\text{NEDE} )  \theta_\text{NEDE}^{*}  \,.
\end{align}  
\end{subequations}
The initial NEDE perturbations $\delta_\text{NEDE}^*$ and $\theta_\text{NEDE}^*$ can now be derived from Einstein's equations. To be precise, the discontinuity of~\eqref{eq:Einstein} together with \eqref{eq:matching}, \eqref{eq:JumpHdot}, and \eqref{eq:matching_EDE} results in our main result of this section:
\begin{subequations}
\label{eq:EDE_ini}
\begin{align}
\delta^*_\text{NEDE} &= - 3 \left( 1 + w_\text{NEDE}^*\right) H_* \frac{\delta q_*}{\dot{\bar{q}}_*} \; ,\\
\theta_\text{NEDE}^* &= \frac{k^2}{a_*} \frac{\delta q_*}{\dot{\bar{q}}_*} \,.
\end{align}
\end{subequations}
Physically, these equations tell us that for a general transition it would be inconsistent to set the perturbations in the NEDE sector to zero after the transition, as they are linked to $\delta q_*$. Note, however, that no such condition applies to the shear $\sigma_\text{NEDE}^*$ (or {any} higher multipoles $F_{\text{NEDE}\,l>3}$ in the Boltzmann hierarchy). {We can therefore consistently impose} 
\begin{align}\label{eq:EDE_ini2}
\sigma^*_\text{NEDE} = 0 \, \quad \text{and} \quad F_{\text{NEDE}\, l>3}^* = 0\,. 
\end{align}

{We note that the junction conditions apply to both sub- and super-horizon modes. As mentioned before,} this is important as modes we observe in the CMB today with $\ell > \ell_*$ were sub-horizon during the phase transition, where we defined 
\begin{align}\label{eq:l_star}
\ell_* \simeq a_* H_* D_\text{rec} \simeq 160 \times  \frac{1}{\sqrt{1-f_\text{NEDE}}}\,\left(\frac{g_*}{3.9} \right)^{1/2} \left( \frac{0.7}{h}\right) \left( \frac{1+z_*}{5000}\right)\,.
\end{align}
Here, we assumed that the transition occurs during radiation domination and approximated $D_\text{rec}$ as in \eqref{eq:D_rec}.
 Specifically, if the phase transition takes place at $z_* = 5000$ and {$f_\text{NEDE} = 0.1$, modes with $\ell > 180$} are on sub-horizon scales during the transition. 

Finally, we check that the comoving curvature perturbation is conserved during the phase transition. For different fluid components $i$, it is defined in the synchronous gauge as
\begin{align}\label{eq:zeta}
\zeta_i = - \eta - H \frac{\delta \rho_i}{\dot{\bar{\rho}}_i} \,.
\end{align}    
Its conservation amounts to the statement
\begin{align}\label{eq:jump_zetai}
 \left[ \zeta_i \right]_\pm &= - \left[ \eta \right]_\pm - H_* \left[ \frac{\delta \rho_i}{\dot{\bar{\rho}}} \right]_\pm \nonumber \\
&= - \left[ \eta \right]_\pm + \frac{1}{3 \left(1 +w_i \right)} \left[ \delta_i \right]_\pm \nonumber \\
&= 0\,,
\end{align}
where we used the matching conditions \eqref{eq:jump_delta} and \eqref{eq:jump_eta}. Note that the same result was also derived in \cite{Deruelle:1995kd} in an inflationary context. In the special case of adiabatic super-horizon perturbations, i.e., $\delta \rho_i / \dot{\bar{\rho}}_i=\delta \rho_{j} / \dot{\bar{\rho}}_{j }$ for $j \neq i$ and $k \ll a H$, definition \eqref{eq:zeta} implies 
\begin{align}\label{eq:adia}
\zeta_i = \zeta_{j\neq i} \equiv \zeta \quad \text{for} \quad  k^2/(aH)^2 \ll 1\,,
\end{align}
where $\zeta = const + \mathcal{O}(k^2/(aH)^2)$. Things are slightly more complicated for NEDE perturbations, {though}. Before the transition $\zeta_\text{NEDE}$ is not well defined as the second term \eqref{eq:zeta} is indefinite, but directly afterwards we derive
\begin{align}\label{eq:zeta_EDE}
\zeta_\text{NEDE}^* 	&= - \eta_* + \frac{1}{3 \left(1 +w_\text{NEDE}^* \right) }\delta_\text{NEDE}^* \nonumber \\
				&= - \eta_* - H_*  \frac{\delta q_*}{\dot{\bar{q}}_*}  \;. 
\end{align}
The question {as to} whether NEDE perturbations fulfill the adiabaticity condition in \eqref{eq:adia} depends on the choice of $q(t,\mathbf{x})$. Their non-conservation would correspond to the presence of entropy perturbations, of which a possible gauge-invariant definition reads\footnote{This definition is independent of the index $i$ provided there are no entropy perturbations before the transition, implying $\zeta_i = \zeta_{j\neq i}$.}
\begin{align}\label{eq:S_EDE3}
S^*_{NEDE,i} 	&= \zeta_i^* - \zeta_\text{NEDE}^* \nonumber \\
			&=  \frac{1}{3 \left( 1 +w_i \right)} \delta_i^* +H_*  \frac{\delta q_*}{\dot{\bar{q}}_*} \,.
\end{align}

\subsection{Transition Surface} \label{sec:transition_surface}

In our particular model, the transition surface $\Sigma$ is fixed by surfaces of constant $\phi$. Recall that the $\phi$ dependence of the Euclidian action in \eqref{eq:SEeff} controls the tunneling rate, which becomes efficient when $\phi$ drops below the threshold value $\phi_*$, implicitly determined by $p(\phi_*) \simeq1$.  We therefore have
\begin{align}\label{eq:phi=q}
q(t,\mathbf{x}) = \phi(t,\mathbf{x})\,.
\end{align}
First, note that this is a gauge-invariant statement and, hence, valid in any frame. This can be seen easily by considering the transformation of $\delta q$ under general diffeomorphisms:
\begin{align}
\Delta_\xi \delta q = \dot{{q}} \, \xi_0 = \dot{\phi} \, \xi_0 = \Delta_\xi \delta \phi \;,
\end{align}
where we used that $\dot{{\phi}} = \dot{{q}}$. In other words, perturbations in the sub-dominant field $\phi$ set the initial conditions for perturbations in the NEDE fluid after the phase transition. In the synchronous gauge, they fulfill
\begin{align}
\delta \phi '' + 2 \mathcal{H} \delta \phi' + (k^2 + a^2 m^2) \delta \phi & = -\frac{1}{2} h' {\phi}'  \,, \label{eq:pert_phi}
\end{align}
where we found it convenient to work in conformal time $\tau$. We also introduced the conformal Hubble parameter $\mathcal{H} \equiv a H = 1 / \tau$ and used the shorthand  $'=d/d\tau$. On the right-hand side, we substitute the adiabatic solution for super-horizon metric perturbations~\cite{Ma:1995ey}, $h = - \zeta \left(k \tau\right)^2 / 2 $, which is not affected by the presence of the sub-dominant scalar $\phi$. In the limit where $k \to 0$ the corresponding equation becomes
\begin{align}\label{eq:eq_delta_phi_k_0}
\delta \phi '' + 2 \mathcal{H} \delta \phi' + a^2 m^2 \delta \phi & = \mathcal{O}(k^2/\mathcal{H}^2)  \,.
\end{align}
{This} limit is useful as it allows us to extract the adiabatic and entropy modes. Also note that the fluctuation equation becomes formally equivalent to the background equation in \eqref{eq:eom_clock} expressed in conformal time
\begin{align}\label{eq:phi_bg_conf}
\phi'' + 2 \mathcal{H} \phi' + a^2 m^2 \phi &= 0\,.
\end{align}


\subsubsection{Adiabatic Case}

The adiabatic mode corresponds to the trivial solution, which is vanishing at leading order, i.e., $\delta \phi = \mathcal{O}(k^2/\mathcal{H}^2)$.
If we further use\footnote{This result can also be found in~\cite{Ma:1995ey} by identifying $\zeta = -2 \, C + \mathcal{O}(k^2/\mathcal{H}^2)$. }
\begin{align}\label{eq:adiabatic_delta}
\delta_i / (1+w_i) = \frac{1}{4} \, \zeta \, k^2 \tau^2 +\mathcal{O}(k^4/\mathcal{H}^4)  \,,
\end{align}
we can derive from \eqref{eq:S_EDE3} that 
\begin{align}
S_{NEDE,i}^{*} \xrightarrow[k \to 0]{} 0\;.
\end{align}
In other words, there are no super-horizon entropy perturbations created by the transition provided we assume adiabatic initial conditions for $\phi$. 
We also derive the first non-vanishing correction to $\delta \phi$ in the limit $m ^2 \ll H^2 $. To that end, we further evaluate the source term in \eqref{eq:pert_phi} by substituting the background solution
\begin{align} \label{eq:phi_attr}
{\phi} &= {\phi}_\text{ini} \left( 1 - \frac{1}{20} \frac{m^2}{H(t)^2}   \right) + \mathcal{O}\left( m^4 / H^4 \right)\,,
\end{align}
which we obtained by expanding \eqref{eq:phi_attr_parabolic}. Recall that this {expression} is valid only during radiation domination. This is consistent as $\phi$ is a sub-dominant field and, thus, we can neglect its gravitational back-reaction. There is also a second branch, which we set to zero as it becomes quickly negligible. At leading order, we {then} obtain
\begin{align}\label{eq:delta_phi_attr}
\delta \phi  &=  - \frac{1}{420}\, \phi_\text{ini}  \frac{m^2 k^2}{a(t)^2 H(t)^4}\, \zeta  + \mathcal{O}\left(m^4/H^4, k^4/(a H)^4 \right) \,.
\end{align}
In particular, this expression vanishes in the limit $\zeta \to 0$, as expected for an adiabatic (or curvature) mode. We will use it to define adiabatic initial conditions when we implement our model in a Boltzmann code. 

\begin{figure}
 \centering 
{\includegraphics[width=13.cm]{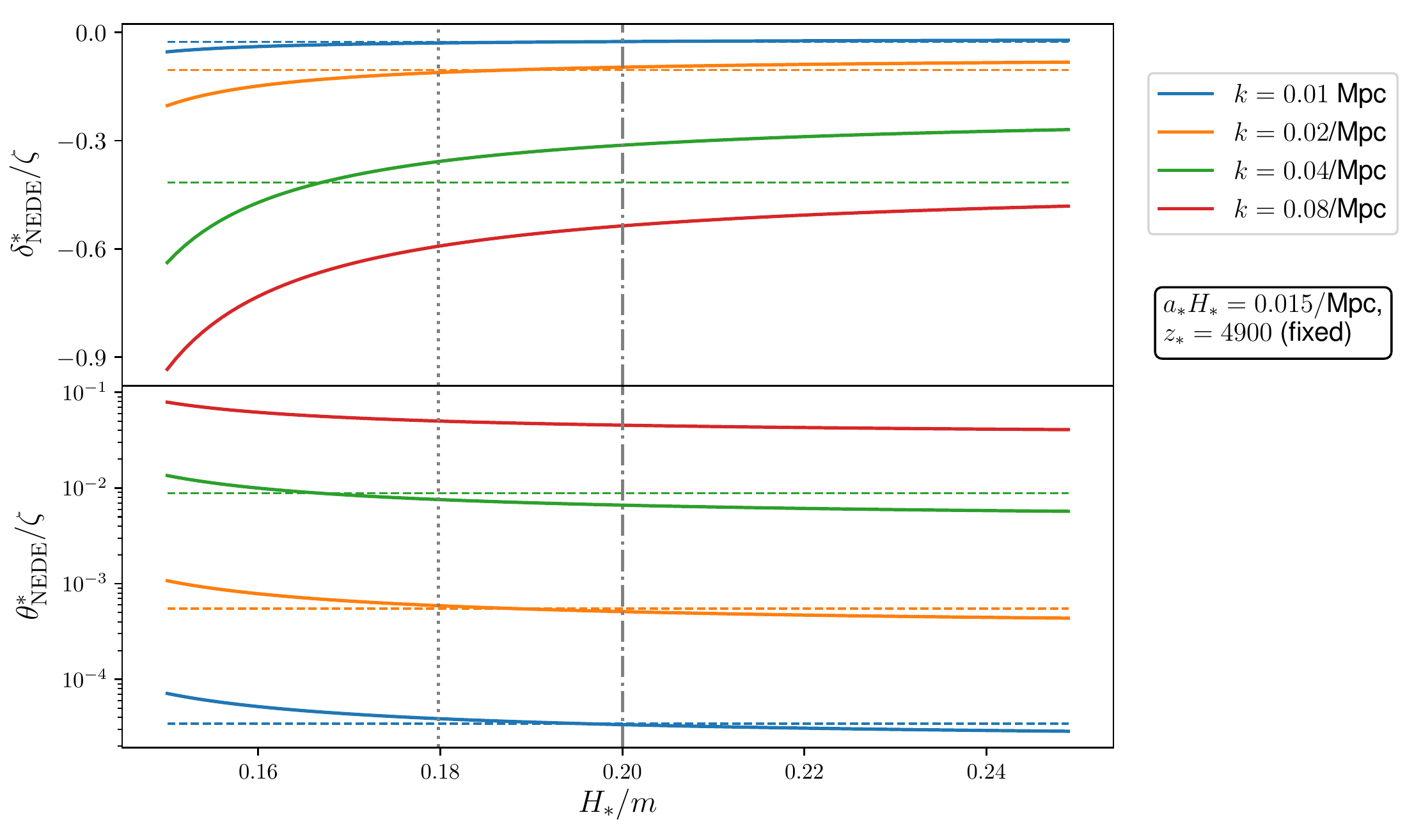}}
 \caption{Dependence of initial fluctuations in NEDE fluid on `trigger parameter' $H_*/m$ for best-fit NEDE as in Tab.~\ref{tab:means_NEDE_LCDM} (w/ SH$_0$ES) and $z_*$ fixed. The dotted line correspond to the case where the phase transition coincides with the zero  of  $\phi$. The dash-dotted line indicates $H_*/m =0.2$ as used in our base model. The dashed {horizontal} lines depict the analytical approximation \eqref{eq:delta_star} valid for super-horizon modes with $k < a_* H_*$ (orange and blue).}
\label{fig:delta_EDE_star}
\end{figure}

Next, we can use \eqref{eq:phi_attr} and \eqref{eq:delta_phi_attr} to derive an expression for $\delta_\text{NEDE}^*$ on super-horizon scales, provided the transition takes place when $H_*/m > 1$. Even though this is not the most interesting case for our model where a typical scenario suggests $H_*/m \simeq 0.2 $, it will still turn out to give a good approximation on super-horizon scales. Plugging the perturbative expressions back into \eqref{eq:EDE_ini} yields at leading order in $k^2 / (aH)^2 $
\begin{align}\label{eq:delta_star}
\delta_\text{NEDE}^* = - \frac{1+w^*_\text{NEDE}}{28} \, k^2 \tau_*^2 \, \zeta \quad \text{and} \quad \theta_\text{NEDE}^* = \frac{1}{84} k^4 \tau_*^3 \, \zeta \,,
\end{align}
which we can compare to adiabatic super-horizon perturbations in the neutrino fluid at the same time~\cite{Ma:1995ey},
\begin{align}
\delta_\nu^* = \frac{1}{3} \, k^2 \tau_*^2 \, \zeta \quad \text{and} \quad \theta_{\nu}^* = \frac{1}{36} \frac{23+4R_\nu}{15+4R_\nu} k^4 \tau_*^3 \, \zeta \,,
\end{align}
where $R_\nu = \rho_\nu / (\rho_\gamma + \rho_\nu) = const$ and $\zeta$ is conserved at leading order. This shows us that even if we have a radiation fluid in both cases, i.e.,  $w^*_\text{NEDE} = 1/3$, perturbations are different already at order $k^2$. While this difference is negligible for super-horizon modes, it becomes important when modes enter the horizon  ($k \tau \geq 1$). For a transition that happens around $z =5000$ these are modes with $k \gtrsim 0.015/\text{Mpc}$ (or ${\ell \gtrsim 180}$). This is exactly the fingerprint left by the trigger field when setting the initial conditions for $\delta_\text{NEDE}$ and $\theta_\text{NEDE}$. In this more general case, we will use our Boltzmann code, which will be introduced in Sec.~\ref{sec:param_extraction}, to numerically determine the evolution of $\delta \phi$. Using our best-fit cosmology, we obtain the curves in Fig.~\ref{fig:delta_EDE_star}, where we plot $\delta_{NEDE,\, k}^*/ \zeta_{k=0}$ and $\theta_{NEDE,\, k}^* (k)/\zeta_{k=0}$ as a function of $H_*/  m$ for different values of $k$. Note that we keep the decay redshift $z_*$ fixed, so variations in the initial fluctuations  result only from the fact that the transition happens at a different stage of the trigger field evolution (recall that $\phi(t)$ depends only on the ratio $H/m$). For example, the dotted line corresponds to the zero of $\phi(t)$ and its first minimum sits towards the left at $H_*/m \simeq 0.12$. We did not include it in the plot because the fluctuations diverge at this point due to the vanishing of  $\dot{\bar{q}}(t)=\dot{\phi}(t)$ in the denominator of \eqref{eq:EDE_ini}. The plot shows that the analytic result in \eqref{eq:delta_star} constitutes a good approximation only for super-horizon modes. In particular, we see that the initial fluctuations after the phase transition become sizable on sub-horizon scales. 

In summary, we find that the dynamics of the trigger field on both the background and perturbation level control the size of initial perturbations in the NEDE fluid after the phase transition. The crucial parameter is given by the ratio $H_*/m$ to which we refer {to} as the trigger parameter. From \eqref{eq:phi_star}, we derive that it is  related to our microscopic model via 
\begin{align}
H_*/m = \phi^{-1} \left( \frac{1}{3} \, M \alpha \frac{\sqrt{ \delta_\text{eff}^*  - \delta} }{ \sqrt{\lambda \tilde \lambda}} \right)\, ,
\end{align}
where $\phi^{-1}$ is obtained by inverting \eqref{eq:phi_attr_parabolic} on the interval before its first minimum, and, in accordance with our previous discussion in Sec.~\ref{sec:percolation}, $\delta_\text{eff}^*$ is implicitly determined by the percolation condition $p|_{\delta_\text{eff}^*} \simeq 1$. To get an idea of its actual value, we can expand $\phi(x)$ around its zero at $x_c \simeq 5.56$, which results in 
\begin{align}\label{eq:taylor}
\phi_* = \frac{d\phi_c}{ dx} (x_*-x_c) \,,
\end{align}
where we remind that $x = m / H $. A quick numerical evaluation of the derivative of \eqref{eq:phi_attr_parabolic} gives ${d \phi_c}/{dx} \simeq  -0.28 \, \phi_\text{ini}$. The bound in \eqref{eq:bound_phi_star} together with \eqref{eq:taylor} then translates to the upper bound 
\begin{align}\label{eq:bound_trigger}
\frac{H_*}{m} < \frac{H_c}{m} \left( 1+ 0.14 \sqrt{\frac{\delta_\text{eff}^* - \delta}{0.1}}\right) \, ,
\end{align}
where we assumed $\delta_\text{eff}^*- \delta_* \leq 0.1$ for the expansion to be valid.
In particular, if we saturate the upper bound in \eqref{eq:cond8}, i.e., $\delta_\text{eff}^*- \delta_* \simeq 0.1$, we obtain $H_* / m < 0.21$.
At the same time, the transition has to happen before the tunneling probability becomes maximal when $\phi \to 0$, leaving only a rather narrow window:
\begin{align}\label{eq:trigger_range}
0.18 < H_*/m \lesssim {0.21} \,.
\end{align}
{The only way to relax the upper bound would be to allow for $\delta_\text{eff}-\delta = \mathcal{O}(1)$, which would, however, lead to sizable oscillations of the trigger field around the true vacuum, incompatible with the assumptions made in our previous analysis. As a result, the range \eqref{eq:trigger_range} constitutes a distinct prediction of our model, which we will confront with data in Sec.~\ref{sec:param_extraction}. }

\subsubsection{Non-Adiabatic Case}

Evidently, \eqref{eq:delta_phi_attr} is only a particular solution of \eqref{eq:pert_phi} to which we can still add a homogeneous one, corresponding to an entropy (or isocurvature) mode. After all, entropy modes are expected to be generated during inflation, because $\phi$ is a sub-dominant field component. 

In the $k \to 0$ limit, the isocurvature solution can be derived from \eqref{eq:eq_delta_phi_k_0}. In fact, we have solved the formally equivalent background equation \eqref{eq:phi_bg_conf} before when we derived \eqref{eq:phi_attr_parabolic}.  We therefore have (valid during radiation domination)
\begin{align}
\delta \phi = \sqrt{2} \, \Gamma(5/4) \, \delta \phi_\text{ini} \, x^{-1/4} \,  J_{1/4}\left(x / 2\right) + \mathcal{O}\left(k^2/\mathcal{H}^2\right)\,.
\end{align}
Recall that $x=m/H$ and $\delta \phi \to \delta \phi_\text{ini}$ as $x \to 0$ (corresponding to early times). Again, we have picked the attractor branch of the more general (two-parameter) family of solutions.  We therefore find
\begin{align}\label{eq:rel_delta_bg_phi}
\delta \phi = \frac{\delta \phi_\text{ini}}{\phi_\text{ini}}\, \phi \,,
\end{align}
where the adiabatic case corresponds to $\delta \phi_\text{ini} = 0$.  The phase transition occurs when ${\phi} \to 0$ (as $H/m \to 0.18$). As $\delta \phi$ tracks the same evolution (up to a rescaling), we also have $\delta \phi \to 0$ at the same time, which, in turn, leads to a suppression of $S_\text{NEDE}^*$ in \eqref{eq:S_EDE3} depending on how close $\phi_*$ (and, hence, $\delta \phi_*$) is to its zero. More explicitly,
\begin{align}
S^{*}_\text{NEDE} 	&= {H_*} \frac{\delta \phi_*}{\dot{{\phi}}_*} \nonumber \\
				&= \frac{1}{2} \, \frac{H_*}{m} \frac{\delta \phi_\text{ini}}{\phi_\text{ini}} \frac{\phi_*}{\frac{d}{dx} {\phi}_*}\,,
\end{align}
where we used \eqref{eq:rel_delta_bg_phi} to derive the last line. {Approximating $d\phi_c/ dx \simeq d\phi_*/ dx \simeq - 0.28 \phi_\text{ini}$ and using \eqref{eq:taylor}, we find}
\begin{align}\label{eq:S_NEDE}
\left|S^{*}_\text{NEDE}\right| 	 \simeq  10^{-5} \left(\frac{m\left(H_c^{-1} - H_*^{-1}\right)}{0.1} \right) \left( \frac{H_*/m}{0.2} \right) \left( \frac{\delta \phi_\text{ini} / \phi_\text{ini}}{10^{-3}}\right) \,.
\end{align}
For the above suggested parameter choice, we would get an entropy mode of the order of $\sim 10^{-5}$, almost matching the size of the adiabatic mode $\zeta \simeq 5 \times 10^{-5}$. However, as NEDE provides only a fraction $\sim 0.1$ of the energy budget, the effect of the entropy mode on the CMB power spectrum is further suppressed. 
To still get an idea of a reasonable phenomenological bound on $S_\text{NEDE}^*$, we can look at the CMB bound on neutrino density isocurvature, which is found to be of the order of $\mathcal{P_{\nu \nu}}/ \left(\mathcal{P_{\nu \nu}} + \mathcal{P_{\zeta \zeta}} \right) \simeq 0.1 $~\cite{Akrami:2018odb}, where $\mathcal{P}_{xx}$ denotes the power spectrum for the mode $x$. Somewhat naively it can be translated to $S_{\nu} / \zeta \lesssim \sqrt{0.1}\simeq 0.3$. 
In our case, this bound will be even weaker by a factor of $f_\text{NEDE} \sim 0.1 $, resulting in $S^*_\text{NEDE}/\zeta \leq \mathcal{O}(1)$. 
This, in turn, might allow modes as large as $S_\text{NEDE}^* \sim 10^{-4}$ without leaving an observable imprint in the CMB. A more precise statement would require that we explicitly track the evolution of NEDE entropy modes in a Boltzmann code when doing a cosmological parameter fit. However, the message here is that we can easily go to a regime where entropy perturbations are phenomenologically irrelevant. 
This can be achieved by either having $\delta \phi_\text{ini} / \phi_\text{ini} \ll 1$ or imposing $m\left(H_c^{-1} - H_*^{-1}\right) \ll 1$. Because of \eqref{eq:bound_trigger}, the latter condition requires $( {\delta_\text{eff}^*-\delta} ) \ll 1$.  
Provided $\delta \phi_\text{ini} / \phi_\text{ini} < 10^{-3}$, the bound derived in \eqref{eq:cond8} is already strong enough to guarantee a sufficient suppression of \eqref{eq:S_NEDE}, as it ensures $m\left(H_c^{-1} - H_*^{-1}\right) \lesssim 0.1$ due to \eqref{eq:bound_trigger}. 

Alternatively, we can also enhance the entropy perturbation by increasing $ \delta \phi_\text{ini}$, at least as long $\delta \phi_\text{ini}/\phi_\text{ini} \lesssim 0.1$ to make the perturbative treatment applicable. In fact, it might be possible that including entropy perturbations to our model even helps with resolving the Hubble tension. We leave a more detailed discussion of their phenomenology within NEDE to future work.  

Finally, if we assume a standard inflationary scenario, the amplitude of the isocurvature perturbation in our trigger field is set by the Hubble scale during inflation, i.e., $\delta \phi_\text{ini} \sim H_{I}$. The absence of entropy perturbations then provides an upper bound on the scale of inflation; explicitly,
\begin{align}
\frac{H_{I}}{M_{pl}} < \mathcal{O}(1) \times 10^{-6} \left(\frac{0.1}{m\left(H_c^{-1} - H_*^{-1}\right)} \right) \left(  \frac{0.2}{H_*/m}  \right) \left( \frac{\phi_\text{ini} / M_{pl}}{10^{-4}} \right)\,,
\end{align}
where the $\mathcal{O}(1)$ factor accounts for the uncertainty in the phenomenological bound on $S_\text{NEDE}^*$.
Given the before mentioned freedom in tuning the phase transition to be close to $\phi=0$ and, hence, $(H_* - H_c) \ll 1 $, this is not a particularly strong bound. 

In summary, our model allows for both natural scenarios with and without a sizable contribution of entropy modes. Whether they are favored by the data remains to be seen. For the remainder of this paper we simply assume {they are} sufficiently suppressed and set $\delta \phi_\text{ini} = 0$.

\subsubsection{Hybrid Early Dark Energy}\label{sec:HEDE}

\begin{figure}
 \centering 
{\includegraphics[width=8.cm]{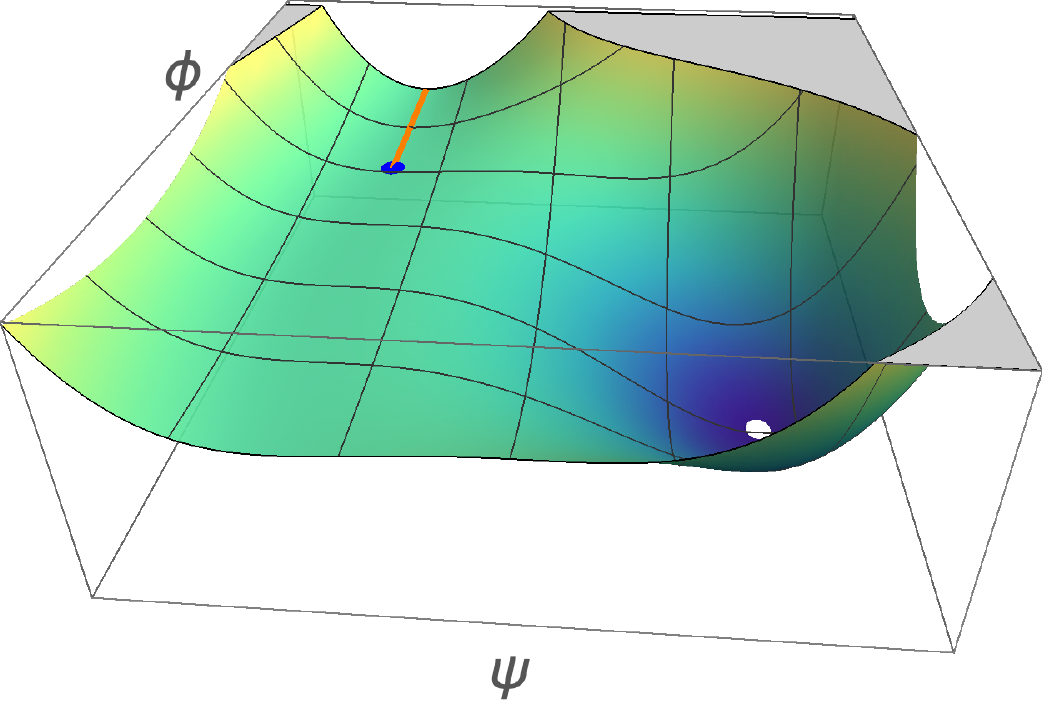}}
 \caption{The two-field potential in \eqref{eq:action2} can also give rise to a second-order phase transition. Similar to the first-order case, the field is initially frozen high up in a potential valley. When the Hubble drag is released around matter-radiation equality, it rolls along the almost flat bottom of the valley (orange line) until it crosses the inflection point (blue dot). Here its trajectory gets deflected towards a deep potential sink around which it will oscillate while dissipating its potential energy. }   
\label{fig:plot3D_hybrid}
\end{figure}

It is also worth pointing out that our cosmological model can capture the physics of a second-order phase transition within a hybrid scenario. For example, this can be achieved in our two-field model \eqref{eq:action2} by setting $\beta = 0$ (this is only for simplicity as $\beta=\gamma \leq 0$ and $\gamma \neq 0$ would also be compatible with this scenario). A schematic picture of such a potential is provided in Fig.~\ref{fig:plot3D_hybrid}. It gives rise to a smooth crossover from an early phase dominated by the vacuum energy stored in $\psi$ to another phase where the field oscillates around its true vacuum. As before, this second phase is triggered when the Hubble drag is released and the sub-dominant field $\phi$ passes the inflection point (blue dot). The crucial observation is that a hierarchy of scales $m \ll M \sim \text{eV}$ is again required to have the transition occur around matter-radiation equality.  As a result, the transition takes places on a cosmologically short timescale $1/ M \ll 1/H \sim 1/m$, which then justifies the use of our cosmological model. Also in this case, \eqref{eq:EDE_ini} can be used to initialize fluctuations in the $\psi$ fluid after the transition. The main difference with NEDE is that the oscillations around the true vacuum correspond to a relatively weak decay. In fact, when averaging over oscillation cycles, the field condensate gives rise to a fluid with a vanishing effective equation of state parameter, which cannot help with the Hubble tension.  The only way of avoiding that conclusion and achieving a quicker decay would consist in tuning the potential such that close to its minimum $\psi_2$ there is an anharmonic direction approximately described by a monomial potential with $\left(\psi - \psi_2\right)^{2n}$ and $n \geq 2$. This, however, would constitute a fine-tuning similar to the one imposed in single-field EDE (see the discussion in Sec.~\ref{sec:EDE}). We therefore believe that the first-order scenario is more natural as it explains the quick decay of NEDE in terms of radiation and anisotropic stress. 

\subsection{Perturbation Equations}\label{sec:pert_eqs}

We will model NEDE {after the phase transition} as a generic fluid, which at the level of perturbations is characterized by its density perturbation $\delta_\text{NEDE}$, velocity divergence $\theta_\text{NEDE}$ and anisotropic stress $\sigma_\text{NEDE}$ (higher moments in the Boltzmann hierarchy {we set to zero,} $F_{\text{NEDE }l\geq 3} =0$). For a time-dependent equation of state parameter $w(t)$, the perturbation equations in the synchronous gauge read~\cite{Poulin:2018dzj,Hu_1998} (for vanishing spatial curvature)
\begin{subequations}
\label{eq:pert_dyn}
\begin{align}
\dot \delta_\text{NEDE} & = - \left( 1+ w \right) \left( \frac{\theta_\text{NEDE}}{a} + \dot h / 2   \right) - 3 \left( c_s^2 - w\right) H \delta_\text{NEDE}  \nonumber \\
&  \hspace{5cm}- 9 \left( 1+ w \right) \left( c_s^2 - c_a^2 \right) H^2 \frac{\theta_\text{NEDE}}{a } \frac{a^2}{k^2} \, ,\\
\dot \theta_\text{NEDE} &= - \left( 1 - 3 c_s^2\right) H \theta_\text{NEDE} + \frac{c_s^2 k^2/a}{1 + w} \delta_\text{NEDE} -  \frac{k^2}{a} \sigma_\text{NEDE}\,.
\end{align}
\end{subequations}
{Here,} we introduced the effective sound speed in the fluid rest-frame, $c_s^2 = \delta p^{(rest)}_\text{NEDE} / \delta \rho^{(rest)}_\text{NEDE}$, as well as the adiabatic sound speed
\begin{align}
c_a^2 = w - \frac{1}{3} \frac{\dot w}{1 +w}\frac{1}{H} \,.
\end{align}
With our assumption $w(t)\equiv w_\text{NEDE}^*=const$, we obtain $c_a^2 = w^*_\text{NEDE}$. In the general approach of \cite{Hu_1998}, these equations are supplemented by (valid for $w \neq -1 $)
\begin{align}\label{eq:shear}
 \left( \dot \sigma_\text{NEDE} + 3 H \sigma_\text{NEDE} \right) = \frac{8}{3} \frac{1}{1+w} c_\text{vis}^2 \left( \frac{\theta_\text{NEDE}}{a} + \frac{\dot h}{2} + 3 \dot \eta \right)\,,
\end{align}
describing the evolution of the anisotropic stress $\sigma_\text{NEDE}$, defined as
\begin{align}
\left(\bar{\rho} + \bar{p}\right)_\text{NEDE} \, \sigma_\text{NEDE} = - \left( \frac{k_i k^j}{\mathbf{k}^2} - \frac{1}{3} \delta_{i}^{\,j} \right) \Pi_{\psi, j}^i \,,
\end{align}
where $\Pi_{\psi,j}^i$ is the traceless part of the energy momentum tensor associated with NEDE.
The above system is then subjected to the initial conditions specified in \eqref{eq:EDE_ini} and \eqref{eq:EDE_ini2}. Here, we will assume the vanishing of the viscosity parameter, $c_\text{vis}=0$, which together with the initial condition \eqref{eq:EDE_ini2} implies the trivial solution $\sigma_\text{NEDE} =0$. Note that this is not in contradiction with having a non-vanishing anisotropic stress arising from the bubble condensate. The point is that the latter contribution is present only on scales that cannot be probed by CMB measurements provided condition \eqref{eq:constraint_beta} on the maximal bubble size $\simeq \bar{\beta}^{-1}$ is satisfied. As a possible avenue for future work, introducing anisotropic stress by having $c_\text{vis} \neq 0 $ and setting the initial value to $\sigma^*_\text{NEDE} \neq 0 $ will enable us to probe the parameter range of our model for which bubble signatures are imprinted in the observable CMB signal. In fact, we will provide a preliminary discussion of this possibility later in Sec.~\ref{sec:extensions} when we allow $c_\text{vis} \neq 0 $ (but still have $\sigma^*_\text{NEDE} = 0 $).

{Finally, as discussed in \cite{Hu_1998}, this parametrization can also mimic the behavior of free-streaming particles for the special choice $(w^*_\text{NEDE},c_{s}^2, c_\text{vis}^2) = (1/3,1/3,1/3)$. However, as we will briefly discuss later, this possibility is not favored by the data, and, therefore we will mostly explore the parameter regime for which $c_\text{vis} = 0$ and $\ w^*_\text{NEDE}=c_s^2=2/3$. As discussed in Sec.~\ref{sec:coalesence}, this is motivated by our expectation that the scalar field condensate is dominated by kinetic energy (rather than potential energy) and small-scale anisotropic stress, which drives the effective equation of state of the NEDE fluid towards stiffer values. Also, we expect $c_s^2$ and $w^*_\text{NEDE}$ to be well described by a constant because the NEDE fluid is relevant only for a narrow redshift window around $z_*$, making cosmological observables rather insensitive to its time dependence.}

\subsection{NEDE Phenomenology}\label{sec:phenomenology}

The primary effect of NEDE is to lift $H(z)$ to higher values before recombination. This changes the anchoring of the inverse distance ladder through a reduction of the (comoving) sound horizon:
\begin{align}\label{eq:r_s}
r_s(z_\text{rec}) = \int_{z_\text{rec}}^{\infty} dz \,  \frac{v(z)}{H(z)}\,,
\end{align}
where $v(z)$ denotes the sound speed in the photon-baryon fluid. In order to keep the angular scales we observe through CMB and BAO measurements fixed, the corresponding distance scales have to be reduced, which is achieved by increasing $H_0$. To be explicit, the CMB observation constrains the angular position of the first peak in the CMB power spectrum:
\begin{align}\label{eq:theta_rec}
\theta_\text{rec} = \frac{r_s(z_\text{rec})}{  D_\text{rec}} \,.
\end{align}
A smaller value of the sound horizon at recombination, $r_s(z_\text{rec})$, then needs to be balanced by reducing the comoving distance $D_\text{rec}$, between us and the surface of last scattering, which due to \eqref{eq:D_rec} is achieved by anchoring the expansion history at a higher value of $H_0$. This, in turn, alleviates the Hubble tension (see also \cite{Poulin:2018cxd, Knox:2019rjx,Lin:2019qug} for different implementations of the same mechanism).

A similar argument holds in the case of BAO measurements. There we observe the imprint  the sound horizon at radiation drag time $z_d \lesssim z_\text{rec}$ leaves in the visible matter spectrum in our low-redshift environment ($0.1 \lesssim z \lesssim 1$). Again, the NEDE modification together with an increased value of $H_0$ lowers both $r_s(z_d)$ and the comoving distance of the observed galaxies in a way which leaves the angular position of the peak in the matter power spectrum approximately invariant.  It is this balanced interplay between the shortening of the sound horizon and the reduction in the angular distances where late-time modifications fall short. As they only modify the expansion history after recombination, they leave the sound horizon invariant (or change it too little by changing the inferences we draw on other $\Lambda$CDM parameters), which leads to unwanted shifts in the angular positions of both CMB and BAO peaks (for a more detailed discussion of this BAO obstruction, see, for example, \cite{Bernal:2016gxb,Evslin:2017qdn,Lemos:2018smw,Aylor_2019,Knox:2019rjx}). 

\begin{figure}
 \centering 
{\includegraphics[width=16.5cm]{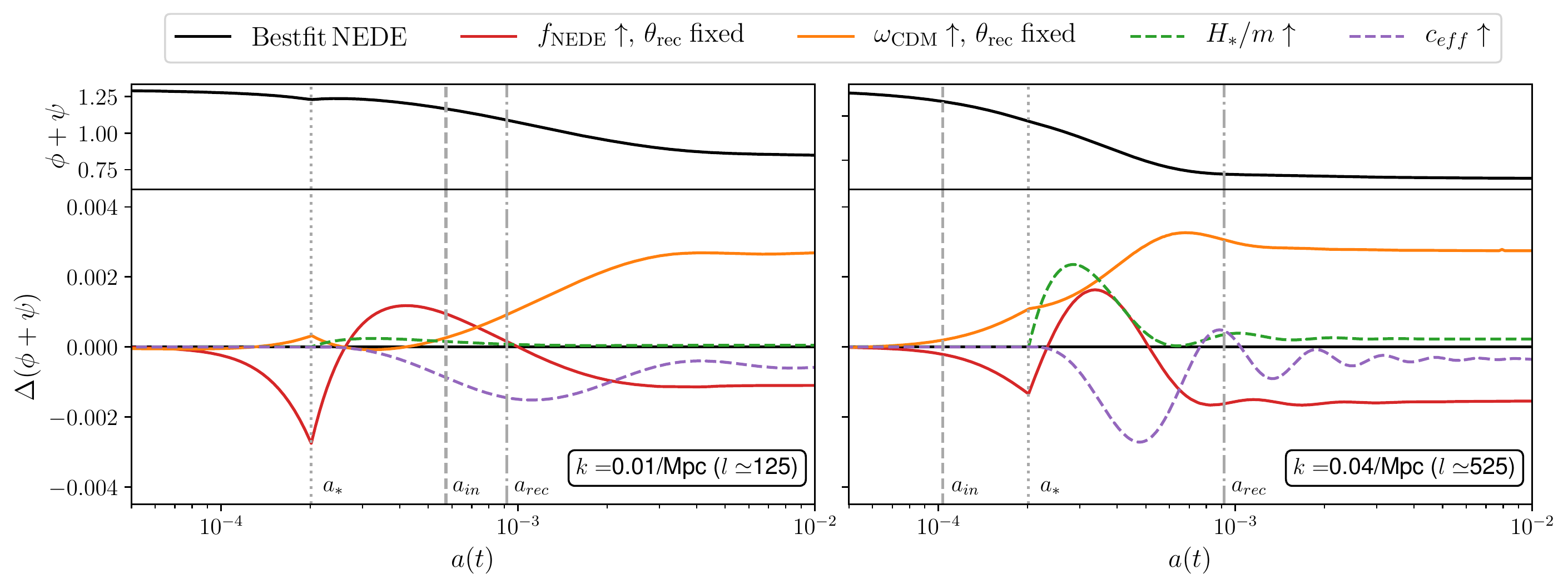}}
 \caption{Relative change in the Weyl potential with our best-fit NEDE as baseline; see Tab.~\ref{tab:means_NEDE_LCDM}.  The dashed line corresponds to the time where modes in the photon-baryon fluid enter their sound horizon and become sensitive to changes in the Weyl potential due to driving effects arising from its decay.  In the left plot this happens after and in the right one before the decay at $a_*$. An increase in $f_\text{NEDE}$ is approximately balanced by an increase in $\omega_\text{cdm}$. The effect of the `trigger parameter' $H_*/m$ cannot be fully compensated by a change in a $k$-\textit{independent} sound speed and, therefore, mimics the effect of a  $k$-\textit{dependent} sound speed as in single-field EDE.  }
\label{fig:Weyl}
\end{figure}

The above early-time mechanism is shared by different modifications of $\Lambda$CDM, most notably DR (or extra sterile neutrinos), EDE and ADE as well as our proposed model.  However, this is not enough to explain the phenomenological success of these models. After all, the CMB measurement is constraining not only the position of the first angular peak but also the shape of the angular power spectrum in a wide $k$ range. As photon-baryon acoustic oscillations are sensitive to the evolution of the new energy component at both the background and perturbation level through their gravitational coupling, any modification at that time risks to alter the CMB power spectrum beyond what is phenomenologically allowed.  To understand how New Early Dark Energy and similar proposals get around this obstacle, we follow the discussion in \cite{Lin:2019qug} and consider {the} Weyl potential:
\begin{align}
\psi + \phi = \eta + \frac{1}{2 k^2} \left[ \frac{d^2 h }{d\tau^2} + 6 \frac{d^2 \eta }{d \tau^2} \right]\,, 
\end{align}
where $\eta$ and $h$ are the metric fluctuations in the synchronous gauge introduced in \eqref{eq:sync}. We plot it in the upper two panels in Fig.~\ref{fig:Weyl} for a mode which enters the CMB sound horizon before (right) and after (left) the phase transition,  indicated by the vertical dashed lines.  

\begin{figure}
 \centering 
{\includegraphics[width=16.5 cm]{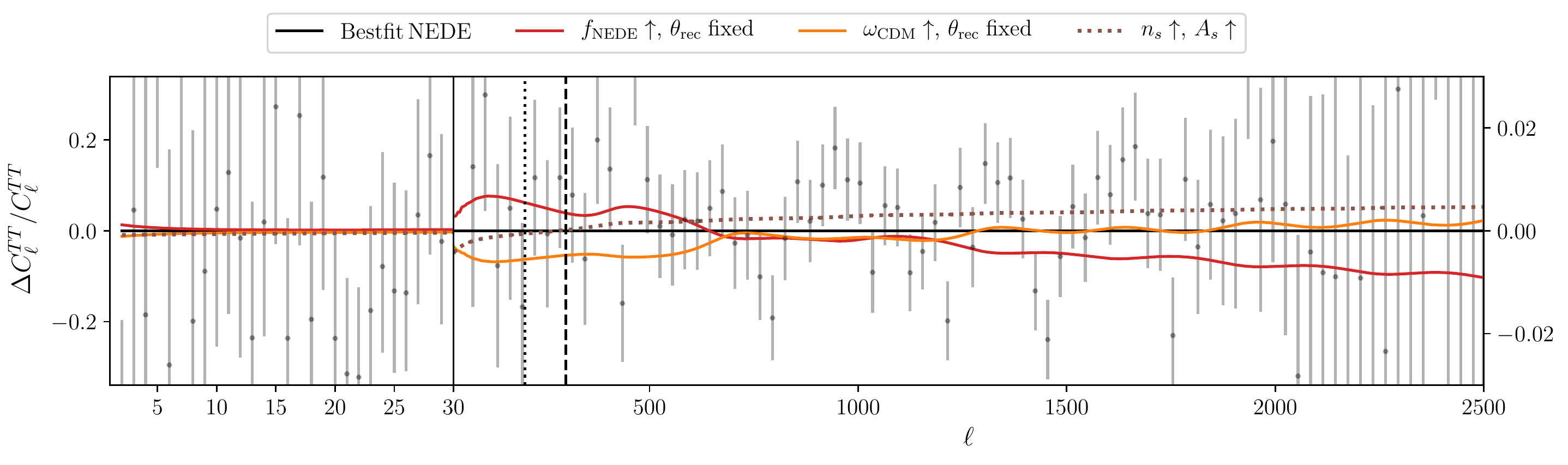}}
 \caption{{\sl Planck} data residuals (gray dots) with respect to the best-fit NEDE model as in Tab.~\ref{tab:means_NEDE_LCDM} (combined analysis with SH$_0$ES). Increasing $f_\text{NEDE}$ while keeping the angular position of the first CMB peak fixed leads to more power on large and less power on small scales due to changes in the Weyl potential and stronger damping effects, respectively. The first effect can be balanced by increasing $\omega_\text{cdm}$ and the second one by reducing the spectral tilt of the primordial spectrum.  The dashed and dotted vertical lines highlight the modes that, respectively, entered the sound and particle horizon at the time of the NEDE phase transition. }
\label{fig:residuals_params}
\end{figure}

The decay of the Weyl potential during the radiation-dominated epoch leads to driving effects in the photon-baryon fluid~\cite{Hu:2000ti}. Modes that have entered the horizon before matter domination are most sensitive to this effect. This can be seen in the upper two panels in Fig.~\ref{fig:Weyl}, showing a stronger decay for the larger $k$ mode.  As NEDE makes its most significant contribution when radiation still dominates (around the dotted line in Fig.~\ref{fig:Weyl}), it affects the driving of CMB acoustic oscillations through the changes it introduces to the expansion history and through its own acoustic oscillations. In the two bottom panels in Fig.~\ref{fig:Weyl}, we depict the effect different NEDE parameters have on the Weyl potential. The best-fit NEDE model serves as the baseline. The red line corresponds to a {10\%} increase in $f_\text{NEDE} $. Here, the corresponding shift in the CMB peak is compensated  by increasing $H_0$ such that $\theta_\text{rec} = const$. That way we make sure that the leading-order degeneracy has been accounted for. Before the transition at $a=a_*$, NEDE behaves as a cosmological constant and the corresponding repulsive boost leads to a quicker decay of the Weyl potential. This behavior is reverted at the point of the phase transition because there NEDE is converted from vacuum energy to a decaying fluid with equation of state parameter $w_\text{NEDE}^* > 1/3$. This fluid then supports its own acoustic oscillations that themselves source the gravitational potential. As has been discussed in the context of ADE, their subsequent Jeans stabilization then leads to an overall suppression of the potential. This excess decay can be compensated at leading order by increasing $\omega_\text{cdm}$, which is depicted by the yellow line. In Fig.~\ref{fig:residuals_params}, we show how this cancellation plays out on the level of the CMB power spectrum. While an increase in $w_\text{cdm}$ is able to balance the additional power on angular scales $\ell < 700$, this no longer works on scales $> 700$. Here, the dominant effect arises from changes in the diffusion damping scale $r_d$. It controls the suppression of the power spectrum at large $\ell$, which goes as $\exp\left[ - 2 (r_d/\lambda)^2\right]$~\cite{Hou_2013}. If we replace the wavelength $\lambda$ of a given mode with its angular scale $\ell \simeq 2 \pi D_\text{rec} / \lambda$ and use \eqref{eq:theta_rec}, we obtain a diffusion-induced suppression factor
\begin{align}\label{eq:damping}
\exp\left[ - 2 \left(\frac{r_d}{r_s}\, \frac{\theta_\text{rec}}{2 \pi} \ell \right)^2\right]\,.
\end{align}
As $\theta_\text{rec}$ is directly (and very precisely) fixed by CMB measurements, the damping is solely controlled by the ratio $r_d/r_s$, where both $r_s[H]$ and $r_d[H]$ are functionals of $H(z)$. Now we use that the mean squared diffusion distance is given by~\cite{Zaldarriaga_1995,Hu:1995en,Hu:1996mn}  $r^2_d \simeq \int^{\infty}_{z_\text{rec}} dz F_d(z)/H $, where $F_d(z)$ {depends on the fraction of free electrons, the baryon-photon ratio, and the cross-section for Thomson scattering}. From \eqref{eq:r_s}, we then derive the response to a functional change $H(z) \to H(z) + \Delta H(z)$,
\begin{align}\label{eq:r_d_over_r_s}
\Delta \left( \frac{r_d}{r_s} \right) &\simeq \frac{r_d}{r^2_s} \frac{v}{2 } \int_{z_\text{rec}}^{\infty} dz \, \frac{\Delta H}{H^2(z)}  \,,\nonumber\\
						 & \simeq  - \frac{r_d}{r_s}\, \frac{\Delta r_s}{2\,r_s} \,,
\end{align}
where we assumed that both functions $v(z)$ and $F_d(z)$ are approximately constant in the redshift window in which the integral picks up its main contribution. This shows that increasing $H(z)$ before recombination (or, equivalently, lowering $r_s$) generically increases the ratio $r_d/r_s$ which, in turn, dampens the power spectrum at large $\ell$ (in agreement with \cite{Hu_1996,Poulin:2018cxd}).\footnote{Note that our analytic estimate agrees with the numerical result obtained in \cite{Poulin:2018cxd} {(see their Fig.~2)}.} As can be see in Fig.~\ref{fig:residuals_params}, we cannot compensate this loss of power by increasing $w_\text{cdm}$, which has too little influence on small scales.  Instead, we can eliminate it by reducing the spectral tilt, corresponding to an increase in $n_s$ towards $1$, and slightly adapting the amplitude of primordial perturbations, $A_s$ (dotted line).

In summary, we have identified the leading-order degeneracies in the CMB measurement introduced by NEDE:

\begin{enumerate}
\item $f_\text{NEDE} > 0$ introduces a shift in the angular peaks which is compensated by increasing $H_0$.

\item {Acoustic oscillations in the NEDE fluid lead to an excess decay of the Weyl potential, which, in turn, affects} the driving of acoustic CMB modes that enter the sound horizon around the decay time {($\ell \simeq 300$)}. This can be counteracted by increasing $w_\text{cdm}$. 

\item A shortening of the CMB damping scale $r_D$, due to the modified expansion history, affects modes on short scales and needs to be compensated by reducing the spectral tilt through $n_s \to 1$ and increasing the amplitude $A_s$. 
\end{enumerate}

We will confirm these relations when we perform the cosmological parameter extraction in Sec.~\ref{sec:param_extraction}. Of course, these degeneracies are not perfect {and, therefore, lead} to {\sl Planck} data residuals, which we will discuss when comparing our best-fit model to its EDE-type competitors in Sec.~\ref{sec:competitor} (see also Fig.~\ref{fig:residuals}).

Finally, we discuss the effect of the trigger parameter, $H_*/m$, which is a distinctive feature of our model. In Fig.~\ref{fig:delta_EDE_star}, we have seen that varying $H_*/m$ while keeping $z_*$ fixed changes the initial amplitude of the NEDE perturbations. This effect is relevant only for modes that have entered the particle horizon before the decay (for $z_* = 5000$, this corresponds to ${\ell \gtrsim 180}$ ). This can also be seen in Fig.~\ref{fig:Weyl}, where only the larger $k$ mode responds to the increase in $H_*/m$ (green dashed line).  The corresponding deformation of the Weyl potential at large $k$ (right panel) can then be balanced by also increasing the effective sound speed, $c_\text{eff}$. However, this balancing does not work for the smaller $k$ mode (left panel). From this we conclude that the effect of the trigger parameter cannot be captured by a $k$-\textit{independent} sound speed. This resonates with the observation in \cite{Smith:2019ihp,Lin:2019qug}, where it was found that the phenomenological success of single-field EDE relies on having a $k$-\textit{dependent} sound speed.

\section{Cosmological Parameter Extraction}
\label{sec:param_extraction}

\subsection{Methodology} \label{sec:metholdology}

The cosmological parameter extraction is performed with the Metropolis-Hastings algorithm using its implementation in  the Monte Carlo Markov chain (MCMC) code {\tt MontePython}~\cite{Audren:2012wb,Brinckmann:2018cvx}. We assume that chains are converged if the Gelman-Rubin criterion~\cite{Gelman:1992zz} yields $R-1< 0.01$. To integrate the background cosmology and solve the linear perturbation equations, we use the Boltzmann code {\tt CLASS} (Cosmic Linear Anisotropy Solving System) \cite{Blas:2011rf} supplemented by our NEDE modifications, {publicly} available on GitHub.\footnote{\url{https://github.com/flo1984/TriggerCLASS}} Also, we add the ADE  model to {\tt CLASS} following \cite{Lin:2019qug} and use the publicly available implementation of EDE.\footnote{\url{https://github.com/PoulinV/AxiCLASS}}

We will use different combinations of the following datasets:

\begin{itemize}

\item {\bf Local $H_0$ measurement } (SH$_0$ES): The most recent SH$_0$ES value \cite{Riess:2019cxk}, $H_0 = 74.03 \pm 1.42 \, {\text{km}\,  \text{s}^{-1} \text{Mpc}^{-1}}$   $(68 \%$  C.L.), using a Gaussian prior.

\item {\bf High-$z$ Supernovae} ({\sl Pantheon}): The {\sl Pantheon} dataset \cite{Scolnic:2017caz} comprised of 1048 SNe Ia in a range $0.01 < z < 2.3$. Since the absolute magnitude scale $M$ is treated as a nuisance parameter, this dataset constrains the shape of the expansion history in the corresponding redshift range rather than $H_0$ {itself}. 

\item {\bf Baryonic Acoustic Oscillations} (large-$z$ and small-$z$ BAO): Anisotropic BAO measurements at redshift $z = 0.38$, $0.51$ and $0.61$ based on the {\sl CMASS} and {\sl LOWZ} galaxy samples of {\sl BOSS DR 12} \cite{Alam:2016hwk}, as well as small-$z$, isotropic BAO measurements of the 6dF Galaxy Survey \cite{6dF} and the {\sl SDSS DR7} main Galaxy sample \cite{MGS} at $z=0.106$ and $z=0.15$, respectively.

\item{\bf Large-Scale Structure} (LSS): Constraints on the growth of structure, quantified by $f \sigma_8(z)$, from redshift-space distortions at redshift $z = 0.38$, $0.51$ and $0.61$ based on the {\sl CMASS} and {\sl LOWZ} galaxy samples of {\sl BOSS DR 12} \cite{Alam:2016hwk}. We note that this measurement is implemented in  {\tt MontePython} in conjunction with the large-$z$ BAO measurement, taking into account their (strong) covariance. When we calculate the tension with LSS, we use the $S_8 = \sigma_8 \sqrt{\Omega_m/0.3}$ measurement  from a combined tomographic weak gravitational lensing analysis of the {\sl Kilo Degree Survey} and the {\sl Dark Energy Survey}~\cite{Joudaki:2019pmv},  $S_8 =  0.762^{+0.025}_{-0.024}$ $(68 \%$  C.L.).

\item {\bf Cosmic Microwave Background} ({\sl Planck} 2018): The {\sl Planck 2018 TT,TE,EE} and lensing likelihood with all nuisance parameters included \cite{Aghanim:2019ame}. 

\item {\bf Big Bang Nucleosynthesis} (BBN): The primordial helium abundance from \cite{Aver:2015iza}, $Y_{p} = 0.2449 \pm 0.0040$ $(68 \%$  C.L.), via a Gaussian likelihood. This is the conservative choice given the tighter bounds from \cite{Peimbert:2016bdg},  $Y_{p} = 0.2446 \pm 0.0029$ $(68 \%$  C.L.). As the helium abundance is sensitive to the early expansion history, it mainly constrains the effective number of relativistic degrees of freedom $N_\text{eff}$ during the BBN epoch. \CLASS \, infers the value of $Y_p$   as a function of $(\omega_b, N_\text{eff})$ from an interpolation table created with {\tt PArthENoPE v1.2}~\cite{Pisanti:2007hk} with a nuclear rate $d(p,\gamma)\prescript{3}{}{{\text{He}}}$ extrapolated from observations by~\cite{Adelberger:2010qa}.
\end{itemize}

We will use two different combinations of the above datasets for our parameter extraction: 

\begin{enumerate}

\item Combined analysis without SH$_0$ES: {\sl Pantheon}, BAO, LSS, {\sl Planck} 2018 and BBN

\item Combined analysis with SH$_0$ES: {\sl Pantheon}, BAO, LSS, {\sl Planck} 2018, SH$_0$ES and BBN

\end{enumerate}

We introduce two new components to \CLASS, the trigger field $\phi$ and the NEDE fluid. The former we describe in terms of a massive scalar field with mass $m$ and sub-dominant energy density, $\bar{\rho}_{\phi}$, and the latter is described on the background level as a fluid with time dependence {as} in \eqref{eq:rho_EDE_bg}. Here, we discuss the implementation of both components in turn. 

\paragraph{\textbf{Trigger field:}}
This sector requires us to introduce three new input parameters to \CLASS, the field's mass $m$, its initial value $\phi_\text{ini}$ as well as the trigger parameter $x_*^{-1}=H_*/m$, defining the transition time with respect to the evolution of $\phi$. The background value of the scalar field is initialized at the same time as the other background quantities.\footnote{For concreteness, we use the {\tt CLASS} standard value $a = 10^{-14}$ as the initial time.} Specifically, we put $\phi$ and  $\dot{\phi}$ on the attractor~\eqref{eq:phi_attr}. We further require $\phi$ to be sub-dominant, which due to \eqref{eq:cond_4} is guaranteed if $\phi_\text{ini}/M_{pl} \ll 1$.  To be specific, in our baseline model we will set ${\phi}_\text{ini} / M_{pl} = 10^{-4}$, which is also compatible with the lower bound in~\eqref{eq:phi_bound}. In any case, its  precise initial value does not matter from  a phenomenological perspective -- at least if it is sub-dominant -- because $\phi_\text{ini}$ drops out of the matching conditions in \eqref{eq:EDE_ini}. {This is because they are proportional to} the ratio $\delta \phi / \dot \phi$, which due to \eqref{eq:phi_attr} and \eqref{eq:delta_phi_attr} is indeed independent of $\phi_\text{ini}$.   We also checked {explicitly} that the  power spectrum  shows a negligible dependence on $\phi_\text{ini}$ (small compared to numerical uncertainties). The field is then propagated forward in time via the background equation \eqref{eq:phi_bg_conf} until the decay of NEDE is triggered when the condition $H =  m/x_* $ is met. As argued before in Sec.~\ref{sec:transition_surface}, our model predicts a rather narrow range of possible values for $x^{-1}_* = H_* / m$ detailed in \eqref{eq:trigger_range}. For our base model, we will therefore fix it to $H_* /m =0.2$, which is the most natural expectation. In fact, a smaller value would require an increasingly delicate parameter tuning. We will also perform one analysis where we allow it to vary within its {allowed} range.

\paragraph{\textbf{NEDE field:}} The NEDE component is straightforwardly implemented in {\tt CLASS} by using the parametrization in~\eqref{eq:rho_EDE_bg}. We use $\bar{\rho}_\text{NEDE}^*$, i.e., the constant value of NEDE before its decay, as new input parameter. The fluctuations in the NEDE fluid are set to zero before the transition and initialized right after by using our matching equations \eqref{eq:EDE_ini} together with the identification $\delta q_* / \dot{\bar{q}}_* = \delta \phi_*/ \dot{{\phi}}_*$, which closes the dynamical system. To properly resolve the matching dynamics and avoid numerical artifacts, we increase the numerical precision by lowering the step size around the transition.  After the transition the fluctuations are propagated forward in time using the  equations \eqref{eq:pert_dyn} describing the dynamics of generic fluid perturbations, fully characterized by the three parameters $(w_\text{NEDE}^*, c_{s}^2,c_\text{vis}^2)$. Following the discussion in Sec.~\ref{sec:coalesence}, we will make the choice $(2/3,2/3,0)$ for our base model. However, we will also consider different extensions that include the case of free-streaming radiation, which according to \cite{Hu_1998} can be approximated by $(1/3,1/3,1/3)$, and {different other cases with} freely varying fluid parameters.

With these choices our base model comes {equipped} with only two undetermined parameters,  $\{ \bar{\rho}^*_\text{NEDE}, m \}$, which will be inferred from a cosmological parameter extraction. To be specific, we vary $\{ f_\text{NEDE}, \log_{10} (m/m_0) \}$, where $m_0 = 1/ \text{Mpc}$ in order to keep with the \CLASS\, convention, and  $f_\text{NEDE}$ is the fraction of NEDE at the time of decay, related to  $\bar{\rho}^*_\text{NEDE} $ via \eqref{eq:def_frac}. Rather than using the {\tt CLASS} input parameter $m$, we will present our results in terms of the more intuitive quantity $z_*$, the redshift at decay time. Its value depends on the the expansion history and needs to be inferred numerically from $H(z_*)$ (which is fixed by the product $ x_*^{-1} \times m$ ). During radiation domination it is approximately given by
\begin{align}\label{eq:zstar}
z_* \simeq 1.6 \times 10^4 \left( 1 - f_\text{NEDE} \right)^{1/4} \left( \frac{3.9}{g_*} \right)^{1/4} \left( \frac{H_*/m}{0.2} \right) \, 10^{ 0.5 \log_{10}(m/m_0)- 1.75}\,,
\end{align}
{where we used \eqref{eq:m}}.
 As expected the main dependence is due to $m$, but also $f_\text{NEDE}$  (and $x_*^{-1}=H_*/m$) go into its determination.
The closure of the budget equation is finally ensured by fixing the cosmological constant, $\Lambda$ [using iterative runs to find an accurate estimate for $\bar{\rho}_\text{NEDE}(t_0)$]. 

\begin{figure}
 \centering 
{\includegraphics[width=16.5cm]{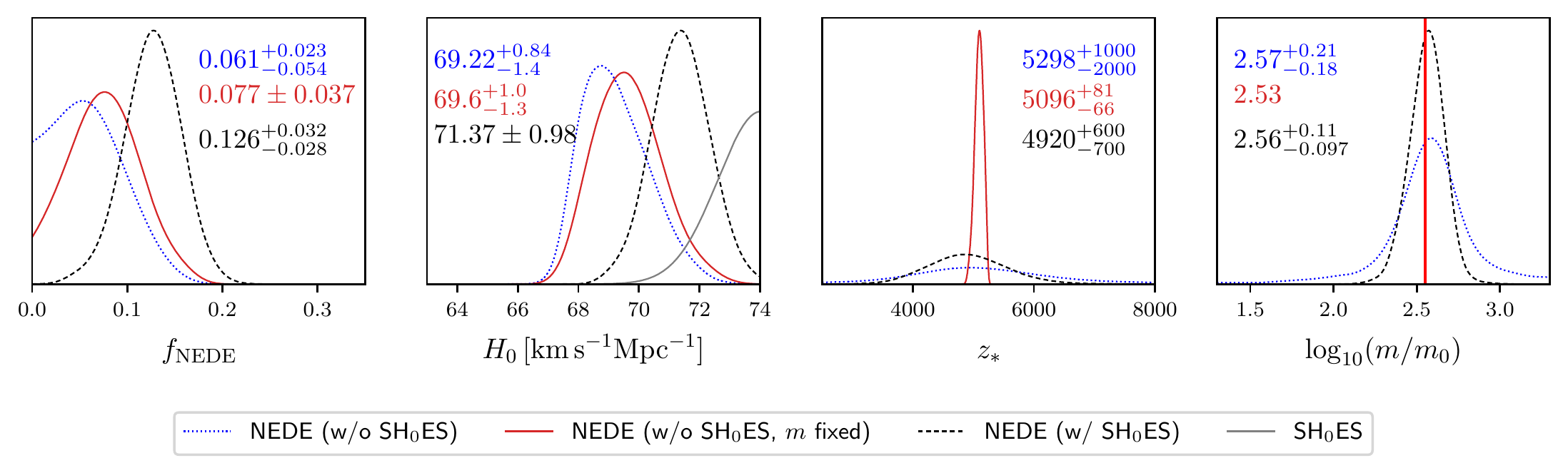}}%
 \caption{The marginalized posterior distribution for our combined analysis without SH$_0$ES. If we allow the mass parameter $m$, and hence the decay time $z_*$, to vary, the increased sampling volume as $f_\text{NEDE} \to 0 $ leads to an unphysical enhancement for $f_\text{NEDE} \ll 1$ (blue dotted line). We avoid this problem by fixing $\log_{10}\left(  m /m _0\right) = 2.58$ (red solid line), which leads to a Gaussian distribution with $1.9 \, \sigma$ evidence for $f_\text{NEDE} > 0$.}
\label{fig:mass_prior}
\end{figure}

We will compare NEDE to the $\Lambda$CDM model and its most prominent early-time competitors DR, ADE and single-field EDE. In all cases, we vary the six $\Lambda$CDM parameters on which we impose flat priors {with standard boundaries}, specifically
$\{ \omega_{b }, \omega_{cdm }, h, \ln10^{10}A_{s }, n_{s }, \tau_\text{reio } \}$, where $ \omega_{b } = \Omega_b h^2$ and $ \omega_\text{cdm} = \Omega_\text{cdm} h^2$ are the dimensionless baryon and cold dark matter density, respectively, $h$  is the dimensionless Hubble constant defined via $H_0 \equiv h \times 100 \, \kmsMpc$, $\tau_\text{reio}$ is the reionization optical depth, and $\ln10^{10}A_{s }$ and $n_s$ are the initial amplitude and tilt of the primordial curvature spectrum at $k = 0.05 \, \text{Mpc}^{-1} $, respectively. For the neutrino sector, we use the {\sl Planck} convention and model it in terms of two massless  and one massive species with $M_\nu = 0.06 \, \text{eV}$, where we ensure that the effective number of relativistic degrees of freedom is $N_\text{eff} = 3.046$ (except for dark radiation where $N_\text{eff}$ is promoted to a free parameter). We impose the flat prior $0 < f_\text{NEDE} < 0.3 $ where the upper boundary could easily be removed as it is never probed as shown in Fig.~\ref{fig:mass_prior}. 

There is a subtlety related to the mass parameter (or the decay time equivalently), which applies equally to all models of the EDE type~(also see the discussion in section III of \cite{Smith:2019ihp}). When $f_\text{NEDE} \to 0 $, the $\Lambda$CDM model is recovered and all values of $m$ (or $z_*$) are exactly equivalent, corresponding to an enhancement of the relevant sampling volume. This, in turn, leads to unphysical artifacts in the posterior distributions which are ``pulled'' towards the degenerate point at $f_\text{NEDE} = 0$ and start to deviate from a Gaussian shape. This can be seen  in Fig.~\ref{fig:mass_prior}, where we depict different posteriors for our combined analysis without SH$_0$ES. In particular, if we vary both parameters, $m$ and $f_\text{NEDE}$, corresponding to the blue dotted curve, there is a probability enhancement for $f_\text{NEDE} < 0.06$. This is due to the fact that the constraints are no longer mainly driven by  data but by the prior volume. 

In order to get around this technical obstruction, we perform a two-step analysis where we first vary both $m$ and $f_\text{NEDE}$ to obtain the best-fit values (which are, of course, not affected by this volume effect). Then we perform a second run where we fix $m$ close to its best-fit value and vary only $f_\text{NEDE}$. To be precise, the prior {boundaries} we impose for the first and second run are\footnote{{We use different prior boundaries in the case of free-streaming radiation where the decay happens significantly earlier at $10^{-3}\, z_* = 35_{-20}^{+8}$.}}
\begin{subequations}
\begin{align}
 1.3<& \log_{10}(m/m_0)<3.3 \,, \label{eq:prior1}\\
 	& \log_{10}(m/m_0)= 2.58\,, \label{eq:prior2}
\end{align}
\end{subequations}
corresponding to the dotted blue and solid red line, respectively. Because of \eqref{eq:zstar}, the first prior probes the rather wide range $1.3< 10^{-3} z_*< 13.1 $, whereas the second prior corresponds to a narrow redshift range around {$5100$}. This completely avoids the issue as can be seen from the red curves in Fig.~\ref{fig:mass_prior}, which are nearly Gaussian for both $f_\text{NEDE}$ and $H_0$.  Strictly speaking, this analysis assumes that our underlying theory fixes $m$ to a very specific value, and, therefore, it can provide us only with the posteriors valid in this rather special case. Nevertheless, since $f_\text{NEDE}=0$ is still included in the allowed parameter range, we believe it is a consistent way of estimating the statistical preference for our model. We will apply  the same procedure when we analyze ADE and EDE without SH$_0$ES in order to ensure a fair comparison. The more general case, where  $\log_{10}\left[ m / m_0\right]$ is allowed to vary freely, will still be investigated when we include the SH$_0$ES measurement. In that case, corresponding to the black dashed curve, $f_\text{NEDE}$ never probes the degenerate parameter region due to the statistical ``pull'' towards higher values of $H_0$, and conventional $\chi^2$ statistics can be used to assess the quality of the NEDE fit.

Finally, we will also discuss three different extensions of our base model where we allow $(w_\text{NEDE}^*,\, c_s,\, f_\text{NEDE}, \log_{10}\left[ m / m_0\right])$, $(c_\text{vis}, \, f_\text{NEDE}, \log_{10}\left[ m / m_0\right])$ and $(H_*/m,\, f_\text{NEDE})$ to vary while keeping the other parameters fixed.

\begin{figure*}
 \centering 
{\includegraphics[width=16.5cm]{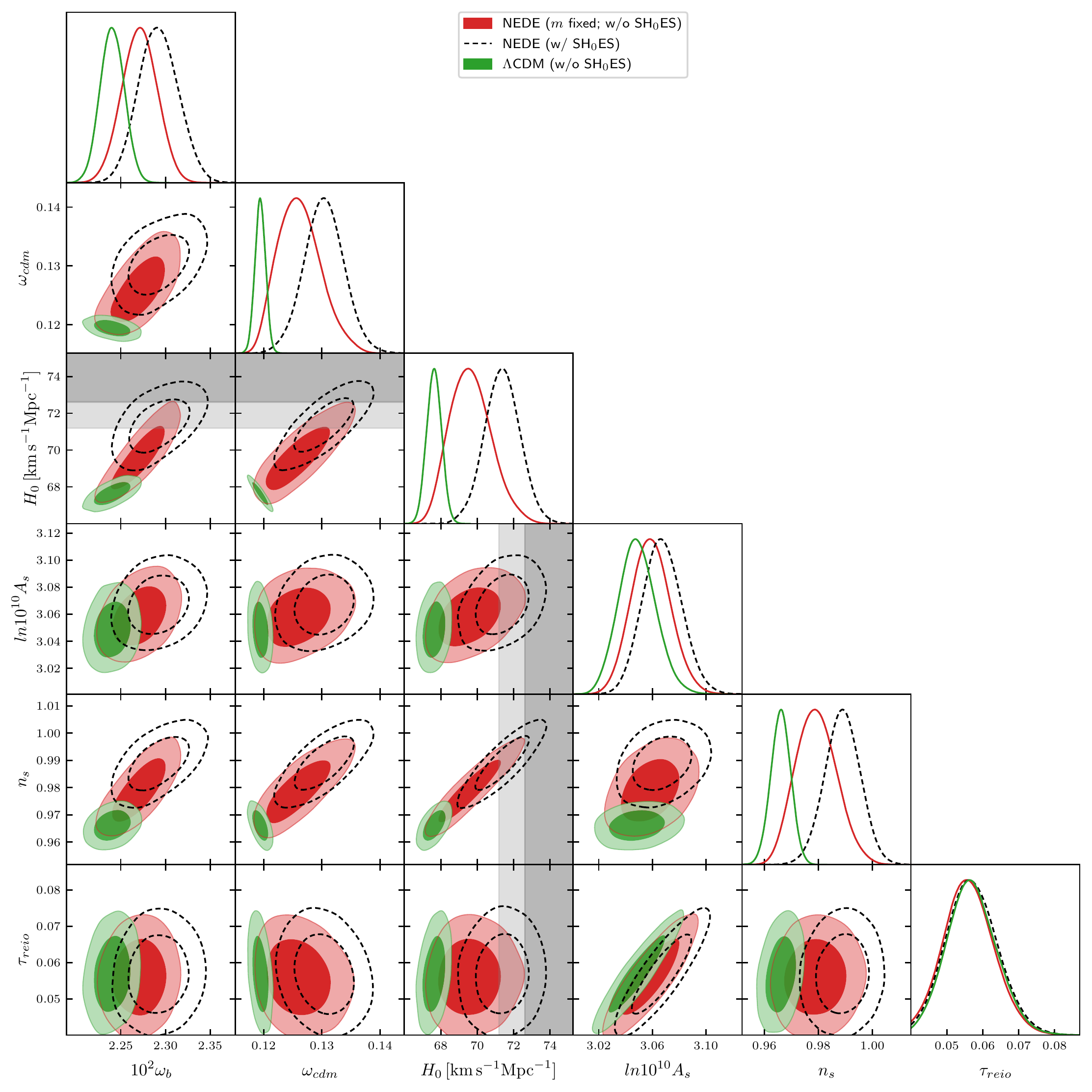}}%
 \caption{Covariances and marginalized distributions of standard cosmological parameters for NEDE (red) and $\Lambda$CDM (green) from a run to the combined datasets without SH$_0$ES. The result of the analysis with  SH$_0$ES included is presented as the dashed contours. Here and henceforth, the darker and lighter shades correspond to the $68 \%$ C.L.\ and the $95 \%$ C.L., respectively. The gray bands depict the constraints from the  SH$_0$ES measurement. The $95 \%$ C.L.\ contours of  SH$_0$ES and NEDE largely overlap. }
\label{fig:triangle}
\end{figure*}

\subsection{Discussion of Results}\label{sec:results}

\subsubsection{Base Model}

\begin{figure}
 \centering 
{\includegraphics[width=16.5cm]{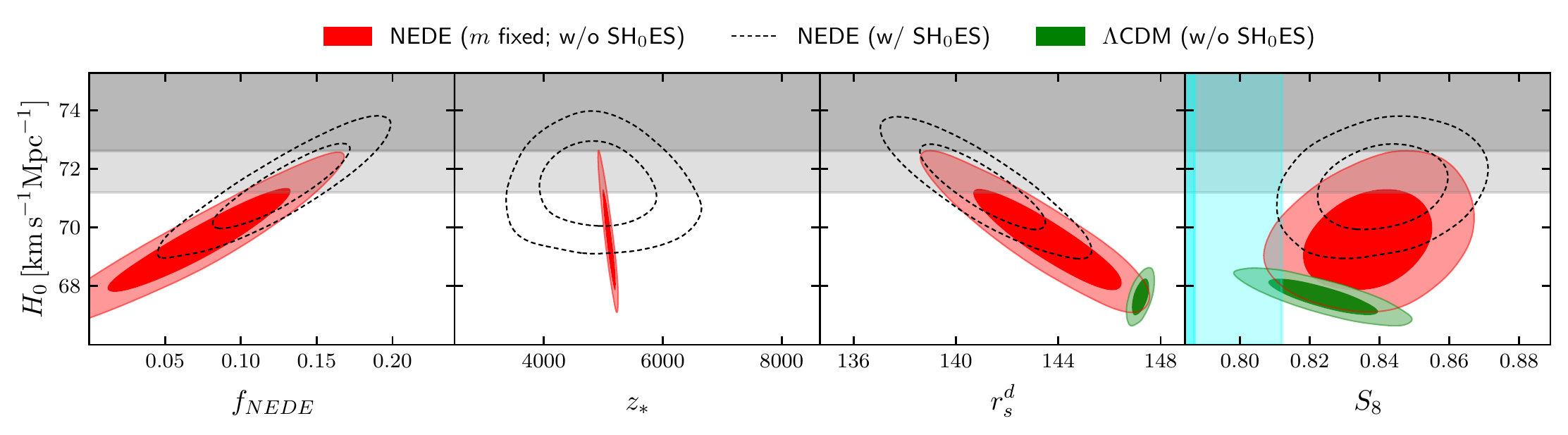}}%
 \caption{The Hubble parameter today vs a subset of parameters for the combined analysis with (dashed contour) and without (red filled contour) SH$_0$ES.
 The light blue band corresponds to the constraint on $S_8$ from weak gravitational lensing combining different cosmic shear surveys~\cite{Joudaki:2019pmv}. The gray band represents the SH$_0$ES measurement. {Without SH$_0$ES the decay time is constrained to a narrow redshift range to avoid sampling volume artifacts.}}
\label{fig:NEDE_LCDM_rect}
\end{figure}

The standard cosmological parameters and their covariances as inferred from our joined analysis (with and without SH$_0$ES) are presented in Tab.~\ref{tab:means_NEDE_LCDM}  and Fig.~\ref{fig:triangle}, respectively.   In Fig.~\ref{fig:NEDE_LCDM_rect} we depict the two parameters of our baseline model as well as the sound horizon at radiation drag, $r_s^d=r_s(z_d)$, and the growth parameter, $S_8 = \sigma_8 \sqrt{\Omega_m/0.3}$ . The main effect of NEDE is to increase the mean value of $H_0$  while also enlarging its uncertainties. As a result, we find an overlap with the local  SH$_0$ES measurement at the $95 \%$ confidence level (light gray and red contours); more specifically, we find {$H_0 = 71.4 \pm 1.0 \kmsMpc$} when we include the local measurement, which leads to a total $\chi^2$ improvement of {15.6}. When we do not include the local measurement, we find {$ H_0 =  69.6^{+1.0}_{-1.3}\kmsMpc$}, which corresponds to a reduced tension of {$2.5 \, \sigma$} (down from $4.3 \, \sigma$). 

As explained before, the main effect of NEDE is to lower the sound horizon, $r_s(z)$, which is balanced by a higher value of $H_0$.  To be specific, we obtain a  ${4 \%}$ reduction of  $r_s^d$, from {$147.20 \pm 0.23\, \text{Mpc}$} ($\Lambda$CDM w/o SH$_0ES$) down to {$141.0^{+1.6}_{-1.7} \,\text{Mpc}$} (NEDE w/ SH$_0$ES), which is then approximately balanced by a $6 \%$ increase in $H_0$.  We also report a {$4.3 \, \sigma$} evidence for a non-vanishing NEDE component when we include the local measurement of $H_0$, corresponding to $f_\text{NEDE} = 12.6^{+3.2}_{-2.9}\, \%$, and if we do not include the local measurement, the evidence for NEDE still reaches $1.9 \, \sigma$, corresponding to a mild preference for our model {driven mostly by {\sl Planck} data}. This resonates with our $\chi^2$ results detailed in Tab.~\ref{tab:chi2}. They show that the improvement is almost entirely due to the SH$_0$ES dataset with $\Delta \chi^2 (H_0) = -13.8$.  However, the important point is that the quality of the fit to the other datasets does not become worse; in fact, it even improves by $\Delta \chi^2 (\backslash H_0)=-1.8$. This beneficial effect is even more pronounced for our analysis without SH$_0$ES, where we obtain $\Delta \chi^2(\text{total}) = -2.9$. In other words, the evidence for NEDE is mainly but not entirely driven by the local  measurement of $H_0$. More importantly, none of the fits to other datasets  shows any significant deterioration {let alone} the emergence of a new tension. On closer inspection, we see that NEDE improves the fit to the CMB temperature and polarization spectrum at both low  and high $\ell$, specifically {$\Delta \chi^2 (\text{TT, EE, TE}) = -3.6$}, at the prize of slightly worsening the lensing fit with {$\Delta \chi^2(\text{lens}) = + 0.4$}, provided we include SH$_0$ES. We will see later when discussing Fig.~\ref{fig:residuals} that this improvement can be attributed to specific kinky and oscillatory features in the power spectrum, which also provides a potential means of detecting NEDE in future surveys or discriminating it from its competitors.

 \input{./tab/means_NEDE_LCDM.tex}

The decay time turns out to be a relatively solid prediction with little sensitivity to the prior on $H_0$. Specifically, we obtain {$z_*= 4920^{+620}_{-730}$} (w/ SH$_0$ES) which is compatible with our assumption that the transition occurs during radiation domination, but close enough to recombination that it can still influence the value of the sound horizon (note that the integral in \eqref{eq:r_s} picks up its main contribution at its lower boundary). 

Looking at different $\Lambda$CDM parameters, we can identify the approximate degeneracies predicted in the last sections: First, there is a strong positive (negative) correlation between $H_0$ and $f_\text{NEDE}$ ($r_s^d$) as expected. The effect of a larger value of $H_0$ is then balanced by a larger value of $\omega_\text{cdm}$ in order to compensate a stronger decay of the Weyl potential. At the same time, the value of $n_s$ moves towards scale invariance, and the amplitude of primordial perturbations, $A_s$, is slightly enhanced to compensate the stronger damping on small scales. These degeneracies can be identified in both cases with and without SH$_0$ES. As a peculiar feature of these early-time modifications note that $n_s$ becomes $2 \, \sigma$ compatible with scale invariance.

\input{./tab/chi2.tex}

Finally, NEDE also affects the matter power spectrum. It increases the amplitude of local structures parametrized by $S_8= \sigma_8 \sqrt{\Omega_m/0.3}$  from {$S_8 = 0.824 \pm 0.010$}  ($\Lambda$CDM w/o SH$_0$ES) to {$S_8 = 0.841 \pm 0.012$}  (NEDE w/ SH$_0$ES). This slightly raises the tension with weak lensing measurements, already present within $\Lambda$CDM at the $2.3 \, \sigma$ level\footnote{The tension  goes down to $1.9 \, \sigma$ when including SH$_0$ES. The reason is that within $\Lambda$CDM $\omega_\text{cdm}$ decreases when $H_0$ goes up. This is not a {valid} way of relieving the tension though because it relies on combining incompatible datasets.}, to $2.8 \, \sigma$ {(see also Tab.~\ref{tab:means_NEDE_LCDM})}. This is based on a recent study in~\cite{Joudaki:2019pmv} reporting $S_8 =  0.762^{+0.025}_{-0.024}$ from the combined tomographic weak gravitational lensing analysis of the {\sl Kilo Degree Survey} and the {\sl Dark Energy Survey}. We depict this constraint as the blue band in Fig.~\ref{fig:NEDE_LCDM_rect}. The conclusion we draw from this is that whatever physics or systematics is responsible for this so-called $S_8$ tension~\cite{Raveri:2018wln}, it seems to be equally affecting NEDE and $\Lambda$CDM. In other words, while NEDE in its simple form does not help with the $S_8$ tension it also does not exacerbate it significantly. In fact, in \cite{Chudaykin:2020acu} it was recently argued that the $S_8$ tension might be a result of the ``lensing tension'', which questions the internal consistency of the {\sl Planck} dataset. Taking that into account  (essentially by neglecting high-$\ell$ {\sl Planck} data), they found a $3 \, \sigma$ preference for single-field EDE over $\Lambda$CDM despite the inclusion of different LSS probes. We also note that we cannot reproduce the somewhat bleak picture given in~\cite{Hill:2020osr}. There, the authors use for one of their analyses the same datasets as we do (except for BBN, which has a negligible impact anyhow), corresponding to our joint analysis with SH$_0$ES. They then find a lower evidence for single-field EDE compared with the {$4.3\, \sigma$} evidence we report for NEDE.  Whether this tendency towards stronger evidence within NEDE persists when including other LSS probes remains to be seen. Let us stress, though, that such an analysis needs to carefully take into account the prior volume effects discussed in Sec.~\ref{sec:metholdology} when not including SH$_0$ES (or another local $H_0$ probe).

\subsubsection{Minimal Extensions}\label{sec:extensions}

\begin{figure*}
 	\subfloat[NEDE + $H_*/m$  \label{fig:ext_trigger} ]
	{\includegraphics[width=3.8cm]{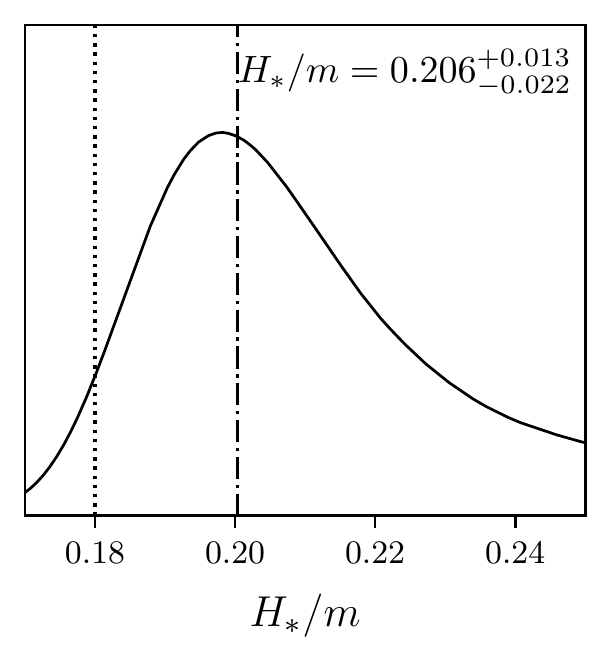}}
	 	\,
 	\subfloat[NEDE + $w_\text{NEDE}^*$ + $c_s$ \label{fig:ext_cs_w}]
	{\includegraphics[width=8.3cm]{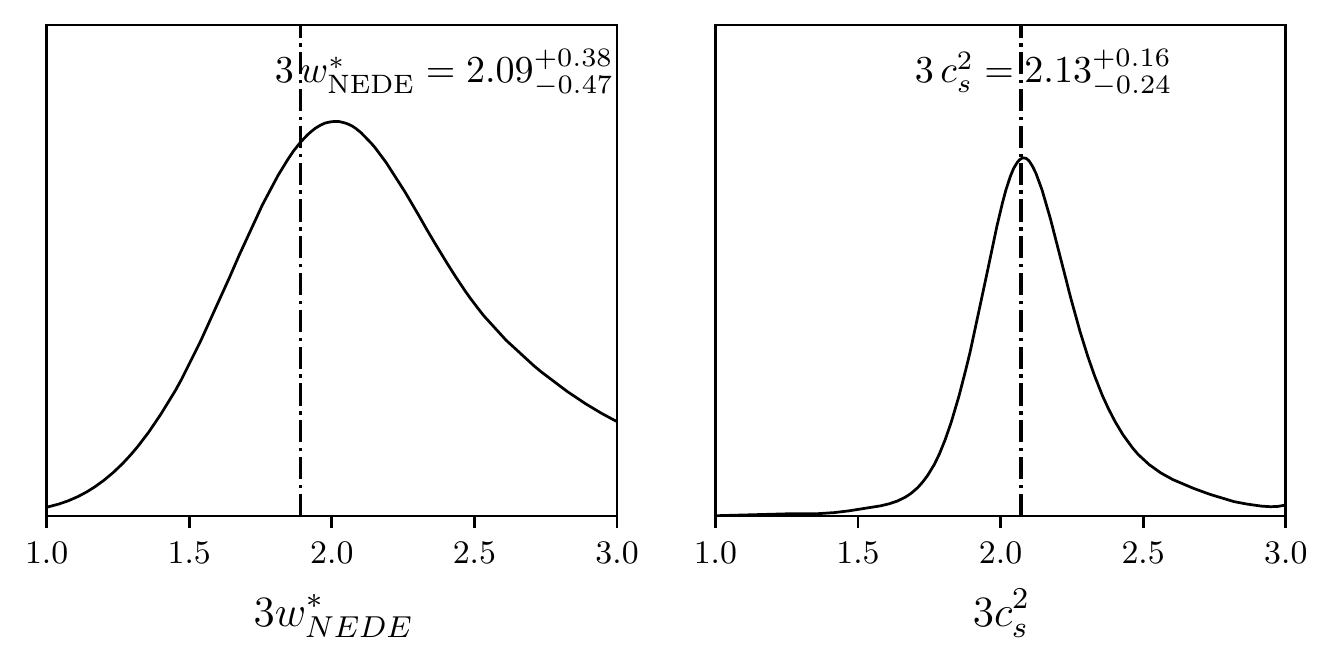}}%
	  	\,
 	\subfloat[ NEDE + $ c_\text{vis}$ \label{fig:ext_vis}]
	{\includegraphics[width=3.8cm]{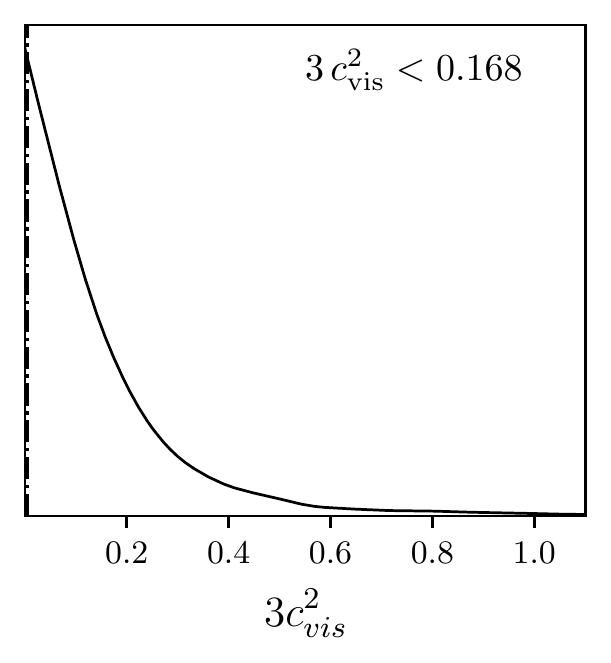}}%
 \caption{Posterior distributions of three different NEDE extensions for our joint analysis with SH$_0$ES. In each case the {listed} parameters (and the $\Lambda$CDM parameters) are allowed to vary while the other NEDE parameters are fixed.  The dash-dotted line represents the best-fit value. The dotted line in (a) corresponds to  $\phi_* = 0$. }
\label{fig:ext}
\end{figure*}

In our base model we fixed the trigger parameter, $x^{-1}_* = H_*/m = 0.2$, as well as the the fluid parameters, $w_\text{NEDE}^* = 2/3 $, $c_s^2 = 2/3$ and $c_{vis} = 0$. Their values correspond to generic choices in agreement with our microscopic model {and expectations from other {EDE-type} models}. Here, we allow them to vary, which constitutes a non-trivial phenomenological check of NEDE and its underlying microphysics. The MCMC analyses in this section include the local measurement of $H_0$ to create more constraining power and achieve quicker convergence. In all cases, we found that the mean values are $1 \, \sigma$ compatible with the ones obtained in our base model in  Tab.~\ref{tab:means_NEDE_LCDM}. In order not to overload this work with data, we therefore cite them only if relevant.   We do, however, provide the results of our  $\chi^2$ analysis in Tab.~\ref{tab:chi2} and present the posterior distributions of the additional model parameters in Fig.~\ref{fig:ext}.  Finally, we will also study the possibility that NEDE decays completely into free-streaming radiation. 

\paragraph{\textbf{Trigger parameter:}}

The trigger parameter determines where exactly in the evolution of $\phi$ the phase transition is initiated, corresponding to the ratio $x_*^{-1} = H_* / m$. As illustrated in Fig.~\ref{fig:delta_EDE_star}, it controls the amplitude of initial fluctuations in the NEDE fluid right after the decay.  The particular dynamics of our two-field model requires it to be within the narrow range {$\eqref{eq:trigger_range}$, explicitly} $0.18 < H_*/m \lesssim {0.21} $, provided we prohibit a fine-tuning of the model's fundamental parameters (see the discussion before Eq.~\ref{eq:trigger_range}).  In order to test this prediction, we allowed the trigger parameter to vary in the slightly larger range  $0.17 < H_*/m \lesssim 0.25 $.\footnote{We fix $log_{10}(m/m_0)$ = 2.63 {(close to the base model best-fit in Tab.~\ref{tab:means_NEDE_LCDM})} to avoid sampling problems that otherwise occur due to a background degeneracy along $m x_*^{-1} = const$ (which is broken at the perturbation level, though). } As a result of this analysis, we report $H_*/m = 0.206^{+0.013}_{-0.022}$, where the posterior distribution is depicted in Fig.~\ref{fig:ext_trigger}. This value lies nicely within the predicted interval. In particular, we find that it is larger than $0.18$ (dotted vertical line), implying that the transition occurs before the tunneling probability becomes maximal when $\phi$ crosses zero for the first time, in accordance with a triggered phase transition.  We consider this strong {additional} evidence for our two-field vacuum decay model.

\paragraph{\textbf{Fluid parameters:}}

The fluid sector has been discussed in Sec.~\ref{sec:coalesence} and~\ref{sec:pert_eqs}. The upshot is that a condensate of colliding bubble walls is dominated by kinetic rather than potential energy on large scales. Together with the radiation arising from the decay of the condensate we therefore expect an effective value of the equation of state parameter within the range $1/3 < w_\text{NEDE}^* < 1$. Our base model prior,  $w_\text{NEDE} =2/3 $, was then simply chosen to be the central value.  Moreover, we assumed the effective sound speed to equal the adiabatic one, implying $c_s^2 = w_\text{NEDE} = 2/3$. To check the phenomenological viability of these choices, we allowed both parameters to vary at the same time. The results are depicted in Fig.~\ref{fig:ext_cs_w}, indeed showing that our base model prior was justified.\footnote{This is also in agreement with  the analysis in \cite{Lin:2019qug}, where in the case of ADE it was shown that data prefer the choice $c_s^2 \simeq w_\text{ADE}^*$.} We also see that the viable range, $w_{NEDE}^* = 0.70^{+0.12}_{-0.16}$, is pretty large, allowing for a possibly wide range of collision and decay scenarios. The constraints on the rest-frame sound speed, $ c_s^2 = 0.711^{+0.057}_{-0.081}$, are significantly tighter but still $1 \, \sigma$ compatible with $c_s^2 = w_\text{NEDE}$. It remains to be seen if such a value can be accommodated in a detailed collision model. We also note that adding two additional parameters to the fit reduces the evidence for NEDE down to $3.4\, \sigma$, corresponding to $f_\text{NEDE} = 11.3^{+3.7}_{-3.3}\, \% $ and $H_0=71.0^{+1.1}_{-1.0} \kmsMpc$.

\paragraph{\textbf{Viscosity parameter:}}

Here, we investigate the possibility that a non-vanishing viscosity parameter, $c_\text{vis}$, leads to the build-up of anisotropic stress $\sigma_\text{NEDE}$ through the sourcing of velocity and metric shear on the right-hand side of \eqref{eq:shear}. As we have argued in Sec~\ref{sec:pert_eqs}, anisotropic stress arises as a generic  component of a perturbed ideal fluid and as such needs to be included in a comprehensive analysis. {Moreover,} we are not discussing the possibility that the initial value $\sigma_\text{NEDE}^* \neq 0$, which would be another way of generating anisotropic stress even if $c_\text{vis}=0$ [see \eqref{eq:shear}]. As is obvious from Fig.~\ref{fig:ext_vis}, we find no evidence for $c_\text{vis} \neq 0$ when we allow it to vary in the range $0 < c_\text{vis}^2 < 1$; instead, we obtain the upper bound $c_\text{vis}^2 < 0.56$. 

In the context of our particular model, this result makes it appear less likely that bubbles grow large and, therefore, leave a sizable imprint in the effective fluid in the form of anisotropic stress (which would otherwise average to zero). This, however, is only an indication as the effect of these large bubbles should be modeled via a $k$ dependent initial value for $\sigma_\text{NEDE}^*(k) \neq 0$. In any case, bubbles never reach cosmological sizes if the bound \eqref{eq:constraint_beta} on $\bar{\beta}$ is satisfied, which is also needed to justify the modeling of the phase transition as an instantaneous event.

\begin{figure}
 \centering 
{\includegraphics[width=15cm]{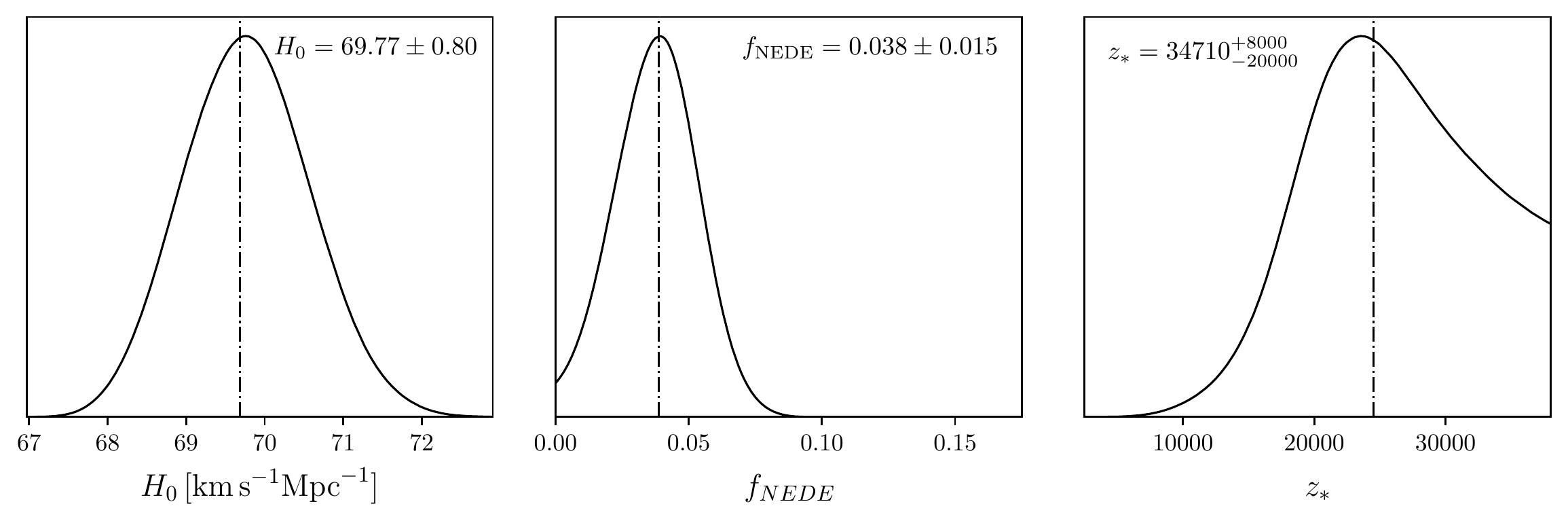}}%
 \caption{Posterior distributions of our combined analysis with SH$_0$ES where we model NEDE after the phase transition as free-streaming radiation, corresponding to the choice $(w^*_\text{NEDE},c_{s}^2, c_\text{vis}^2) = (1/3,1/3,1/3)$. best-fit values are represented by the dash-dotted vertical lines.}
\label{fig:ext_fs}
\end{figure}

\paragraph{\textbf{Free-streaming radiation:}}

Rather than an extension of the base model, we consider a modification where NEDE after the phase transition is converted to free-streaming radiation. In our model this can be realized through a quick decay of the bubble wall condensate into light particles. On a  purely phenomenological level, this case should also be close to the axion proposal in \cite{Kaloper:2019lpl}. As explained before in Sec.~\ref{sec:pert_eqs}, we model the free streaming through the choice $(w^*_\text{NEDE},c_{s}^2, c_\text{vis}^2) = (1/3,1/3,1/3)$. This approximation relies on an early truncation of the Boltzmann hierarchy at order $l = 2$~\cite{Hu_1998}.\footnote{We also devised an independent code where we solved the Boltzmann hierarchy up to higher order, yielding compatible but slightly stronger results, $H_0 = 70.1 \pm 0.9 \kmsMpc$, $f_\text{NEDE} = 0.046_{-0.016}^{+0.017}$ and $z_* = 20130^{+3600}_{-3300}$.}

Our results are presented in Fig.~\ref{fig:ext_fs}. In particular, we find  $f_\text{NEDE} = 3.8 \pm 1.5\, \%$, corresponding to a $2.5 \, \sigma$ evidence for free-streaming NEDE. This has to be contrasted with a {$4.3 \, \sigma$} evidence for the base model relying on a stiffer fluid. This is also in accordance with a smaller overall fit improvement of $\Delta \chi^2 (\text{total})= -5.3$. 
Moreover, the transition occurs significantly earlier for free-streaming radiation, $z_* = 35000^{+8000}_{-20000} $, which implies that changes to the Weyl potential, and, hence, the driving of acoustic CMB oscillations, are limited to rather {small scales in the range $ 420 < \ell < 1300$}.  We conclude that free-streaming NEDE allows one only to alleviate rather than fully resolve the tension. {However, }we will see  that {with an overall improvement of  $\Delta \chi^2  (\text{total})= -5.3$} it is still superior to dark radiation (or sterile neutrinos) with $\Delta \chi^2  (\text{total})= -3.2$.

\subsection{Early-Time Competitors} \label{sec:competitor}

NEDE competes with other early-time modifications that similarly rely on reducing the sound horizon by adding a new energy component that is relevant before matter-radiation equality. Specifically, we will consider dark radiation (DR), Acoustic Dark Energy (ADE) and single-field Early Dark Energy (EDE).  Here, we briefly discuss each proposal and compare it with NEDE.  In each case we perform two combined analyses, one with and one without the local measurement of $H_0$. The corresponding results are collected in Appendix~\ref{app:comp}, in Tab.~\ref{tab:means_comp} we list the mean values together with their $1 \sigma$ uncertainties, and in Tab.~\ref{tab:Delta_chi_2} the $\chi^2$ improvements are detailed. 

\begin{figure*}
 	\subfloat[Combined analysis with SH$_0$ES]
	{\includegraphics[width=16.5cm]{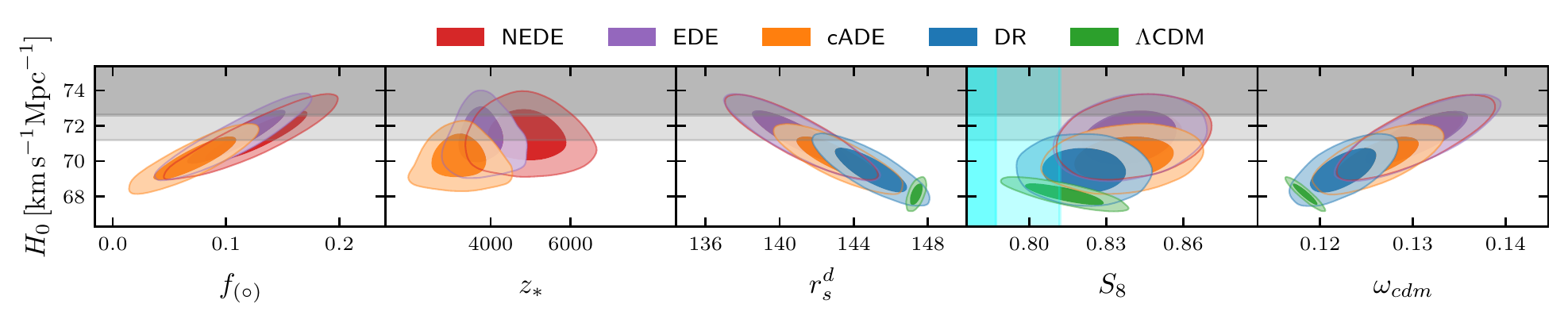}}
	 \label{fig:Competitors_A_w_H0} 	
	  	\\
 	\subfloat[Combined analysis without SH$_0$ES]
	{\includegraphics[width=13cm]{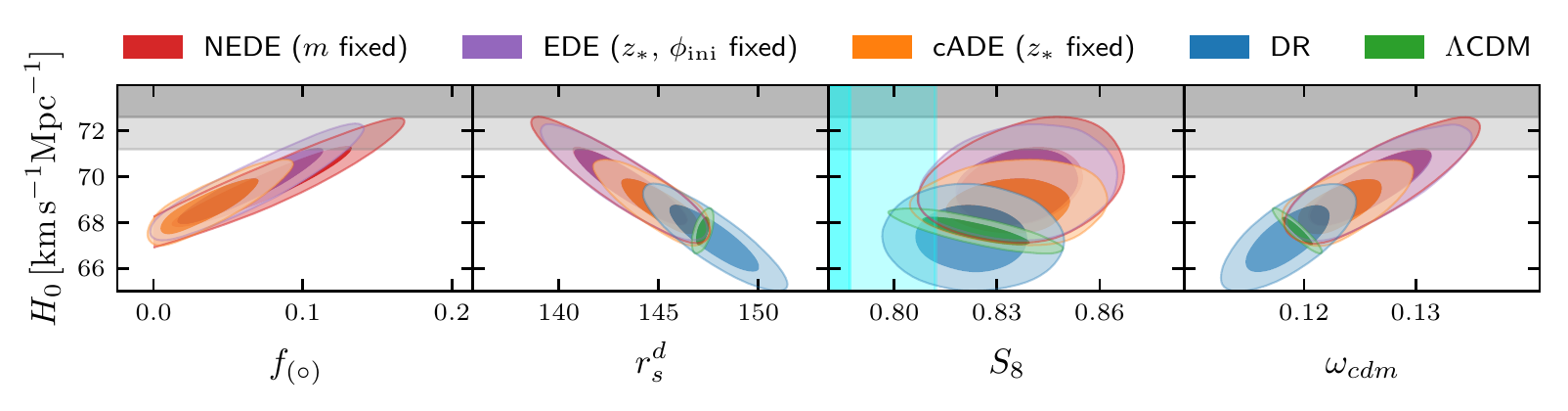}}%
	\label{fig:Competitors_A_wo_H0}
 \caption{The Hubble parameter today vs a subset of parameters for NEDE and its competitors.
 The blue and gray bands corresponds to the $95 \%$ C.L. constraints on $H_0$ (SH$_0$ES) and $S_8$~\cite{Joudaki:2019pmv}, respectively.}
\label{fig:Competitors_A}
\end{figure*}

\begin{figure*} 
	{\includegraphics[width=7.6cm]{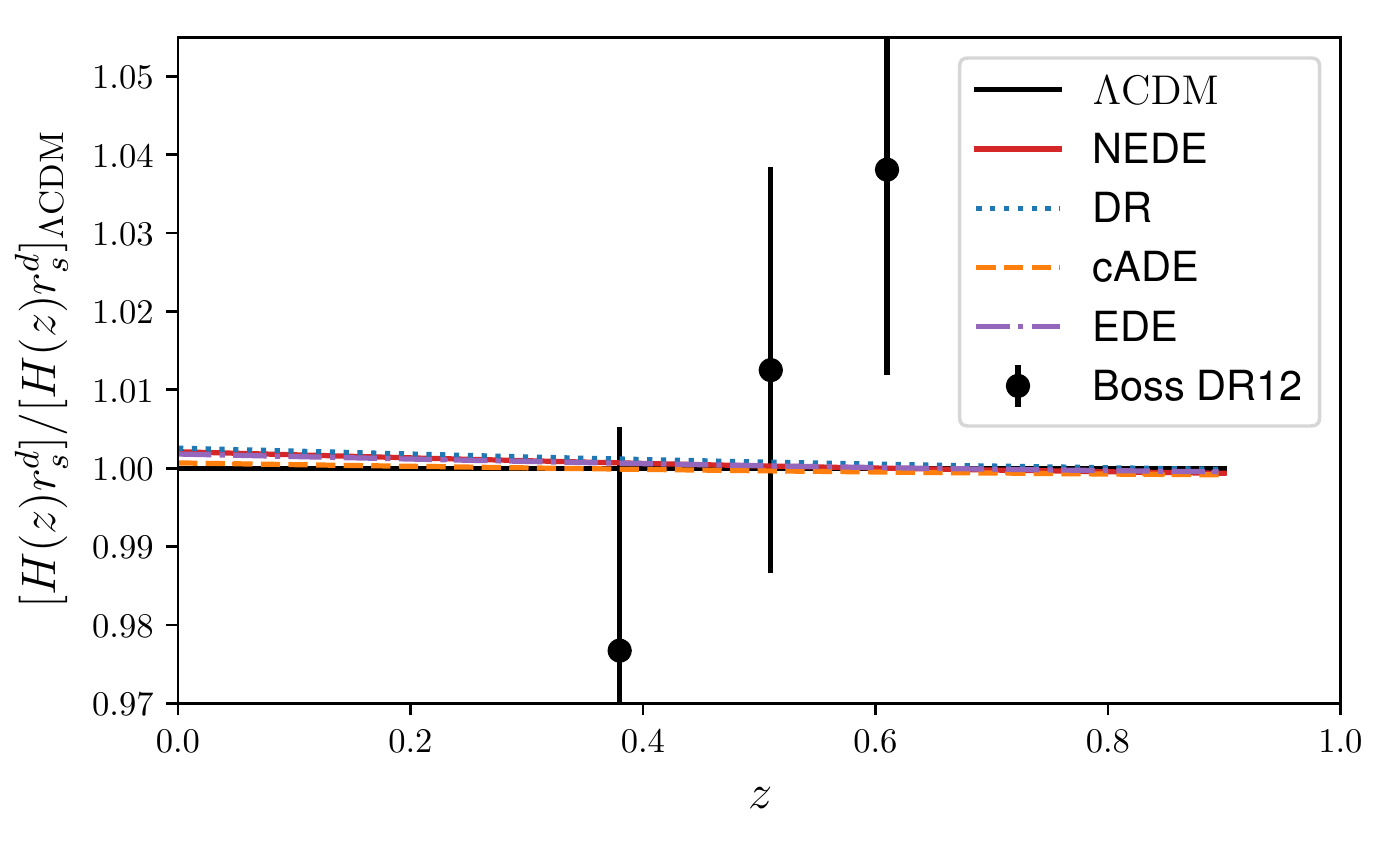}}
	 \label{fig:BAO1} 	
	 \quad
	{\includegraphics[width=7.6cm]{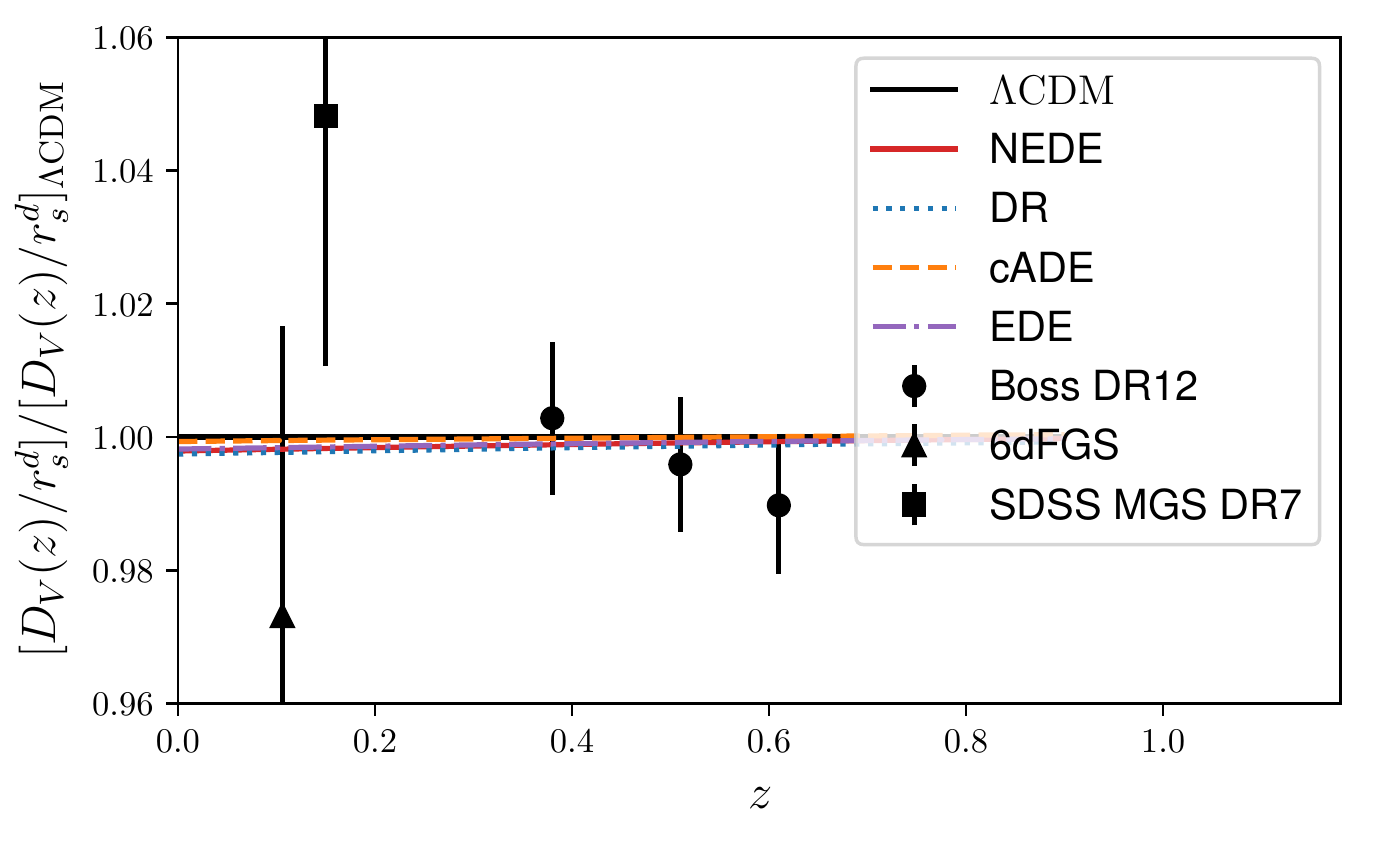}}%
	\label{fig:BAO2}
 \caption{Different BAO measurements represented as black dots. We compare to the best-fit models obtained from our combined data analysis with SH$_0$ES. As opposed to late time modifications, all the early-time modifications studied in this work provide a similarly good fit as $\Lambda$CDM.}
\label{fig:BAO}
\end{figure*}

{These proposals have a common} phenomenological feature which explains their previous success and sets them apart from late-time modifications. They {all} keep the product $r_s^d H_0 \simeq const $, which we argued in Sec.~\ref{sec:phenomenology} is key to accommodating BAO measurements. This is demonstrated in Fig.~\ref{fig:BAO} where we see that all best-fit models provide a consistent fit to BAO data even when we use the local measurement to ``pull'' $H_0$ to higher values. {The left panel depicts $H(z) r_s^d$ directly, and in the right panel we plot the combination $D_V(z) / r_s^d= \left[z D(z)^2 /H(z)\right]^{1/3} \! / r_s^d$, where $D(z)$ is the comoving angular diameter distance at redshift $z$ generalized from \eqref{eq:D_rec}. Both plots are normalized with respect to the $\Lambda$CDM curve. }

\subsubsection{Dark Radiation}

DR (or extra sterile neutrinos) was an early and widely popular attempt to address the Hubble tension~(e.g.~\cite{Wyman:2013lza,Lancaster:2017ksf,Aghanim:2018eyx}).   It relies on introducing new relativistic (or mildly relativistic) degrees of freedom in a dark sector that behave as free-streaming radiation.  It can be parametrized by promoting the effective number of relativistic degrees of freedom, $N_\text{eff}$, to a free parameter. The deformation parameter is then given by $\Delta N_\text{eff} =N_\text{eff} - 3.046 $, where we subtracted the effective number or relativistic degrees of freedom  corresponding to three neutrino species. This modification has been studied extensively in the literature and in its simple form also been ruled out as a full resolution of the Hubble tension (see, for example, \cite{Leistedt:2014sia, Raveri:2017jto,Aghanim:2018eyx} but also more promising recent attempts based on interacting dark radiation~\cite{Kreisch:2019yzn,Blinov:2020hmc}). This statement is also backed by our MCMC analysis, which shows only little evidence for $\Delta N_\text{eff} \neq 0$ ($1.9 \, \sigma$) leading to a small overall improvement of the fit ($\Delta \chi^2 = -3.2$) when the local measurement is included {(in agreement with the recent analysis in \cite{Blinov:2020hmc})}. Similarly, without the SH$_0$ES value we find that the Hubble tension is still at the $\simeq 4\, \sigma$ level.  In the following, we will mainly use DR as a reference modification to infer the efficiency of the other EDE-type proposals. 

From a phenomenological perspective there are two major differences. First, DR is changing the expansion history not only in a small window around radiation-matter equality but also at much earlier times. This is problematic because it affects BBN, causing an increase of the primordial helium abundance, $Y_p$. This worsens the corresponding fit by {$\Delta \chi^2(\text{BBN}) = 0.8$} when we include the local $H_0$ measurement. This argument could be refined by also taking into account the deuterium abundance as done in \cite{Schoneberg_2019}, with the qualitatively same outcome. As we can also see from Tab.~\ref{tab:Delta_chi_2}, this problem is completely avoided by modifications of the EDE type which have a negligible effect during BBN. A second background effect of DR is the lowering of $r_s$, which due to \eqref{eq:r_d_over_r_s} also implies an enhancement of the ratio $r_d/r_s$. This leads through \eqref{eq:damping} to stronger diffusion damping effects. So far this is analogous to NEDE. However, as DR is only a one-parameter extension of $\Lambda$CDM, it does not offer enough freedom to fully compensate this effect by, for example, increasing $n_s$ as explained in Sec.~\ref{sec:phenomenology}. This becomes also apparent from the residuals depicted in Fig.~\ref{fig:residuals} where we detect a stronger loss of power at small scales than for the EDE-type models (see also \cite{Hou_2013}).  Accordingly, DR causes the strongest degradation of the high-$\ell$ {\sl Planck} fit when SH$_0$ES is included, amounting to {$\Delta \chi^2(\text{high-}\ell) = 3.2$}. Besides this damping problem, at the perturbation level, DR corresponds to an additional free-streaming radiation component which sources anisotropic stress. This has been shown to lead to a shift in the phase of CMB oscillations~\cite{Baumann_2016} which is also adversely affecting the CMB fit. {However, as recently argued in \cite{Brust:2017nmv,Blinov:2020hmc} both problems can be improved upon} by introducing an interacting radiation component.

In summary, we find that DR cannot resolve the Hubble tension mainly because it is constrained by the BBN probe, but also because it does not offer enough flexibility to balance the additional damping and phase shift effects it introduces to the photon-baryon acoustic oscillations.

\subsubsection{Acoustic Dark Energy}

ADE was developed to mitigate the before-mentioned detrimental CMB effects by introducing acoustic oscillations in a dark fluid (rather than free-streaming particles)~\cite{Lin:2019qug}. At the background level it is very similar to other EDE implementations as it describes a smooth but relatively sudden transition from an equation of state parameter close to (but slightly above) $-1$ at early times to $w^*_\text{ADE} > 1/3$ at late times, parametrized as
\begin{align}
w_\text{ADE}(a) +1 = \frac{1+ w_\text{ADE}^*}{\left[ 1 + \left( a_*/a\right)^{6 \left(1+ w_\text{ADE}^*\right)}\right]^{1/2}} \,,
\end{align}
{where $a_* \sim 1/ 3000 $ leads to a transition close to matter-radiation equality. As with NEDE, the amount of ADE is implicitly defined via its relative abundance at `decay time', $f_\text{ADE} = \rho_{ADE}^* / (3 M_{pl}^2 H_*^2)$.}

On the perturbation level, ADE is described as a fluid with time-dependent equation of state where the corresponding fluctuations $\delta_\text{ADE}$ and $\theta_\text{ADE}$ are controlled by the system \eqref{eq:pert_dyn}. Here, we will focus on its canonical implementation (cADE) for which $(w_\text{ADE}^*,c_s^2,c_\text{vis}^2) = (1,1,0)$. In~\cite{Lin:2019qug} it was argued that this can be realized with a canonical scalar field that converts all of its potential to kinetic energy around $a_*$ (without converting it back). An example is provided by a field that after the release of the Hubble drag explores a flat direction in its potential. In particular, this excludes scenarios where the field oscillates around a minimum like the ones discussed in~\cite{Poulin:2018cxd,Agrawal:2019lmo}. The parametrization also covers more exotic models with a non-standard kinetic term, which, in principle, allow for  $w_\text{ADE}^* = c_s^2 < 1$~\cite{Lin:2019qug}. As it is generically harder to justify these models from a fundamental perspective, we will focus here on the stiff fluid case only.

We implemented cADE in {\CLASS } and performed a MCMC likelihood analysis imposing flat priors on the two parameters $-4.2 < log_{10} \left( a_* \right) < -2.6$ and  $0 < f_\text{ADE} < 0.3$. The resulting parameter covariances are shown as the orange contours in  Fig.~\ref{fig:Competitors_A}. For the analysis without the local measurement of $H_0$, we fixed  $log_{10} \left( a_* \right) = -3.46$  (close to the best-fit determined from a previous run) to avoid the sampling problems outlined in Sec.~\ref{sec:metholdology}. In this case, cADE predicts  $H_0 = 68.94^{+0.68}_{-0.93}\kmsMpc$ which reduces the tension down to $ \simeq \, 3.2\, \sigma$ (including $w^*_\text{ADE}$ and $c_s^2$ as free parameters would presumably further reduce the tension). The evidence for cADE is rather low with $f_\text{ADE} = 0.038_{-0.027}^{+0.018}$.  If we include the local prior on $H_0$, on the other hand, we reach a higher  value, $H_0 = 70.15^{+0.86}_{-0.81} \kmsMpc$, together with a significant fit improvement of $ \Delta \chi^2(\text{total}) = -9.9$ and a $3.2 \, \sigma$ evidence for $f_\text{ADE} \neq 0$, which places it somewhere in the middle between DR and NEDE.\footnote{cADE performs slightly better in~\cite{Lin:2019qug} with $H_0 =  70.60 \pm 0.85 \kmsMpc$. We attribute this to the difference in datasets as their analysis used {\sl Planck} 2015 (rather than {\sl Planck} 2018) data.} It appears that cADE is held back by the large-$\ell$ temperature and polarization data, which indeed deteriorate significantly by {$\Delta \chi^2 (\text{high-}\ell)= 2.9$} if we {apply more} statistical ``pull'' towards higher values of $H_0$ by including the SH$_0$ES value.

What explains the difference between cADE and NEDE? Besides the stronger decay of the background energy density, the main difference lies in the perturbation sector. In contrast to NEDE, the fluid perturbations are initialized long before the decay when they are still frozen outside the horizon, i.e., $a H \gg k$. In the case of NEDE, on the other hand,  modes are initialized much later at $z_* \sim 5000$, right after the phase transition, implying that for  ${\ell \gtrsim 180}$ they are inside the particle horizon. As we discussed in Sec.~\ref{sec:phenomenology}, that makes them particularly sensitive to fluctuations in the transition surface, which is a distinct $k$ dependent feature of NEDE not present in ADE. Another notable difference between cADE and both EDE and NEDE is its later decay time, which roughly coincides with radiation matter equality; to be specific,  $z_* = 3230^{+400}_{-500}$. This is due to the quicker decay of cADE, which prevents adversary driving effects which would otherwise prevent the decay from happening too late. 

\begin{figure}
 \centering 
{\includegraphics[width=16.5cm]{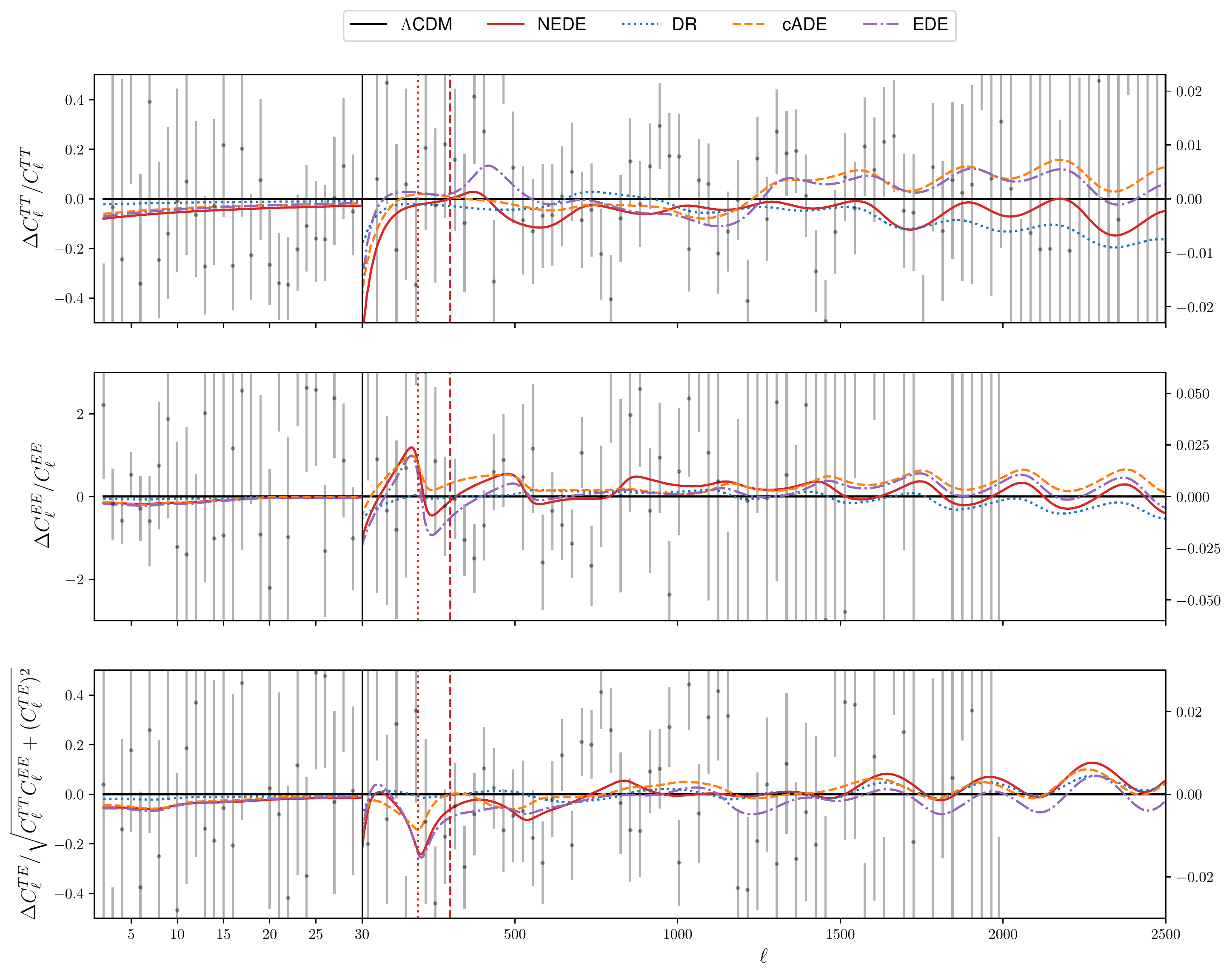}}%
 \caption{{\sl Planck} data residuals (gray dots) of NEDE (red solid), DR (blue dotted), and cADE (orange dashed) and EDE (purple dash-dotted) with respect to $\Lambda$CDM.  Best-fit values are taken from combined analysis with SH$_0$ES (Tab.~\ref{tab:means_NEDE_LCDM}). The plots use a different scaling for low $\ell$ ($<30$) and high $\ell$ ($>30$). The red dashed and dotted vertical lines highlight the modes that respectively entered the sound and particle horizon at the time of the NEDE phase transition.}
\label{fig:residuals}
\end{figure}
 
 In summary, cADE shows the same approximate degeneracies as NEDE but is not able to fully resolve the tension, even though it leads to a clear reduction of it {by introducing only two new parameters}. This general picture is confirmed in Fig.~\ref{fig:residuals}, where cADE shows larger residuals in the temperature power spectrum for $600 < \ell < 1000$.

\subsubsection{Early Dark Energy} \label{sec:EDE}

Finally, there is single-field EDE~\cite{Poulin:2018dzj,Poulin:2018cxd,Smith:2019ihp}. It introduces an ultra-light axion field with decay constant $f_a \, (\sim M_{pl})$ and potential
\begin{align}\label{eq:pot_EDE}
V^{(n)}(\varphi) = \Lambda_a^4 \left[ 1 - \cos\left( \varphi / f_a \right) \right]^n \,,
\end{align}
where $\Lambda_a \, ( \sim eV)$ is a dark sector energy scale. At early times when $H^2 \gg d^2V/d\varphi^2$, the field is stuck at  $\varphi \simeq \varphi_\text{ini} \, (< \pi f_a)$. Once the Hubble drag is released around  $z= z_*$, implicitly defined  via $H^2_* \simeq  d^2V/d\varphi^2$, $\varphi$ starts to oscillate in a near power law potential, $ \varphi^{2n}$, which leads to a quick decay of EDE.  By averaging over oscillation cycles, the decay can be described by an ideal fluid with equation of state parameter  $w^*_{EDE} \simeq (n-1)/(n+1) $, where $n=1$ corresponds to a decay like matter, $n=2$ to one like radiation and $n \geq 3$ to an even quicker decay.  In fact, the first two cases were shown to be disfavored by data, whereas the model with $n=3$, where kinetic dominates over potential energy, is most efficient in addressing the Hubble tension, implying $w^*_{EDE} = 1/2$.   Such a shallow potential term, scaling like $\phi^6$, has been argued to arise in axion scenarios when taking into account higher-order instanton corrections~\cite{Kamionkowski:2014zda,Poulin:2018dzj,Kappl:2015esy}.  The problem with this argument, however, is that within the low-energy theory \textit{all} terms are expected to be present, i.e.,  in general, $ V(\varphi)  = \sum_n c_n V^{(n)}$ with dimensionless coefficients $c_n = \mathcal{O}(1)$. The model, therefore, relies on the assumption $c_1,c_2 \ll c_3$, which from a low-energy perspective appears to be a non-generic choice~\cite{Kaloper:2019lpl}.  

Besides this naturalness issue, the  above model constitutes an interesting proposal for resolving the Hubble tension. On a cosmological level, it introduces four new parameters: the decay time $z_*$, the fraction of EDE at decay time, $f_\text{EDE}= \rho_\text{EDE}^* / (3 M_{pl}^2 H_*^2)$, the initial field value $\varphi_\text{ini}  \, (< \pi f_a)$ and the integer power $n$. Here, we set $n=3$ and, therefore, study a three-parameter extension of $\Lambda$CDM. Like with ADE, when described as an ideal fluid, EDE shows a very similar behavior to NEDE on the background level even though it relies on a second- rather than a first-order phase transition. Differences, however, arise on the perturbation level, where the higher-order potential in \eqref{eq:pot_EDE} leads to an effectively $k$ dependent sound speed $c_s(k)$, which was argued to be crucial for relaxing the Hubble tension~\cite{Smith:2019ihp}. In particular, $c_s(k)$ depends on the initial field value $\varphi_\text{ini}$, which needs to be set close to the  Planck scale for the mechanism to work properly, thereby probing the flat part of the potential. Our cosmological model, on the other hand, uses a $k$ independent sound speed. Also, {NEDE} can accommodate an initial field value of its trigger field in a wide range~\eqref{eq:phi_bound}. The reason is that {in NEDE}  the trigger field, $\phi$, is sub-dominant throughout its whole evolution; moreover, $\phi_\text{ini}$ drops out of the junction conditions rendering cosmological observables insensitive to its value.  Another crucial difference is that the transition in EDE is a smooth process at both the background and perturbation level, whereas it is of a sudden nature in NEDE. To be precise, in NEDE it is initiated when $\phi/\phi_\text{ini}$ drops below a certain threshold determined by the parameter $H_*/m$ {alone}.  As we have argued in Sec.~\ref{sec:phenomenology}, it is this {trigger} mechanism which can mimic a $k$ dependent sound speed.

In order to perform our MCMC analysis of EDE, we used an existing {\CLASS} implementation of EDE and scanned in addition to the $\Lambda$CDM parameters over $(\log_{10}(a_*), f_\text{EDE}, \varphi_\text{ini})$. We also tracked the full evolution of $\varphi$ including its late-time oscillations. Everything else was implemented as detailed in Sec.~\ref{sec:metholdology}. The posterior distributions turn out to be slightly bimodal, where the dominant mode is centered around $\log_{10}(a_*) \simeq - 3.6$ and the less significant one around $\log_{10}(a_*) \simeq -3.8$. In fact, in~\cite{Smith:2019ihp} it was argued that this can be attributed to the constraining effect of the high-$\ell$ polarization data, which is also {what we find here}. 
As we want to focus on the more significant mode only, we chose rather tight prior boundaries for our final MCMC analysis, $ -3.7 < \log_{10}(a_*) < -3.2 $ (and $ 2.0 < \varphi_\text{ini} < 3.14 $). When we performed our analysis without SH$_0$ES, we encountered again the sampling problem detailed in Sec.~\ref{sec:metholdology}. In fact, it is even worse compared to NEDE, as two (rather than one) parameters, $ f_\text{EDE}$ and $\varphi_\text{ini}$, become degenerate in the limit $f_\text{EDE} \to 0 $ eliminating any evidence for EDE. We were again able to mitigate this problem by fixing both $ f_\text{EDE}$ and $\varphi_\text{ini}$ to their near best-fit values as detailed in Tab.~\ref{tab:means_comp}. 

The resulting covariances of EDE are presented in Fig.~\ref{fig:Competitors_A}. It is striking to see that the EDE contours (purple) are largely overlapping with the ones of NEDE (red) despite the differences of both models at the perturbation level. This picture is solidified when we look at the results listed in Tab.~\ref{tab:means_comp} and \ref{tab:Delta_chi_2}. Specifically, EDE achieves {$H_0 = 69.6^{+1.0}_{-1.2} \kmsMpc$ } without SH$_0$ES, which brings the Hubble tension down to {$2.5 \, \sigma$} and leads to a weak detection of EDE at the {$2 \, \sigma$} level. With SH$_0$ES, our MCMC analysis yields a stronger $ {4 \, \sigma}$ detection\footnote{This value is  compatible with the $\simeq 3.5\, \sigma$ reported in~\cite{Smith:2019ihp}. } and {$H_0 = 71.3 \pm 1.0 \kmsMpc$ },\footnote{As a cross-check note that this is compatible with the value found in \cite{Smith:2019ihp}, $H_0 = 71.49 \pm  1.20 \kmsMpc$, using the same datasets with the exception of {\sl Planck} 2018 data (instead of 2015 data) and the helium abundance.} corresponding to an improvement of the overall fit of {$\Delta\chi^2(\text{total}) = -16.4$}, which is almost identical with NEDE. The transition times $z_*$ are quite different, though. While EDE prefers to decay at ${z = 3760_{-410}^{+320}}$, {the NEDE phase transition} is considerably earlier at $z_* = 4920^{+620}_{-730} $ (see also Fig.~\ref{fig:Competitors_A_w_H0}).  

On the level of the cosmological parameter extraction, NEDE trades the parameter $\varphi_\text{ini}$ for the trigger parameter $H_*/m$, and achieves a {similarly promising} outcome in terms of cosmological observables. However, there is a crucial difference: While  $\varphi_\text{ini}$ is unconstrained within the axion framework of EDE, we do not have much freedom in choosing the ratio $H_*/m$. In fact, the trigger dynamics of our underlying model fix it to be $H_*/m \simeq 0.2$, which happens to be the value preferred by data. A closer inspection of Fig.~\ref{fig:residuals} also reveals that NEDE leads to {different} oscillatory residuals in both the temperature and polarization  power spectrum (in particular for $\ell \gtrsim 1500$). {We attribute this mainly to different perturbation dynamics related to the trigger sector.} Ultimately, we believe next-generation CMB experiments like the ground-based \textit{CMB-S4}~\cite{Abazajian:2016yjj} or  \textit{The Simons Observatory}~\cite{Ade:2018sbj} will be able to discriminate between both scenarios  (notwithstanding that gravitational waves might provide another distinct feature of NEDE) and detect either of them in CMB data. In order to further quantify these statements, an analysis with mock data similar to the one in \cite{Smith:2019ihp} would be needed.

\section{Conclusion}

NEDE is an early-time modification of the expansion history that relies on a first-order phase transition in a dark sector {to add} a short burst of gravitational repulsion around the eV scale. It introduces a two-field mechanism in order to trigger the phase transition at the right time and avoid problems with premature nucleation events that would otherwise risk a contamination of the CMB. By performing a cosmological parameter extraction of an effective cosmological implementation of NEDE, we find that it shares the phenomenological success with its single-field predecessor despite different perturbation dynamics. We report a $4.3 \, \sigma$ evidence for our NEDE base model and an increased $H_0=71.4 \pm 1.0 \kmsMpc$ with SH$_0$ES included in our analysis. This corresponds to an improved fit with $\Delta \chi^2(\text{total}) = -15.6$ when including the fraction of NEDE $f_\text{NEDE}$ and the trigger mass $m$ as additional free parameters.  Without SH$_0$ES, we still find $1.9 \, \sigma$ evidence and $H_0 = 69.6^{+1.0}_{-1.3} \, \kmsMpc$, which amounts to a reduced Hubble tension of $2.5 \, \sigma$ (down from $4.3\, \sigma$ for $\Lambda$CDM). 

This improvement is achieved for a rather generic choice of microscopic parameters, where the relevant terms in the potential of the NEDE field $\psi$ are set by $M \sim \text{eV}$ and its quartic coupling is weak $\lambda < 0.1$. In fact, only a mild one-digit tuning is applied to suppress isocurvature modes and prevent sizable trigger field oscillations around the true vacuum. The trigger sector, on the other hand, necessitates ultra-light physics with a mass scale $m \sim 10^{-27} \, \text{eV}$, corresponding to a decay around redshift $\sim 5000$.  We demonstrated the technical naturalness of this setup. {In particular, we derive an upper bound on the coupling between both fields to ensure the radiative stability of the ultra-light scale.} 

The crucial trigger parameter $H_*/m$, {controlling the amplitude of sub-horizon modes in the NEDE fluid}, is predicted by the model to be in a very narrow range around $H_*/m=0.2$ for natural, non-fine-tuned, choices of parameters. A fit to data treating it as a free parameter gives a best fit value of $H_*/m={2.06}^{+0.013}_{-0.022}$, which we take as additional strong evidence for the NEDE model.

{We also provide a detailed comparison with single-field EDE and find a similar successful phenomenology.   On the theory side, however, EDE is faced by two challenges. First, a steep anharmonic potential is needed to generate a sufficiently quick decay of the background condensate, and second, the potential has to flatten to avoid detrimental effects in the polarization power spectrum that otherwise arise when the range of sub-horizon modes with sound speed $<0.9$ is too small~\cite{Smith:2019ihp}. While the last challenge can be naturally satisfied within an axion framework, an anharmonic potential constitutes a non-generic choice and requires further theory adaptions. NEDE gets around both obstacles in a different way. We expect the stiff fluid needed for a quick decay to be effectively generated by the small-scale anisotropic stress associated with the colliding bubble wall condensate, and the issue with polarization residuals does not arise in our case which we attribute to the momentum-dependent trigger dynamics.}

While we expect the high-$\ell$ signatures of NEDE to be revealed more clearly through future ground-based CMB experiments, a possible detection through gravitational waves is a unique feature of our first-order scenario. As the transition occurs relatively late in the early Universe as compared to the electroweak phase transition, the expected gravitational wave spectrum is peaked at extremely small frequencies $f_*$, e.g.\  for $H_* \bar{\beta}^{-1} > 10^{-3}$, we have $f_* < 10^{-14} \, \text{Hz}$ . We still find that the spectrum's large frequency tail can overlap with the claimed sensitivity bound of future gravitational wave experiments provided the duration of the phase transition is long enough. We also expect this signal to be strengthened in extended scenarios where we include a coupling between the dark sector and visible matter due to sound waves in the primordial fluid as well as turbulence effects. 

We will end this paper by outlining different directions for future investigations. First, there is the theoretical challenge of finding a more fundamental framework that explains not only the light physics involved in our model but also the extremely weak coupling between both fields needed to keep radiative corrections under control. We observe that the trigger mass lies between a typical mass scale for ultralight DM ($\sim 10^{-22} \, \text{eV}$) and quintessence DE ($\sim 10^{-33} \, \text{eV}$), both of which can be realized in axion scenarios~\cite{Marsh:2015xka,Ferreira:2020fam}, where the mass is protected by a weakly broken shift symmetry. We believe that the two-field mechanism of NEDE can also be obtained from a string-inspired axion setup~\cite{Svrcek:2006yi,Arvanitaki:2009fg} but leave the details to future work. 

We have seen that the phenomenological success of NEDE crucially depends on the properties of the fluid after the decay and, in particular, on its equation of state {parameter}, $w_\text{NEDE}^*$, as well as its rest-frame sound speed, $c_s$. To be specific, by extending our base model we obtained $w_{NEDE}^* = 0.70^{+0.12}_{-0.16}$  and  $ c_s^2 = 0.711^{+0.057}_{-0.081}$, suggestive of a quickly decaying fluid with $c_s^2 = w_\text{NEDE}^*$. Adding two more parameters then also diminishes the evidence for a non-vanishing fraction of NEDE down to $3.4 \, \sigma$ [$f_\text{NEDE} = 11.3^{+3.7}_{-3.3}\, \% $]. While we indeed expect the relativistic bubble wall condensate to manifest itself on cosmological scales as a fluid with $w_\text{NEDE}^* > 1/3$ due to its small-scale anisotropies and kinetic energy domination, we have not yet provided a microscopic calculation.  Instead, we have treated both quantities as phenomenological parameters and fixed them in our base run to $c_s^2 = w_\text{NEDE}^* = 2/3$, as suggested by our extended analysis and previous studies. {We expect to further explore this in future work.}

From a more phenomenological perspective, there remains the issue with the growth of structure as inferred from weak lensing measurements. This so-called $S_8$ tension~\cite{Raveri:2018wln} is slightly increased from $2.3\, \sigma$ to $2.8 \, \sigma$ when compared with $\Lambda$CDM. This means that if the tension gets worse in the future, it will threaten the success not only of $\Lambda$CDM but also of  EDE-type models. We therefore believe that further sophistications of NEDE should be driven by the goal to reduce both tensions at the same time. Such possibilities include considering non-trivial interactions between NEDE and the visible sector or including a sizable amount of scalar field oscillations around the new vacuum after the phase transition (assumed to {vanish} in the present study) which will change the decay properties of the Weyl potential.  In order to test this class of proposal, it will be important to include additional LSS probes in the data analysis.

\acknowledgments
 
We thank {Guido D'Amico, Edmund Copeland, Jos\'e Espinosa, Colin Hill, Nemanja Kaloper, Antonio Padilla, Vivian Poulin, Kari Rummukainen, and {David Weir}} for useful discussions. {We also thank the SDU eScience Center for providing us with the computational resources for the MCMC analysis.}
This work is supported by Villum Fonden grant 13384.

\appendix
\section{Junction Conditions in Cosmological Perturbation Theory}  \label{sec:matching}

{In this appendix, we derive the perturbed matching conditions across the transition surface  $\Sigma$, formally defined in terms of the function $q(t,\mathbf{x})$ via \eqref{eq:q}. 
To that end, we restore the most general form of the metric \eqref{eq:sync}:
\begin{align}
ds^2 = -\left(1+E\right) \, dt^2 +2\, a(t) \partial_i F dt dx^i 
+ a(t)^2 \left[\left( 1 + A \right) \delta_{ij} + \partial_i \partial_j B \right] dx^i dx^j  \;,
\end{align}
with scalar perturbations $A(t, \mathbf{x})$, $B(t, \mathbf{x})$, $E(t, \mathbf{x})$, and $F(t, \mathbf{x})$ (in this work we are not interested in vector and tensor modes).}
The synchronous gauge corresponds to the choice
\begin{align}\label{eq:dict_sync}
A = -2 \eta\,, && k^2 B^2 = - \left(h + 6 \eta \right)\,, && E=F=0\,.
\end{align}
The different fluctuation fields then transform under an infinitesimal coordinate transformation  $x^0 \to x^0 + \xi^0$ and $x^0 \to x^i + g^{ij} \, \partial_j \xi$  as
\begin{subequations}\label{eq:gauge}
\begin{align} 
\Delta_\xi E &= 2 \dot \xi_0 \,, & \Delta_\xi F &= -\frac{1}{a} \left( \xi_0 + \dot \xi - 2 H \xi \right) \,, \\
 \Delta_\xi A &= 2 H \xi_0 \,, &\Delta_\xi B &= -\frac{2}{a^2} \xi \,. 
\end{align}
\end{subequations}
In order to simplify our derivation, we consider only spatial diffeomorphisms $\xi$ that are continuous across the surface, i.e., $\left[ \xi \right]_\pm =0$; otherwise, $\xi$ and $\xi_0$ are arbitrary functions. 
For the function $q(t, \mathbf{x})$ we find
\begin{align}
\Delta_\xi \delta q = \dot{\bar{q}} \, \xi_0 \,.
\end{align}
We now use this gauge freedom to fix
\begin{align}
\delta q \to \delta \hat{q} = \delta q + \dot{\bar{q}} \, \hat{\xi}_0 \stackrel{!}{=}0\, ,
\end{align}
where $\hat{\xi}_0 = - \delta q / \dot{\bar{q}} $. Working in  the gauge ${\delta \hat{q}} =0$, which we henceforth indicate by hatted variables, greatly simplifies the derivation of the matching conditions. 

Next, we introduce the projection tensor 
\begin{align*}
\tilde{g}_{\alpha \beta} = g_{\alpha \beta} + n_\alpha n_\beta \;,
\end{align*}
{where $n$ is the unit normal vector with respect to the transition surface,
\begin{align}\label{eq:n}
n_\alpha = c[g] \, \partial_\alpha q\,
\end{align}
with $c[g]$ a normalization constant determined through  $n_\alpha n^\alpha = -1$. At the background level, we find $\bar{n}_\alpha = (1,0,0,0)$.}
The extrinsic curvature tensor is then defined as 
\begin{align}
K_{\alpha \beta} &= \frac{1}{2} \mathcal{L}_n \tilde{g}_{\alpha \beta} \nonumber\\
		&= \frac{1}{2} n^\gamma \partial_\gamma \tilde{g}_{\alpha \beta} + \tilde{g}_{\gamma(\alpha} \partial_{\beta)} n^\gamma   \;.
\end{align}
{Israel's junction conditions in the absence of matter localized on $\Sigma$ read~\cite{Israel:1966rt}
\begin{align}\label{eq:Israel}
[K_{\alpha \beta}]_\pm = 0\, ,
\end{align}
where we use the notation introduced in \eqref{eq:disc_bracket}. It is straightforward to show that this reduces to \eqref{eq:JumpH} on the level of background quantities.}
Perturbing $K_{\alpha\beta}$ and projecting on the spatial components yields
\begin{align}
\delta K_{ij} = \delta n^0 H g_{ij} - 1/2 \delta \dot{\tilde{g}}_{ij} + a^2 \delta_{k(i}\partial_{j)} \delta n^k\,,
\end{align}
where we used the background expression for the normal vector. Further perturbing \eqref{eq:n} as well as its normalization condition, we derive
\begin{align}
\delta n_i =0 \,, && \delta n^i = \frac{1}{a} \delta^{ik} \partial_k \hat{F} \,, &&  \delta n^0 = \delta n_0 = \frac{1}{2} \hat{E} \;.
\end{align}
From this it follows that $\delta g_{ij} =\delta \tilde{g}_{ij} $. With these definitions, we finally have
\begin{align}
\delta K_{ij} = \left[\frac{1}{2} H \hat{E} - \frac{1}{2} \dot{\hat{A}} - H \hat{A} \right] g_{ij} 
+ a^2 \partial_i \partial_j \left[ - \frac{1}{2} \dot{\hat{B}} - H \hat{B} + \frac{1}{a} \hat{F} \right]\,.
\end{align} 
{The continuity of the hypersurface-induced metric, $\left[ \tilde g_{ij} \right]_\pm = 0$, together with Israel's matching \eqref{eq:Israel} requires at the perturbation level }
\begin{subequations}
\begin{align}
\left[ \delta \tilde{g}_{ij}\right]_\pm &=0 \;, \\
\left[ \delta K_{ij}\right]_\pm &=0 \;,
\end{align}
\end{subequations}
which, in turn, provides us with the following junction conditions in the ``hat''-gauge,
\begin{subequations}
\begin{align}
[\hat{A}]_\pm&=0 \,, \label{eq:hat_A}\\
[\hat{B}]_\pm&=0 \,, \\
H_* [\hat{E}]_\pm &= [\dot{\hat{A}}]_\pm \,,\\
[\dot{\hat{B}}]_\pm &= \frac{2}{a_*} [\hat{F}]_\pm  \label{eq:hat_L}\,. 
\end{align}
\end{subequations}
We now use \eqref{eq:gauge} together with  $\hat{\xi}_0 = - \delta q / \dot{\bar{q}} $ to restore $\delta q$ and, hence, the full gauge invariance,\footnote{Note that this agrees with the result in \cite{Deruelle:1995kd} subject to the dictionary  $E = 2 \phi_\text{there}$, $F = B_\text{there}$, $A = -2 \psi_\text{there}$ and $B = 2 E_\text{there}$. }
\begin{subequations}
\label{eq:israel}
\begin{align}
\left[A \right]_\pm &= 2 H_0 \left[\frac{\delta q}{\dot{\bar{q}}} \right]_\pm\,, \\
[B]_\pm &=0\,, \\
H_* [E]_\pm + 2 \left[\dot H \, \frac{\delta q}{\dot{\bar{q}}}\right]_\pm &= [\dot A]_\pm\,, \\
a_*^2 [\dot B]_\pm - 2 \left[\frac{\delta q}{\dot{\bar{q}}}\right]_\pm &= 2 a_* [F]_\pm\,. 
\end{align}
\end{subequations}
Using \eqref{eq:dict_sync} then yields the junction conditions in the synchronous gauge in \eqref{eq:matching0}. 

Next, we consider a fluid with energy density $\rho_i = \bar{\rho}_i + \delta \rho_i$ and equation of state parameter $w_i=const$. In order to derive the matching equations for the density contrast, $\delta_i = \delta \rho_i / \bar{\rho}_i$, and the divergence of the velocity field, $\theta_i $, we follow a slightly different but entirely equivalent approach to the one used before. The conservation of the energy momentum tensor implies at linear order two independent equations,
\begin{subequations}
\label{eq:fluid_pert}
	\begin{align}
		\begin{multlined}[b]
		\dot{{\delta}}_i  
		+ \left(1 +w_i \right) \left( \frac{{\theta}_i}{a} +   \frac{k^2}{a^2} {L} + \frac{3}{2}  \dot{{A}} \right)  	
		\end{multlined} & =0 \,,
	\label{eq:delta_fluid}\\
	\begin{multlined}[b]
a \, \dot{{\theta}}_i - \frac{w_i}{1 + w_i} k^2   {\delta_i}   + \left( 1 - 3 w_i \right) a H {\theta}_i + k^2 \sigma_i  - k^2 \frac{{E}}{2}  
	\end{multlined} & =0 \,,
\end{align}
\end{subequations}
where we introduced $L = a (F- a \dot{B}/2)$ and assumed that the rest-frame sound speed equals the adiabatic sound speed.  As a consistency check, we recover \eqref{eq:pert_dyn} {in the limit $c_{s,i}^2 = c_{a,i}^2  = w_i = const$} by using the synchronous gauge condition in \eqref{eq:dict_sync}. We now impose the `hat'-gauge (corresponding to $\delta\hat{q} = 0 $), which `straightens' the transition surface. In order to derive the matching equations, we follow a slightly different but entirely equivalent approach to the one used before: We integrate both equations over an infinitesimal time interval; specifically, we perform $\lim_{\epsilon \to 0}\int_{t*-\epsilon}^{t_*+\epsilon} d t\,\circ $, and use  that the terms proportional to $\hat{L}, \hat{\delta_i},\hat{\theta}_i, \sigma_i$ and $\hat{E}$ vanish as they do not contain any distributional part $\propto \delta(t - t_*)$. In other words, we expect perturbations without at least one time derivative acting on them to be regular. This last steps works only because we work in a gauge for which the transition surface is not fluctuating. In any case, we obtain\footnote{As a cross-check note that we could have obtained Israel's equations in \eqref{eq:israel} by performing the same steps with Einstein's equations that are second-order in time derivatives.}
\begin{subequations}
\begin{align}
[\hat\delta_i]_\pm &=  - \frac{3}{2}\left(1 +w_i \right) \left[ {\hat{A}} \right]_\pm  = 0 \,,\\
[\hat\theta_i]_\pm &= 0 \,,
\end{align}
\end{subequations}
where we used \eqref{eq:hat_A} and \eqref{eq:hat_L} to derive the second equality in the first line. We now use the transformation rule
\begin{subequations}
\begin{align}
\Delta_\xi \delta_i &= - 3 H \left( 1 +w_i \right) \xi_0 \,, \\
\Delta_\xi \theta_i &= \frac{k^2}{a} \xi_0 \,
\end{align}
\end{subequations}
together with  $\hat{\xi}_0 = - \delta q / \dot{\bar{q}} $ to restore $\delta q$ and, hence, the full gauge invariance,
\begin{subequations}
\label{eq:matching_fluid}
\begin{align} 
\left[\delta_i +  H \left(1 +w_i \right) \frac{\delta q}{\dot{\bar{q}}} \right]_\pm &=0\,, \\
\left[\theta_i - \frac{k^2}{a} \frac{\delta q}{\dot{\bar{q}}}\right]_\pm &= 0 \,.
\end{align}
\end{subequations}
In the next step, we can also evaluate the discontinuity of \eqref{eq:fluid_pert} directly, which after using \eqref{eq:matching_fluid} and \eqref{eq:israel} yields
\begin{subequations}
\label{eq:matching_fluid_2}
\begin{align}
\left[\dot{\delta}_i + \frac{3}{2} \left(1 +w_i \right)  \dot{A} \right]_\pm &= 0 \,, \\
\left[ a \, \dot{\theta}_i + k^2 \sigma_i  - k^2 \frac{E}{2} \right]_\pm &= 0 \,.
\end{align}
\end{subequations}
This discussion applies directly to cold dark matter perturbations and the first two momenta of the neutrino distribution $F_{\nu l}$, i.e., for $\delta_\nu = F_{\nu 0}$ and $\theta_{\nu} = (3/4) k F_{\nu 1} $. There are several special cases, though. 

First, for massless neutrinos (and photons) we need to solve, in general, the full Boltzmann hierarchy of higher momenta $F_{\nu l}$ where  $F_{\nu 2} \equiv 2 \sigma_i$ is the anisotropic stress. In practical terms this means that Eqs.~\eqref{eq:fluid_pert}  have to be supplemented with~\cite{Ma:1995ey}
\begin{subequations}
\label{eq:fluid_pert_2}
\begin{align}
 \dot{\sigma}_{\nu}  -\frac{4}{15} \left( \frac{ \theta_\nu}{a} + \frac{K^2}{a^2} {L} \right) + \frac{3}{10}  \frac{k}{a} F_{\nu 3} &=0  \;,\\
 \dot F_{\nu l} - \frac{k}{a} \frac{1}{(2l +1)} \left[ l F_{\nu (l-1)} - (l+1) F_{\nu (l+1)} \right] &=0 \;, \quad l \geq 3 \,.
\end{align}
\end{subequations}
Integrating these equations again over an infinitesimal interval implies the continuity $\left[ \sigma_\nu \right]_\pm = \left[ F_{\nu l} \right]_\pm = 0$ (for $l \geq 3$) provided the functions $F_{\nu l}$ are sufficiently regular. Taking the jump, we also find $[ \dot{\sigma}_\nu]_\pm = [\dot{F_{\nu l}}]_\pm =  0$. 

Second, in the case of photons and baryons (and electrons) the dynamical equations \eqref{eq:fluid_pert} (and \eqref{eq:fluid_pert_2} for photons) have to be equipped with interaction terms. However, it is easy to check that the additional terms do not change the matching conditions provided we make the same regularity assumptions as before. In summary, the matching relations \eqref{eq:matching_fluid} and \eqref{eq:matching_fluid_2} are valid for all fluid components as long as the equation of state is not changing during the transition. In the case of relativistic degrees of freedom, including photons and neutrinos, the higher momenta and their first time derivatives are continuous.

\section{Competitor Tables} \label{app:comp}

In this appendix, we provide the result tables from our analysis of DR, cADE and single-field EDE.  While Tab.~\ref{tab:means_comp} lists the means and their uncertainties, Tab.~\ref{tab:Delta_chi_2} displays the $\chi^2$ improvements with respect to $\Lambda$CDM.

\input{./tab/means_comp.tex}

\input{./tab/Delta_chi_2.tex}

\bibliography{NewEDE_PRD}
\end{document}

%% file: tab/means_NEDE_LCDM.tex
\begin{table}
\centering
\begin{footnotesize}
\begin{tabular}{|c|c|c|c|c|} 
 \hline \hline
{\bf Parameter} 			&  \multicolumn{2}{c|}{$\Lambda$CDM} 												&  \multicolumn{2}{c|}{NEDE}  \\ 
 					& w/ $H_0$ 								& w/o $H_0$ 							& w/ $H_0$ 							& w/o $H_0$ \\ 
\hline \hline
$100~\omega_{b }$ 		&$2.251_{-0.013}^{+0.014}$ ($2.251$)			&$2.240_{-0.014}^{+0.014}$ ($2.242$)		& $2.292_{-0.024}^{+0.022}$  ($2.297$) 		&  $2.271_{-0.020}^{+0.020}$ ($2.271$)\\ 
$\omega_\mathrm{cdm }$ 		&$0.1184_{-0.0009}^{+0.0009}$ ($0.1183$)		&$0.1194_{-0.0009}^{+0.0009}$ ($0.1194$)	& $0.1304_{-0.0035}^{+0.0034}$  ($0.1306$)	& $0.1261_{-0.0042}^{+0.0033}$ ($0.1254$) \\
$h$ 					&$0.6813_{-0.0041}^{+0.0041}$ ($0.6816$)		&$0.6763_{-0.0042}^{+0.0042}$  ($0.6764$)	& $0.714_{-0.010}^{+0.010}$ ($0.715$)		& $0.696_{-0.013}^{+0.010}$ ($0.695$)\\
$\ln10^{10}A_{s }$ 		&$3.053_{-0.016}^{+0.014}$ ($3.053$)			&$3.049_{-0.015}^{+0.013}$ ($3.050$)		& $3.067_{-0.015}^{+0.014}$ ($3.068$)		& $3.059_{-0.016}^{+0.014}$ ($3.058$)\\
$n_{s }$ 				& $0.9686_{-0.0037}^{+0.0037}$ ($0.9698$)		&$0.9661_{-0.0037}^{+0.0038}$ ($0.9672$)	& $0.9889_{-0.0066}^{+0.0067}$  ($0.9912$)	& $0.9792_{-0.0082}^{+0.0073}$  ($0.9794$)  \\ 
$\tau_{reio }$ 			&$0.0599_{-0.0078}^{+0.0071}$ ($0.0598$)		&$0.0570_{-0.0075}^{+0.0064}$($0.0572$)	& $0.0571_{-0.0077}^{+0.0068}$  ($0.0572$)	& $0.0562_{-0.0075}^{+0.0065}$ ($0.0558$) \\ 
$f_\text{NEDE }$ 		&--										&--									& $0.126_{-0.029}^{+0.032}$  ($0.1296$)		& $0.077_{-0.040}^{+0.038}$   ($0.072$)\\ 
$\log_{10}(m/m_0)$		&--										&--									& $2.56_{-0.10}^{+0.12}$ ($2.57$)			& 2.58 (fixed)\\
$H_*/m$				&--										&--									& 0.2 (fixed)							& 0.2 (fixed)\\
$w_\text{eff}^*$			&--										&--									& 2/3 (fixed)							& 2/3 (fixed)\\
$c_{s}^2$				&--										&--									& 2/3 (fixed)							& 2/3 (fixed)\\
\hline
$\sigma_8$ 			&$0.8090_{-0.0065}^{+0.0060}$ ($0.8092$)		&$0.8104_{-0.0061}^{+0.0057}$ ($0.8110$)	& $0.841_{-0.010}^{+0.010}$ ($0.841$)		& $0.828_{-0.012}^{+0.0010}$ ($0.826$)\\ 
$S_8$ 				&$0.814_{-0.010}^{+0.010}$ 					&$0.824_{-0.010}^{+0.010}$ 				& {$0.841^{+0.012}_{-0.012}$} 		 		&  {$0.837_{-0.012}^{+0.012}$} 	\\ 
$r_s^{d} $ [Mpc] 		& $147.40_{-0.23}^{+0.23}$ ($147.38$)			&$147.20_{-0.23}^{+0.23}$ ($147.19$)		&  $141.0_{-1.7}^{+1.6}$   ($140.9$)			& $143.5_{-1.8}^{+2.1}$ ($143.8$)\\ 
$z_{*} $				&--										&-- 									& $4920_{-730}^{+620}$ ($4960$)			&  $5100_{-70}^{+80}$  ($5110$)\\ 
\hline 
Tension SH$_0$ES 		& --										&$4.3 \, \sigma$						&--									&$ 2.5 \, \sigma$\\ 
Tension S$_8$ 			& $1.9 \, \sigma$ 							& $2.3 \, \sigma$ 						&$2.8 \, \sigma$ 						& $2.7 \, \sigma$\\
$\Delta \chi^2 	$		& 0										&0									& -15.6								& -2.9\\
$f_{\text{NEDE}} \neq 0$	& --										& --									& $4.3 \, \sigma$						& $1.9 \, \sigma$ \\ 
\hline \hline
 \end{tabular} \\ 
 \end{footnotesize}
\caption{ The mean value and $\pm 1 \,\sigma$ error (with bestfit value in parentheses) of the cosmological parameters from our combined analyses for $\Lambda$CDM and NEDE with and without imposing the SH$_0$ES prior on $H_0$. }
\label{tab:means_NEDE_LCDM}
\end{table}

%% file: tab/chi2.tex
\begin{table}
\centering
\begin{footnotesize}
\begin{tabular}{|l|c|c|c|c|c|c|c|c|} 
 \hline 
 \hline
{\bf Dataset} & \multicolumn{2}{c|}{$\Lambda$CDM} & \multicolumn{2}{c|}{NEDE} & \multicolumn{4}{c|}{NEDE Extensions}\\ 
\hline
\multicolumn{1}{|r|}{SH$_0$ES} 	& w/ $H_0$ 	& w/o $H_0$ 	& w/ $H_0$ 	& w/o $H_0$	& w/ $H_0	$				& w/ $H_0$						& w/ $H_0$				&	w/ $H_0$\\  
\multicolumn{1}{|r|}{Extension} 		&		--	&		--	&		--	&		--	&	$\scriptstyle{H_*/m}$		&$\scriptstyle{w_\text{NEDE},\, c_s}$	&$\scriptstyle{ c_\text{vis}}$	&	FS\footnote{{For this run, NEDE is assumed to be free-streaming radiation with $w_\text{NEDE} = c_s^2 = c_\text{vis}^2=1/3$.}}		\\
\multicolumn{1}{|r|}{$\Delta$DOF}	&0			&0			& 2			&1 (2)\textsuperscript{\ref{fn:prior}}		&	2 (3)\footnote{\label{fn:prior}We fixed $\log_{10}(m/m_0) $ close to its base model bestfit value detailed in Tab.~\ref{tab:means_NEDE_LCDM} to avoid prior volume artifacts; see discussion in Sec.~\ref{sec:metholdology}. This leads to a slight underestimate of the fit improvement.} &4			& 3&2\\
							 \hline\hline
{\emph{Planck}} high-$\ell$ TT, TE, EE & 2,348.9 	& 2,347.1 		& 2,348.2 		&{2345.9}		&2348.6	&2348.4	&2348.7	&2352.1\\  
{\emph{Planck}} low-$\ell$ TT		& 22.7 		& 23.1 		& 20.8 		&21.7		&20.8	&20.7	&20.8	&21.4\\  
{\emph{Planck}} low-$\ell$ EE 		& 397.3 		& 396.7 		& 396.4 		&396.2		&396.1	&396.3	&396.3	&396.8\\  
{\emph{Planck}} lensing 			& 9.1 		& 8.8 		& 9.5 		&9.1			&9.5		&9.6		&9.5		&9.3\\  
 BAO low-$z$					& 1.6 		& 1.2			& 1.8 		&1.4			&1.8		&1.7		&1.8		&1.8\\  
 BAO high-$z$  + LSS				& 5.9 		& 6.7 		& 6.8 		&6.5			&6.8		&6.9		&6.8		&6.3\\  
 Pantheon 					& 1,027.0 		& 1,027.3		&1,027.3 		&1027.1		&1027.3	&1027.0	&1027.0	&1027.1\\  
 BBN 						& $<0.1$ 		& $<0.1$ 		&$<0.1$ 		&$<0.1$		&$<0.1$	&$<0.1$	&$<0.1$	&$<0.1$\\  
 SH$_0$ES 					& 17.1 		& -- 			& 3.3			& --			&3.1		&2.5		&3.1		&9.4\\ \hline \hline 
 $\chi^2$ 						& 3,829.6 		& 3,810.7 		& 3,814.0 		&3,807.8 		&3,813.9	&3,813.1	&3,814.0	&3,824.3\\
 $\Delta \chi^2 $ 				& -- 			& --			& -15.6 		& -2.9		&-15.6	&-16.5	&-15.6	&-5.3\\ \hline \hline
\end{tabular} \\
\end{footnotesize}
\caption{The bestfit $\chi^2 = - 2 \ln(\mathcal{L})$ for each dataset from combined analysis with and without SH$_0$ES. The quality of the overall fit is assessed in terms of $\Delta \chi^2 = \chi^2 (\text{NEDE}) - \chi^2(\Lambda\text{CDM})$, corresponding to the ratios between the maximum likelihoods $\mathcal{L}$. }
\label{tab:chi2}
\end{table} 

%% file: tab/means_comp.tex
\begin{table}[ht]
\centering
\begin{footnotesize}
\begin{tabular}{|c|c|c|c|c|c|c|} 
 \hline \hline
{\bf Parameter} 					&  \multicolumn{2}{c|}{Dark Radiation} 							&  \multicolumn{2}{c|}{cADE} 										&  \multicolumn{2}{c|}{EDE ($n=3$)}  \\ 
 							& w/ $H_0$ 					& w/o $H_0$ 					& w/ $H_0$ 					& w/o $H_0$						& w/ $H_0$ 					& w/o $H_0$ \\ 
\hline \hline
$100~\omega_{b }$ 				&$2.265_{-0.016}^{+0.016}$ 		&$2.236_{-0.017}^{+0.017}$ 		&$2.287_{-0.023}^{+0.023}$ 		&$2.257_{-0.017}^{+0.016}$			& $2.276_{-0.021}^{+0.020}$		& $2.263_{-0.017}^{+0.017}$\\ 
$\omega_\mathrm{cdm }$ 				&$0.1225_{-0.0024}^{+0.0024}$	&$0.1186_{-0.0026}^{+0.0025}$	&$0.1262_{-0.0029}^{+0.0030}$	&$0.1234_{-0.0030}^{+0.0021}$		& $0.1304_{-0.0037}^{+0.0037}$ 	& $0.1262_{-0.0041}^{+0.0032}$\\ 
$h$ 							&$0.6949_{-0.0086}^{+0.0084}$ 	&$0.6732_{-0.0100}^{+0.0096}$	&$0.7015_{-0.0081}^{+0.0086}$	& $0.6894_{-0.0093}^{+0.0068}$		& $0.713_{-0.010}^{+0.010}$ 		&$0.696_{-0.012}^{+0.010}$ \\ 
$\ln10^{10}A_{s }$ 				&$3.062_{-0.016}^{+0.014}$		&$3.047_{-0.017}^{+0.015}$		&$3.065_{-0.015}^{+0.015}$		&$3.056_{-0.015}^{+0.014}$			& $3.069_{-0.016}^{+0.014}$ 		& $3.060_{-0.016}^{+0.014}$\\ 
$n_{s }$ 						&$0.9761_{-0.0055}^{+0.0055}$	&$0.9642_{-0.0062}^{+0.0062}$	&$0.9821_{-0.0064}^{+0.0065}$	& $0.9731_{-0.0061}^{+0.0051}$		& $0.9867_{-0.0068}^{+0.0066}$ 	&$0.9775_{-0.0068}^{+0.0064}$ \\ 
$\tau_{reio }$ 					&$0.0593_{-0.0082}^{+0.0066}$	&$0.0571_{-0.0076}^{+0.0068}$	&$0.0574_{-0.0078}^{+0.0066}$	&$0.0558_{-0.0074}^{+0.0063}$		& $0.0576_{-0.0078}^{+0.0066}$ 	&$0.0568_{-0.0077}^{+0.0064}$ \\  
$f_{\text{ADE}}$, $f_{\text{EDE}}$ 	& -- 							&		--					&$0.073_{-0.023}^{+0.026}$ 		& $0.038_{-0.027}^{+0.018}$			& $0.109_{-0.027}^{+0.031}$ 		&$0.066_{-0.033}^{+0.032}$\\  
$\Delta N_{\text{eff}}$			& {$0.25_{-0.13}^{+0.13}$}		&$-0.05_{-0.15}^{+0.14}$ 			& --							&--								&--							&--\\ 
$\phi_\text{ini}$					&		--					&			--				&		--					&		--						& $2.705_{-0.072}^{+0.13}$		& $2.637$ (fixed)\\
\hline
$\sigma_8$ 					&$0.8204_{-0.0091}^{+0.0083}$	& $0.8080_{-0.0094}^{+0.0088}$	& $0.833_{-0.010}^{+0.010}$ 		&  $0.8226_{-0.0100}^{+0.0087}$	 	&  $0.838_{-0.010}^{+0.010}$ 		& $0.8269_{-0.0110}^{+0.0093}$ \\ 
$S_8$ 						&$0.821_{-0.011}^{+0.011} $		&$0.823_{- 0.011}^{+0.011}$		& $0.837_{-0.013}^{+0.013} $		& $0.834^{+0.012}_{-0.012} $ 			& $0.839_{-0.012}^{0.012}$		&$0.836^{+0.012}_{-0.012}      $\\
$r_s^{d} $ [Mpc] 				&$144.9_{-1.3}^{+1.3}$			&$147.8_{-1.5}^{+1.5}$ 			& $143.1_{-1.5}^{+1.4}$ 			& $145.0_{-1.1}^{+1.6}$ 				& $141.1_{-1.8}^{+1.7}$			& $143.5_{-1.8}^{+2.0}$ \\ 
$ z_{*} $						& -- 							& --							& {$3230^{+400}_{-500}$}			& $2880 $ (fixed)					&$3760^{+320}_{-410} $			&3780 (fixed)\\ 
\hline 
Tension SH$_0$ES 				& -- 							&	$3.9 \, \sigma$				&--							&$3.2 \, \sigma$					&			--				&$2.5 \, \sigma$\\  
Tension S$_8$ 					& $2.2 \, \sigma$  				& $2.2 \, \sigma$ 				& $2.7 \, \sigma$  				& $2.6 \, \sigma$  					& $2.8 \, \sigma$  				& $2.7 \, \sigma$  \\  
$\Delta \chi^2$ 					& -3.2  						&		-0.2					&-9.9						&-1.0							&-16.4						&-2.9\\ 
$f_{(\circ)}$, $\Delta N_\text{eff} \neq 0$	&  $1.9 \, \sigma$			&	$0.4 \, \sigma$				&$3.2 \, \sigma$ 				&$ 1.4\,  \sigma$					&$4.0\, \sigma$					&$2.0 \, \sigma$\\  
\hline \hline
 \end{tabular} \\ 
 \end{footnotesize}
\caption{The mean value and $\pm 1 \,\sigma$ error of the cosmological parameters from our combined analyses with and without  imposing the SH$_0$ES prior on $H_0$ for dark radiation, cADE and single-field EDE.}
\label{tab:means_comp}
\end{table}

%% file: tab/Delta_chi_2.tex

\begin{table}[ht]
\centering
\begin{footnotesize}
\begin{tabular}{|l|c|c|c|c|c|c|c|c|} 
 \hline 
 \hline
{\bf Dataset} & \multicolumn{2}{c|}{Dark Radiation} & \multicolumn{2}{c|}{cADE} & \multicolumn{2}{c|}{EDE}  &  \multicolumn{2}{c|}{NEDE}  \\ 
 \multicolumn{1}{|r|}{SH$_0$ES} 	& w/ $H_0$ 	& w/o $H_0$ 	& w/ $H_0$ 	& w/o $H_0$	& w/ $H_0$ 	& w/o $H_0$	& w/ $H_0$ 	& w/o $H_0$  \\ 
 \multicolumn{1}{|r|}{$\Delta$DOF} 	&	1		&	1 		& 2		& 1 (2)\textsuperscript{\ref{fn:prior2}}			& 3			&	1 (3)\textsuperscript{\ref{fn:prior2}}		&2		& 1 (2)\footnote{ \label{fn:prior2} We  varied  $f_{(\circ)}$ (after setting the other parameters close to their bestfit) only to avoid prior volume artifacts; see discussion in Sec.~\ref{sec:metholdology}. }\\
\hline \hline
{\emph{Planck}} high-$\ell$ TT, TE, EE & 3.2		& -0.3 	& 2.9 	&0.0 		&-2.3	& -1.7	& -0.8	&-1.2\\  
{\emph{Planck}} low-$\ell$ TT 		& -0.7		& 0.2 	& -1.5 	&-0.6 	&-1.5 	& -1.2	& -1.9	&-1.4\\  
{\emph{Planck}} low-$\ell$ EE 		& -0.4 		& -0.4 	& -1.0 	&0.8		&-1.0	&-0.3	&-0.9	&-0.5\\  
{\emph{Planck}} lensing 			& 0.1			& 0.0 	& 0.2 	&0.1		&0.4		&0.2		&0.4		&0.3\\  
 BAO low-z 					& 0.2 		&  -0.1	& 0.0 	&0.0		&0.1		&0.2		&0.1		&0.2\\  
 BAO high-z  	+ LSS			& 0.2 		& 0.1 	& 0.8 	&0.1		&1.0		&-0.2	&0.9		&-0.2\\  
 Pantheon 					& -0.1 		& 0.2		& 0.0		&0.1		&1.5		&-0.2	&0.3		&-0.1\\ 
 BBN 						& 0.8			& 0.0		& 0.0 	&0.0		&0.2		&0.2		&0.0		&0.0\\ 
 SH$_0$ES 					& -6.4		&  -- 		& -11.3  	&--		&-14.7	&--		&-13.8	&--\\ \hline \hline
 Total $\Delta \chi^2 $ 			& -3.2 		& -0.2	& -9.9 	&-1.0	&-16.4	&-2.9	&-15.6	&-2.9\\
  \hline \hline
\end{tabular} \\
\end{footnotesize}
\caption{The bestfit $\Delta \chi^2$ with $\Lambda$CDM as baseline model from combined analysis with and without SH$_0$ES.  The $\Lambda$CDM reference values are provided in Tab.~\ref{tab:chi2}.   }
\label{tab:Delta_chi_2}
\end{table}